\documentclass{aa}
\usepackage{multirow}
\usepackage{amsmath}
\usepackage[colorlinks,linkcolor=blue,anchorcolor=blue,citecolor=blue]{hyperref}
\usepackage{graphicx} 
\usepackage{lscape}
\usepackage{longtable}
\usepackage{txfonts}
\usepackage{natbib}

\begin{document}

\title{Systematic search for lensed X-ray sources in the CLASH fields}

\author{
Ang Liu\inst{1},
Paolo Tozzi\inst{1},
Piero Rosati\inst{2,3},
Pietro Bergamini\inst{3},
Gabriel Bartosch Caminha\inst{4},
Roberto Gilli\inst{3},
Claudio Grillo\inst{5},
Massimo Meneghetti\inst{3}, 
Amata Mercurio\inst{6},
Mario Nonino\inst{7},
Eros Vanzella\inst{3}
}

\institute{
INAF - Osservatorio Astrofisico di Arcetri, Largo E. Fermi, I-50125 Firenze, Italy \\
\email{paolo.tozzi@inaf.it; liuang@mpe.mpg.de}
\and
Dipartimento di Fisica e Scienze della Terra, Universit\'a degli Studi di Ferrara, 
via Saragat 1, I-44122 Ferrara, Italy
\and
INAF - Osservatorio di Astrofisica e Scienza dello Spazio, via Pietro Gobetti 93/3, 40129 Bologna, Italy
\and
Max-Planck-Institut für Astrophysik, Karl-Schwarzschild-Str. 1, D-85748 Garching, Germany 
\and 
Dipartimento di Fisica, Università degli Studi di Milano, via Celoria 16, I-20133 Milano, Italy
\and
INAF-Osservatorio Astronomico di Capodimonte, Salita Moiariello 16, I-80131 Napoli, Italy
\and
INAF - Osservatorio Astronomico di Trieste, via G. B. Tiepolo 11, I-34143 Trieste, Italy
}

\titlerunning{Lensed X-ray sources in CLASH}
\authorrunning{Liu et al.}

\abstract
   {}
{We exploit the high angular resolution of {\sl Chandra} to search for unresolved X-ray emission 
from lensed sources in the field of view of 11 CLASH clusters, whose critical 
lines and amplification maps were previously obtained with accurate strong-lensing models. 
We consider a solid angle in the lens plane corresponding to a magnification 
$\mu>1.5$, which amounts to a total of $\sim 100$ arcmin$^2$, of which only $10$\% corresponds
to $\mu>10$. Our main goal is to assess the efficiency of massive clusters as 
cosmic telescopes to explore the faint end of the X-ray extragalactic source population.}
{The main obstacle to this study is the overwhelming diffuse X-ray emission from the intracluster medium that encompasses the region with the strongest magnification 
power.  To overcome this aspect, we first searched for X-ray emission from strongly 
lensed sources that were previously identified in the optical and then performed an untargeted detection of
lensed X-ray sources. }
{We detect X-ray emission in either in the soft (0.5--2 keV) or
hard (2--7 keV) band in only 9 out of 849 lensed or background optical sources.
The stacked emission of the sources without detection 
does not reveal any signal in any band.
Based on the untargeted detection in the soft, hard, and total band images, 
we find 66 additional X-ray sources without
spectroscopic confirmation that are consistent with being lensed (background) sources.   
Assuming an average redshift distribution consistent with the Chandra Deep Field South
survey (CDFS), we estimate their magnification, and after accounting for 
completeness and sky coverage, 
measure the soft- and hard-band number counts of lensed X-ray sources for the first time.  
The results are consistent with current modeling of the population distribution of active galactic nuclei (AGN).  The distribution 
of delensed fluxes of the sources identified in moderately deep CLASH fields 
reaches a flux limit of $\sim 10^{-16}$ and $\sim 10^{-15}$ erg/s/cm$^{2}$ 
in the soft and hard bands, respectively, therefore approximately 1.5 orders of magnitude 
above the flux limit of the CDFS. } 
{We conclude that in order to match the depth of the CDFS in exploiting
massive clusters as cosmic telescopes, the required number of cluster fields 
is about two orders of magnitude larger than is offered by the 20-year {\sl Chandra} archive. 
At the same time, the discovery of strongly lensed sources close to the critical lines remains an attractive if rare occurrence because the source density in the X-ray sky is low.
A significant step forward in this field will be made when future X-ray facilities an angular resolution of $\sim 1$ arcsec  and a large effective area will allow 
the serendipitous discovery of rare, strongly lensed 
high-$z$ X-ray sources. This will enable studying faint AGN activity in the early Universe and measuring gravitational time delays in the X-ray variability of multiply imaged AGN.}


\maketitle


\section{Introduction}

Strong lensing in massive clusters has been investigated thoroughly in the past ten years. 
High-resolution images from the Hubble Space Telescope (HST) allowed identifying multipe-image families of lensed background galaxies in all clusters that were observed with sufficient 
depth in the optical band. Eventually, spectroscopic follow-up observations provided the 
accurate redshifts of lensed sources, which are required to fully exploit the potential of strong 
lensing, specifically in three diverse science cases: i)  
reconstructing the total mass distribution in galaxy clusters, ii) identifying and characterizing 
intrinsically faint but highly magnified sources at high redshifts, and iii) constraining  
cosmological parameters by measuring the global geometry of the Universe \citep[for a review,
see][]{2011Kneib}.  The weak-lensing regime, at distances larger than 
2--3 arcmin from the cluster center, is also widely used to recover the cluster mass distribution 
at large radii and to statistically constrain cosmological parameters.

We focus on the role of massive clusters as {\sl \textup{cosmic telescopes}}. 
In regions close to the critical lines, the magnification by strong lensing 
enables the investigation of distant and faint objects that would have been 
impossible to detect in blank fields (i.e., without intervening massive systems). 
Deep surveys in the fields of massive clusters are very efficient in finding candidates 
of the most distant galaxies in the Universe up to redshift $z \sim 9$ 
\citep{Coe2013,Bouwens2014,Zitrin2014,Salmon2018}. In addition, studies of background 
faint galaxies at lower redshifts $3<z<6$ 
\citep{Caminha2017b,Vanzella2017a,Vanzella2017b,patricio2018,Vanzella2019} 
have also greatly benefited from the higher spatial resolution and amplified flux
produced by the gravitational lensing effect. This led to some important insights 
into the evolution and characterization of galaxies and has indicated that this optically faint 
population might be important for the reionization of the Universe 
\citep{yue2014,bouwens2015,robertson2015}.

The accurate reconstruction of the intrinsic properties of the lensed sources 
heavily relies on the model of the mass distribution of the lens in order to compute precise 
and high-resolution mass and magnification maps 
\citep[see, e.g.,][]{Grillo2015,Johnson2016,Caminha2017a,Caminha2017b,2017Meneghetti}.
To this aim, it is crucial to have the spectroscopic redshift of the multiple-image families. 
Therefore several extensive spectroscopic campaigns have targeted subsamples of 
massive clusters to simultaneously identify a large number of 
high-redshift candidates, multiple images, and cluster members. 
In this respect, a recent major improvement has been made possible by the Multi 
Unit Spectroscopic Explorer \citep[MUSE, see][]{Bacon2010,Bacon2014} on the\ Very Large Telescope (VLT), which has 
revolutionized these studies and allowed increasing the number of multiply lensed sources 
with respect to those identified in HST images. Its efficiency, field of view of
$1\arcmin\times1\arcmin$, and capability of detecting faint emission lines out to $z\sim 6.7$ 
without any source preselection are employed to expand the spectroscopic
confirmation of multiply lensed sources and cluster members that are then used to constrain 
lens models. The systematic investigation of MUSE data cubes has been successfully applied to 
CLASH \citep[Cluster Lensing And Supernova survey with Hubble,][]{2012Postman} clusters, providing new detections of 
strongly lensed sources and a much higher accuracy in the 
reconstruction of the mass profile, including the position of the critical 
lines and the corresponding magnification as a function of redshift 
\citep[see][]{Grillo2015,Caminha2017a,Caminha2017b,Lagattuta2017,Mahler2018,Jauzac2019,2019Caminha}.

In addition to detailed studies of single, peculiar, and rare sources, it is possible to probe the 
faint end of the high-$z$ luminosity functions of galaxies and push the detection limits 
below the current limits that can be achieved in blank fields with current instrumentation 
\citep{2015Atek,2018Atek,Bouwens2017}. The achromatic nature of gravitational lensing enables searches for lensed sources at any wavelength simply by mapping critical 
lines in well-modeled lensing clusters. The critical-line mapping strategy has been 
successfully applied, for example, in the submillimeter (submm) regime, also because cluster 
member galaxies are not strong submm sources and therefore are transparent to this wavelength. 
The same happened in the mid-IR, where the imaging of lensing clusters with ISO 
pierced through the IR confusion limit and led to the discovery of many lensed sources 
at 70\,$\mu$m \citep[][]{2003Metcalfe}. This revealed a positive evolution with redshift of this 
population, as also confirmed at 100 and 160 $\mu$m with {\sl Herschel} deep imaging of 
massive clusters \citep{2010Altieri}.  

The critical-line mapping strategy can also be applied to high energies such as the X-ray 
band (0.5--10 keV), which is currently accessible mostly with the {\sl Chandra} and {\sl XMM-Newton} observatories. However, several factors made it particularly challenging. The first problem is the 
strong diffuse emission from the hot intracluster medium (ICM) in the core of any massive 
galaxy cluster, which typically encompasses the critical lines. However, because the ICM is optically 
thin, the emission of background sources must still be identified, despite a strong foreground. 
The second problem is that the angular resolution, which is key to increasing the sensitivity,
is usually very limited in the high-energy domain.  The only facility that reaches
subarcsecond resolution is the {\sl Chandra} X-ray telescope. Its $\sim 1$ arcsec angular 
resolution at the aim point and $\leq 2 $ arcsec within 3 arcmin nicely fits the size of 
the critical line regions in massive clusters.  Finally, a third problem is that the density on the sky of X-ray sources 
is far lower than in optical and near-IR sources, therefore the 
probability of crossing a critical line by a background AGN or a star-forming galaxy is much 
lower, and the statistics to measure the depletion effect require many massive clusters 
in order to accumulate sufficient solid angle in the source plane. The effect of cluster 
lensing on the background X-ray sources was discussed for the first time by 
\citet[see][]{1996Refregier,1997Refregier} before the launch of {\sl Chandra}. 
So far, however, a systematic search of lensed or background sources in the 
field of massive cluster has never been attempted, and therefore this effect
has never been observed. 

On the other hand, the interest in rare, strongly lensed X-ray sources 
has recently been revived by the detection of an unresolved star-forming region 
in a highly magnified galaxy in the midst of the ICM X-ray emission of 
the Phoenix cluster at redshift $z\sim 0.6$ \citep{2020Bayliss}.
The lensed galaxy 
at $z\sim 1.52$  is a low-mass, low-metallicity starburst with high X-ray emission, 
likely an analog to the first generation of galaxies.  This exciting finding has placed 
more pressure on searching for such targets. Another source of interest in strongly lensed
X-ray source multiple images is the time variability, which is ubiquitous in AGN at any flux
\citep[see][]{2017Paolillo}. The time-delay measurement among different counterparts of the
same source can be used to constrain the cosmological parameters \citep[see][]{2009Coe}. 
Such studies are currently performed with a small number of known X-ray quasars that are strongly lensed by 
single galaxies \citep[see, e.g.,][]{2012Chen}, and in only two cases, by group-size halos
\citep{2006Ota,2012Ota}. Clearly, extending these studies at fainter fluxes
would dramatically increase the number of lines of sight to be probed and thus improve
the constraints on cosmological parameters.  Massive clusters naturally provide 
a much larger cross section in a single field of view and therefore appear to be more convenient 
for the search of strongly lensed X-ray sources.  

We present for the first time a systematic search for lensed X-ray 
sources in the field of view of massive clusters, with the aim of exploring the potential 
use of cosmic telescopes in finding rare, multiply lensed objects for cosmological tests, 
and to constrain the faint end of the extragalactic X-ray source population. 
Among the most relevant observational 
programs aiming at employing massive clusters as cosmological telescope, we find the CLASH \citep[][]{2012Postman}, the Hubble Frontier 
Fields \citep[HFF,][]{lotz2017}, and more recently, the REionisation Lensing Cluster Survey 
\citep[RELICS,][]{coe2019,salmon2020}, all based on deep imaging with the HST, whose angular 
resolution is key to identifying faint features associated with lensed galaxies. CLASH 
clusters also have high angular resolution {\sl Chandra} imaging with medium-deep exposures 
and {\sl XMM-Newton} observations, mostly focused on the diffuse ICM emission 
to reconstruct the hydrostatic mass profile. {\sl XMM-Newton} data are, however, not as helpful in searching for unresolved X-ray sources at the faintest level because the angular resolution is relatively low. Therefore we chose to perform
a systematic search for lensed X-ray sources for the first time using the {\sl Chandra} archival data of selected CLASH clusters
for which we have developed high-precision strong-lensing models.

The paper is organized as follows. In \S 2 we briefly describe the CLASH sample; the 
optical data set, including the recent results from the MUSE data cubes, which provided an 
updated list of multiply lensed sources; and the X-ray data set, including the details of 
the data reduction.  In \S 3 we summarize the lensing analysis of each cluster that was performed in previous works, and describe the magnification maps used in this work. 
In \S 4 we focus on the optically identified multiply lensed sources to search for X-ray 
emission from single sources or their average emission with the stacking technique. 
In \S 5 we perform untargeted detection of X-ray sources, and carefully quantify the 
completeness and the sky coverage of our sample of new lensed X-ray source candidates. 
In \S 6 we compute the delensed log$N$-log$S$ in the soft and hard bands and compare it to the
deepest results to date  \citep[the Chandra Deep Field South survey, hereafter CDFS,][]{2017Luo}
and recent modeling \citep{2007Gilli}.
In \S 7 we discuss perspectives for similar studies with the current
{\sl Chandra} archive, and future perspectives 
with next-generation X-ray missions.  Finally, in \S 8 we summarize our results.
Throughout this paper, we adopt the seven-year WMAP cosmology with $\Omega_{\Lambda} =0.73 $, 
$\Omega_m =0.27$, and $H_0 = 70.4$ km s$^{-1}$ Mpc$^{-1}$ \citep{2011Komatsu}. Quoted error 
bars correspond to a 1 $\sigma$ confidence level unless noted otherwise.


\section{Optical and X-ray data}

\subsection{CLASH sample: Optical imaging and spectroscopy}

\begin{table}
\caption{Subsample of CLASH clusters. }
\label{clashsample}
\begin{center}
\begin{tabular}[width=0.5\textwidth]{lcccc}
\hline\hline
Cluster & Redshift & $N_{\rm src}$ & $N_{\rm image}$ & $N_{\rm bk}$ \\
\hline
Abell S1063$^2$  & 0.351 & 20 & 55 & 41 \\
MACS J0329.7-0211$^1$  & 0.450 & 9 & 23 & 19 \\
MACS J0416.1-2403$^3$ & 0.397 & 66 & 182 & 59 \\
MACS J0429.6-0253$^1$  & 0.399 & 3 & 11 & 41 \\
MACS J1115.9+0129$^1$  & 0.352 & 3 & 9 & 30 \\
MACS J1206.2-0847$^2$ & 0.441 & 27 & 82 & 23 \\
MACS J1311.0-0310$^1$  & 0.494 & 3 & 8 & 17 \\
MACS J1931.8-2635$^1$  & 0.352 & 7 & 19 & 16 \\
MACS J2129.4-0741$^1$  & 0.587 & 11 & 38 & 40 \\
RX J1347.5-1145$^1$  & 0.451 & 8 & 20 & 18 \\
RX J2129.7+0005$^1$ & 0.234 & 7 & 22 & 76 \\
\hline
\end{tabular}
\tablefoot{The lensing analysis, cluster redshifts, and the number of identified 
counterparts of the
lensed sources are from \citet{2019Caminha} (1), \citet{Bergamini2019} (2), and \citet{2021Vanzella} (3). 
$N_{\rm src}$ is the number of unique sources (single galaxies or resolved clumps
within a galaxy) that are multiply lensed, and 
$N_{\rm image}$ is the total number of images of the lensed sources identified in each field.
In the last column we list the number of additional background sources in the FOV 
that do not have multiple counterparts. }
\end{center}
\end{table}

CLASH \citep{2012Postman} is a 524-orbit multicycle treasury program with the Hubble
Space Telescope to use the gravitational lensing properties of 25 galaxy clusters to
address mainly two science goals: constraining the baryonic mass and dark matter 
distributions in the cluster cores and in the outskirts, and finding highly magnified 
high-$z$ galaxies. A total of 16 broadband filters, spanning the near-UV to near-IR, 
were employed for a 20-orbit campaign on each cluster. All the clusters were also observed 
in the mid-IR with Herschel, in the NIR with Spitzer, and in 
the optical with Subaru/Suprime-Cam, and they were also intensively followed up in the optical 
band to obtain detailed spectra and to securely confirm member galaxies and lensed sources.  
In particular, a panoramic spectroscopic campaign of 13 southern CLASH clusters was 
performed with the VIsible Multi-Object Spectrograph (VIMOS) in a VLT Large program 
(PI P. Rosati) covering $\sim 0.1$ deg$^2$ in each cluster \citep{2015Grillo,2016Balestra,2017Monna}.  
We used a subset of CLASH clusters 
for which a significant number of new confirmed multiple images discovered 
by MUSE/VLT spectroscopy were used to constrain the lens models 
\citep{2019Caminha,Bergamini2019,2020Meneghetti,2021Bergamini,2021Vanzella}. The clusters used in this work and their redshifts are listed in Table \ref{clashsample}.  Our cluster sample 
spans the redshift and virial mass ranges of $0.234<z<0.587$ and $4.6\times 10^{14}<M_{200}<35.4
\times 10^{14} \, M_\odot$,  as obtained from weak-lensing measurements
\citep{2014Meneghetti,2015Merten,2018Umetsu}. With the exception of M2129, R1347, and M0416,
these clusters were selected within the CLASH survey to be
dynamically relaxed based on {\sl Chandra} X-ray observations.  

In our search for X-ray lensed sources, we started from the list of 
sources with redshifts that were optically identified as strongly lensed with multiple images
or simply in the cluster background and therefore potentially magnified to some extent.
The number of unique multiply lensed sources and the total number of images in each field 
is listed in columns 3, and 4, respectively, in Table \ref{clashsample}.
We remark that a {\sl \textup{unique source}}  can be a single galaxy or a resolved clump 
within a galaxy. 
Overall, in 11 fields, 164 multiply lensed unique sources are identified in the optical 
for a total of 469 counterparts. These were used to constrain the cluster mass 
distribution. In addition, from a collection of published redshift catalogs from the CLASH-VLT survey \citep{2013Biviano,2017Monna,Caminha2016,2016Balestra,2017Karman,2019Caminha,Bergamini2019}, we obtained a number of sources with redshifts higher than the 
cluster redshift for each cluster.  This 
subsample of background sources also includes the highly magnified arcs identified in the HST images.  
We selected only those background sources that were potentially relevant to our study.
First, we focused on a field with a size of
$300\arcsec \times 300\arcsec$ (25 arcmin$^2$), then we considered 
only the sources with $z_{\rm s}>z_{\rm cl}+0.1$ and $\mu_{\rm max}>1.5$, where $\mu_{\rm max}$ is the 
maximum magnification expected at the source position.  
This last step ensured that we did not include 
background sources with low magnification (see Sect. 4 for a more detailed discussion).
A total of 380 background sources satisfied these criteria 
and were not included in the list of sources with multiple images, which are distributed in the 
11 CLASH fields as listed in column 5 of Table \ref{clashsample}.
This means a total of 849 optically determined positions in 11 fields 
in which we searched for magnified X-ray emission.

\begin{table*}
\caption{{\sl Chandra} observations. }
\footnotesize
\label{TableX}
\begin{center}
\begin{tabular}[width=0.9\textwidth]{lcccc}
\hline\hline
Cluster name (abbreviation) & ObsID & Exp. & ${\rm ecf}_{\rm soft}$, ${\rm ecf}_{\rm hard}$ [$\Gamma=1.4,\, 1.8$] & $n_{\rm H}$ \\
 &  & [ks] & [$10^{-12} ({\rm erg/s/cm}^{2})/({\rm cts/s})$] & [$10^{20} {\rm cm}^{-2}$] \\
\hline
Abell S1063 (AS1063) & 4966, 18611, 18818 & 123.6 & [9.0--9.7], [31.0--27.2] & 1.29 \\
MACS J0329.7-0211 (M0329) & 3257, 3582, 6108 & 68.5 & [6.7--7.1], [28.9-25.2] & 5.56 \\
MACS J0416.1-2403 (M0416) & 10446, 16236, 16237, 16304, 16523, 17313 & 323.0 & [8.5--9.2], [30.2--26.5] & 3.49 \\
MACS J0429.6-0253 (M0429) & 3271 & 22.6 & [6.4--6.8], [28.9--25.3] & 5.07 \\
MACS J1115.9+0129 (M1115) & 3275, 9375 & 52.9 & [6.9--7.3], [29.1--25.4] & 4.92 \\
MACS J1206.2-0847 (M1206) & 3277, 20544, 20929, 21078, 21079, 21081 & 199.4 & [10.5--11.6], [30.5--26.8] & 5.33 \\
MACS J1311.0-0310 (M1311) & 3258, 6110, 9381 & 44.3 & [6.5--6.8], [29.0--25.4] & 1.98\\
MACS J1931.8-2635 (M1931) & 3282, 9382 & 111.1 & [7.9--8.5], [29.5--25.9] & 11.9 \\
MACS J2129.4-0741 (M2129) & 3199, 3595 & 37.8 & [6.7--7.1], [29.0--25.4] & 5.27 \\
RX J1347.5-1145 (R1347) & 506, 507, 3592, 13516, 13999, 14407 & 232.5 & [9.3--9.8], [36.2--31.7] & 5.83 \\
RX J2129.7+0005 (R2129) & 552, 9370 & 39.2 & [6.8--7.2], [29.5--25.8] & 4.21 \\
\hline
\end{tabular}
\tablefoot{In parentheses we list the short
name of each target that we used in the text for simplicity.  
All the observations were taken with ACIS-I, except for ObsIDs 506 and 507, which were 
taken with ACIS-S. The exposure time in the third column refers to the total exposure after 
data reduction. The energy conversion factor ${\rm ecf}$ from the observed count–rate to the 
unabsorbed flux is measured for the soft 
(0.5--2 to 0.5--2 keV) and hard (2--7 to 2--10 keV) bands for a power law with $\Gamma =1.4$
and $1.8$, absorbed by the Galactic column density measured in each field. This is shown in the last column.}
\end{center}
\end{table*}

\subsection{X-ray data reduction}

The X-ray data reduction was performed using the software {\tt CIAO} 
v4.12, with the latest release of the {\sl Chandra} 
calibration database at the time of writing ({\tt CALDB  v4.9}). 
Time intervals with a high background were filtered out by performing a 
3$\sigma$ clipping of the background level on the light curve, which was 
extracted in the 2.3--7.3 keV band, and binned with a time interval of 200 s. 
For each clusters we produced the X-ray images in the soft (0.5--2 keV) and hard 
(2--7 keV) bands. Data for clusters with multiple observations were merged together.
The choice of using soft- and hard-band images separately was motivated
by the fact that the background or foreground
regime and the sensitivity change significantly below and above 2 keV. 
Background (instrumental and cosmic) is higher in the hard band, while the 
foreground ICM emission is much stronger in the soft band.  At the same time, 
the source emission in the soft band typically has more signal than in the hard band, 
except for $z<1$ strongly absorbed sources.  Because of these properties, the detection 
of strongly lensed X-ray sources is expected to be cleaner in the hard band, but most of the
sources are in any case expected to be detected in the soft band.  However, it is 
not possible to define a general strategy to optimize the detection because the 
foreground and background properties of each field strongly depend on a combination of the
morphology of the cluster, its ICM temperature, the depth of the observation, and because of the
time-dependent contamination of ACIS, also on the epoch of observation.

Normalized exposure maps at the monochromatic energies of 1.5 and 4.5 keV 
were obtained and merged in the same way.  The observation IDs and exposure times (after reduction) 
of the X-ray data we analyzed are listed in Cols. 2 and 3 of Table \ref{TableX}, 
respectively.  Unfortunately, the archival {\sl Chandra} observations were not planned
as a uniform follow-up of the CLASH sample, but consist of a mix of shallow ($\sim 20$ ks) and 
deep  ($>300$ ks) exposures, with exposure times set by the original science goal. This has 
a relevant effect on the depth of our X-ray analysis. 

Although no spectral analysis is involved in this work, we need to estimate 
the energy flux of the unresolved X-ray sources.  As is usually done in the 
regime of low photon counts, the fluxes were estimated directly from the observed net count rate
assuming an average spectral shape that corresponds to average conversion factors.  
We computed the energy conversion factors ${\rm ecf}$ in the soft and hard bands separately
at the aim point in each field, extracting the ARF and RMF files from each Obsid and 
combining them according to the exposure time.  The flux corresponding
to a given count rate was obtained by multiplying it by the 
corresponding conversion factors. Because e the conversion holds at the aim point, 
the observed count rate was rescaled to the rate expected at the aim point 
based on the ratio of the exposure map at the aim point to the exposure map at the 
source position.  Conversion factors were computed for a standard intrinsic spectral shape
described with a power law with slope $\Gamma$ = 1.4--1.8, bracketing a range corresponding to
mildly obscured AGN and unobscured AGN or star-forming galaxies.  We remark that while the
conversion factors in the soft band were applied to the observed 0.5--2 keV count rate to 
obtain the 0.5--2 keV flux, the conversion in the hard band was from the 2--7 keV count rate to the
2--10 keV flux for a better comparison with the values that are commonly quoted in the literature.
We also accounted for Galactic hydrogen absorption as described by the
model {\tt phabs} \citep{phabs1992}, where the Galactic column density $n_{\rm H}$ at the cluster 
position is set as $n_{\rm H,tot}$ following \citet{2013Willingale}. This takes not only the neutral hydrogen into account, but also  molecular and ionized hydrogen. The final 
flux therefore is the value that would be observed in absence of Galactic absorption.
The range of conversion factors (corresponding to $\Gamma$ = 1.4--1.8) 
at the aim point in the soft and hard bands is listed in the Col. 4 of 
Table \ref{TableX}, and the Galactic $n_{\rm H}$ value we used to compute them
is listed in the last column. We note that the conversion factors in the soft band 
can vary up to 70\% in different fields; this percentage is $\sim 10$\% in the hard band.  This is mostly due to the increasing loss of 
effective area caused by the time-dependent contamination corrections of ACIS at low energies 
(below 2 keV), so that sources that were observed more recently have significantly 
larger conversion factors in the soft band (corresponding to a lower sensitivity for a given flux), 
while the hard band is only slightly affected.  All our fields except for one were observed more than 
once, therefore the corresponding conversion factors also depend on the observation dates of each 
exposure. 

We also verified the consistency of the astrometry between the {\sl Chandra} and the 
HST observations. We selected several X-ray bright unresolved 
sources with well-sampled and symmetric point spread functions (PSFs) in each field of view and computed the distance between their 
positions in {\sl Chandra} and the source center in HST images. Unfortunately, only 24 sources
in the 11 fields satisfied these requirements, therefore only computed
an average shift and dispersion in all the fields, which can be used as an estimate of the 
difference among the fields.  The average offset is measured to be $\delta=0.35\arcsec$, 
with an {\sl rms} dispersion of $0.23\arcsec$. The maximum offset is $\leq 0.8\arcsec$. 
These values should be considered upper limits because no error on the X-ray centroid or 
HST centroid was assumed. We conclude that a conservative matching radius 
to identify the optical counterpart of X-ray sources is $1.5\arcsec$. 

As a final note, we remark that we included ACIS-I and ACIS-S data because the field of view 
in both cases ($16\times 16$ arcmin$^2$ and $8\times 8$ arcmin$^2$ for ACIS-I and ACIS-S, respectively)
is large enough to cover the high magnification region we investigated, 
which is a square with a side of 5 arcmin centered on the peak of the X-ray image.
In any case, in our sample ACIS-S data were obtained only for one cluster (R1347).


\section{Gravitational lensing analysis and magnifications maps}

\begin{figure*}
\begin{center}
\includegraphics[width=0.49\textwidth, trim=0 0 20 20, clip]{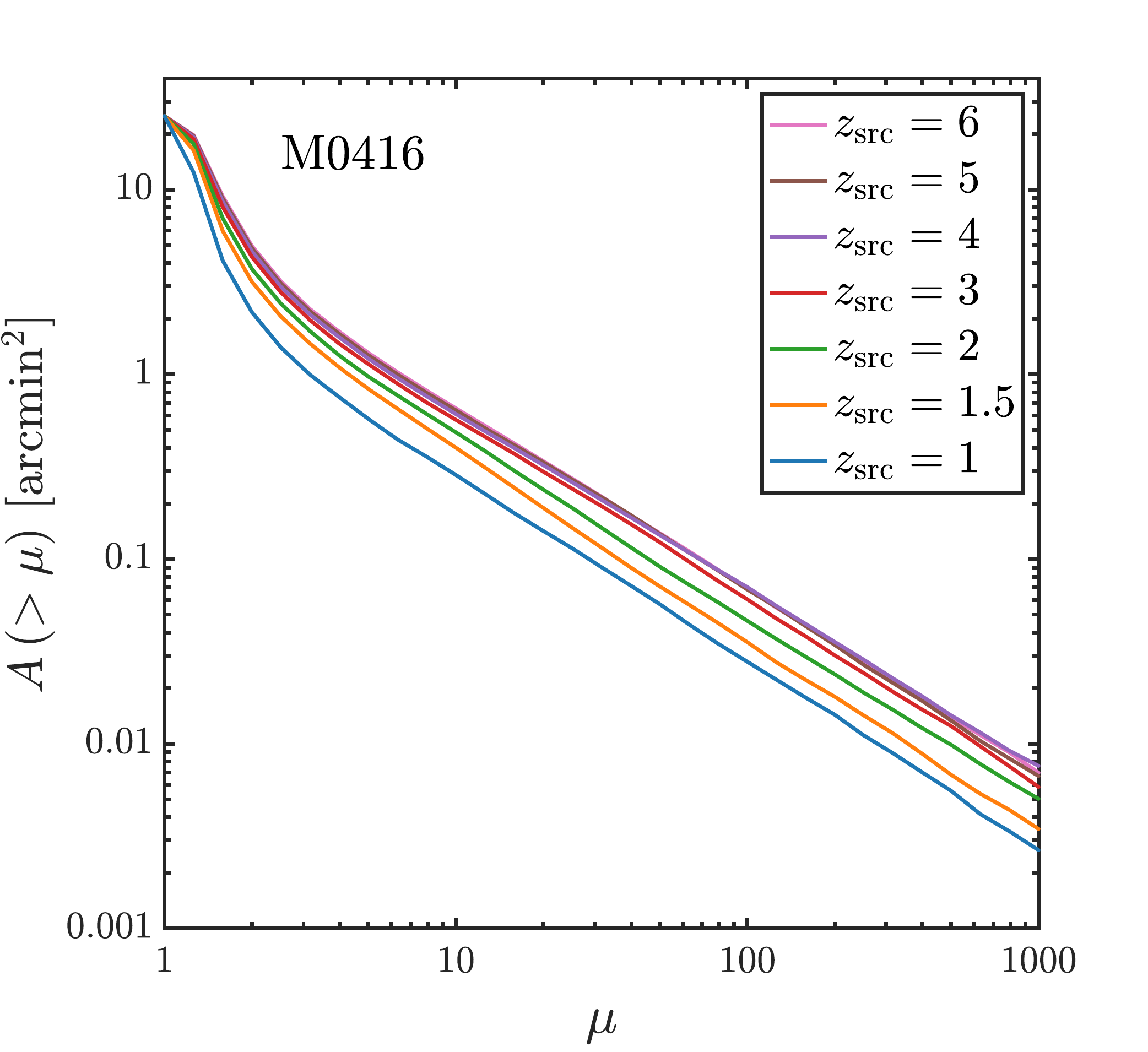}
\includegraphics[width=0.49\textwidth, trim=0 0 20 20, clip]{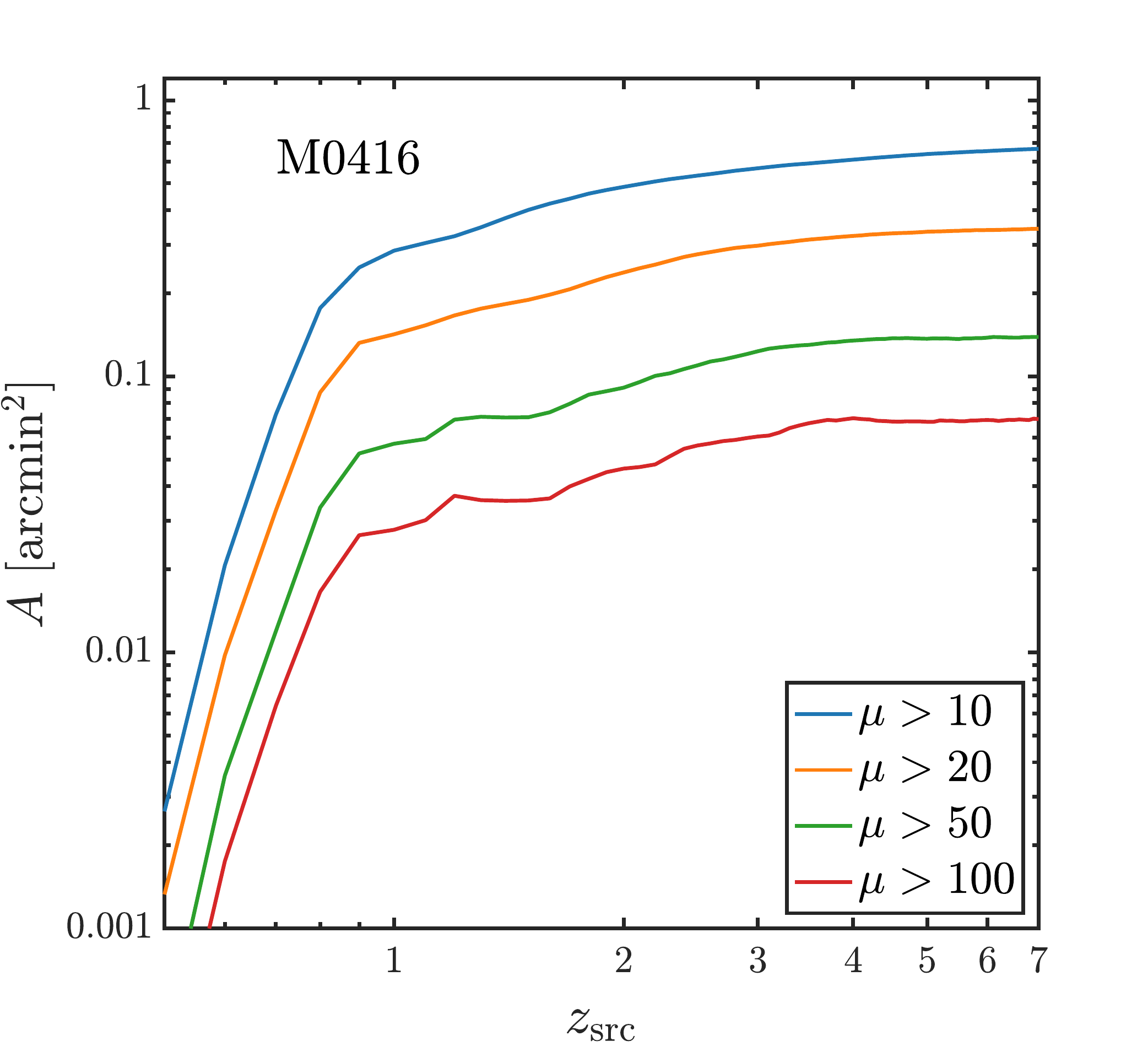}
\caption{Solid angle as a function of magnification and redshift for M0416. {\sl Left panel}: Cumulative solid angle corresponding to a 
magnification higher than a given value $A (>\mu)$ in the lens plane 
plotted as a function of magnification for a set of redshifts from 1 to 6 for the cluster M0416 at $z=0.397$.  The maximum solid angle corresponds 
to the 25 arcmin$^2$ surveyed area in each cluster. {\sl Right panel}: Cumulative 
solid angle $A(>\mu)$ for $\mu= 10, 20, 50,$ and $100$ in the lens plane 
as a function of the redshift of the lensed source for the cluster M0416.  }
\label{mu_angle_m0416}
\end{center}
\end{figure*}

\begin{figure*}
\begin{center}
\includegraphics[width=0.49\textwidth, trim=0 0 20 20, clip]{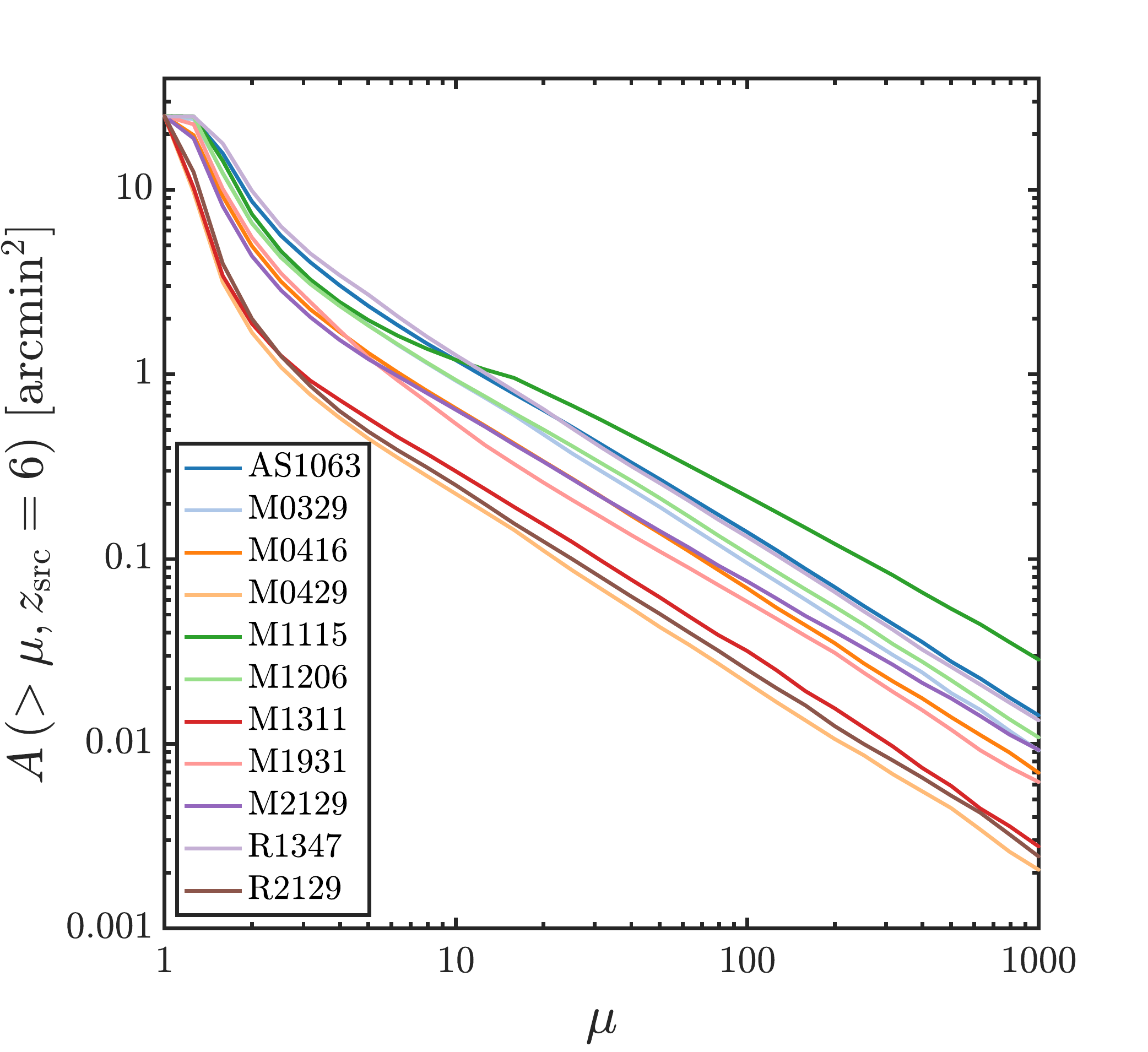}
\includegraphics[width=0.49\textwidth, trim=0 0 20 20, clip]{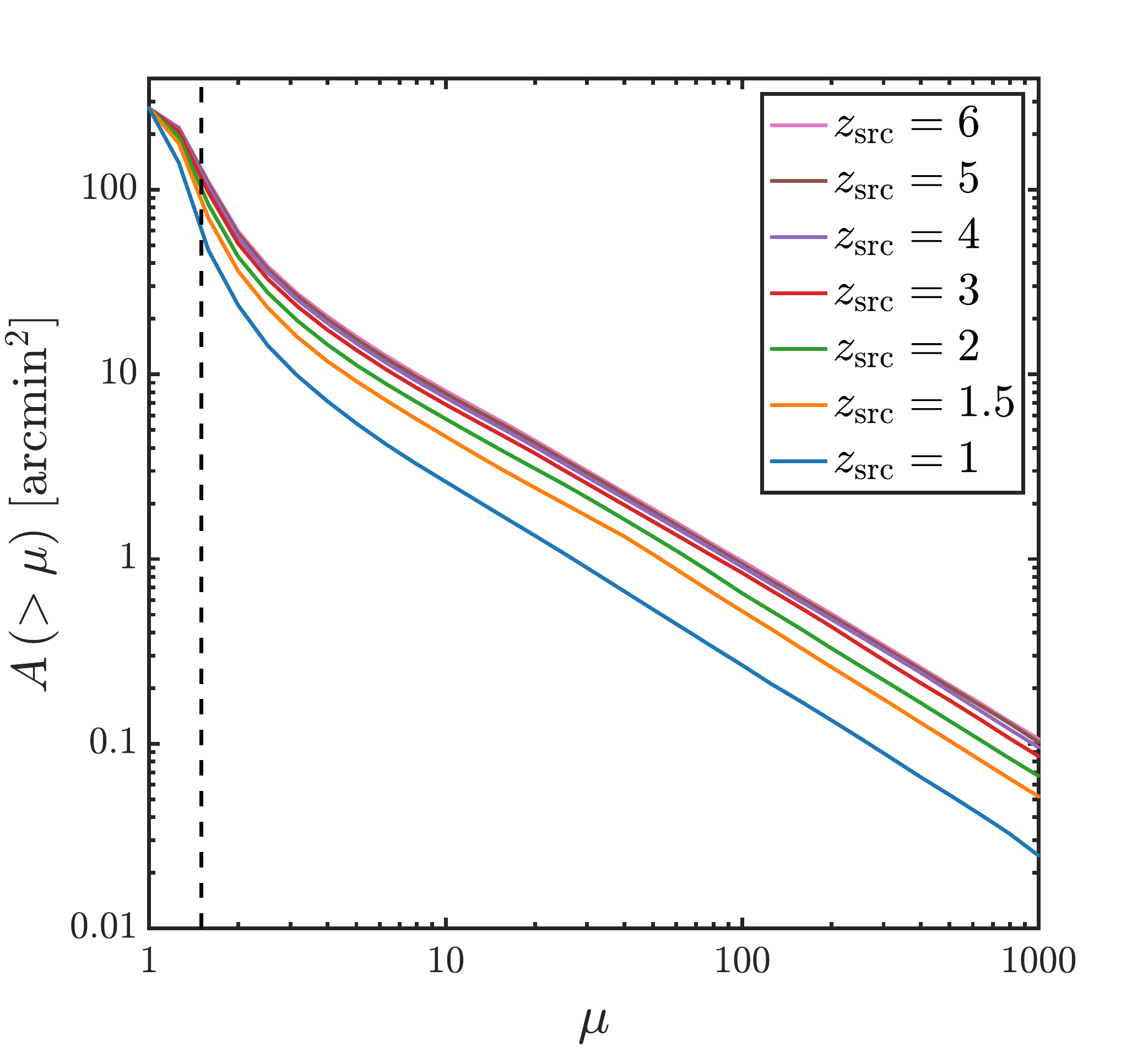}
\caption{Solid angle as a function of magnification and redshift for the 11 clusters. {\sl Left panel}: Solid angle $A (>\mu)$ in the lens plane for each 
of the 11 clusters as a function of $\mu(z_{\rm s}=6)$. {\sl Right panel}: Cumulative solid angle
$A (>\mu)$ in the lens plane for our entire sample for a source redshift from 1 to 6. The vertical dashed line 
shows the lower magnification limit $\mu_{\rm max}=1.5$ that defines the field 
of view we investigated.}
\label{mu_angle}
\end{center}
\end{figure*}

\begin{figure}
\begin{center}
\includegraphics[width=0.49\textwidth, trim=0 0 20 20, clip]{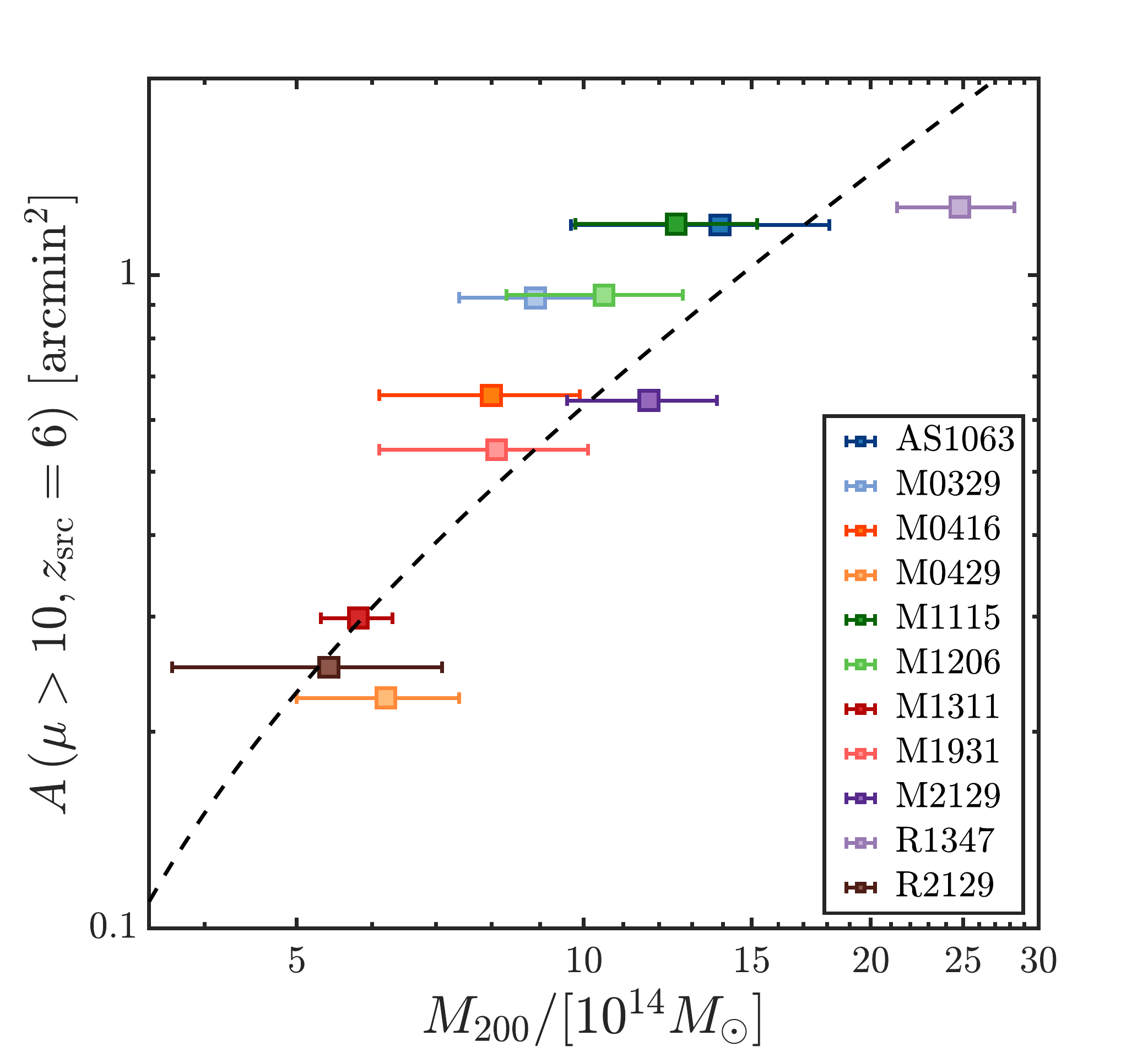}
\caption{Solid angle $A (\mu>10)$ in the lens plane computed for $z=6$, 
plotted against the virial mass computed within $r_{200}$ from the weak-lensing 
measurements of \citet{2018Umetsu}. Masses of M1311 and M2129 are based on 
X-ray scaling relations according to \citet{liu2020}. The dashed line shows the best-fit linear 
relation $A = 0.08\times(M_{200}/10^{14}M_{\odot})-0.17$ as a reference to appreciate the slope of the relation in our sample. }
\label{mu_m200}
\end{center}
\end{figure}

The lens models of all the 11 clusters in our sample have been published in  
\citet{Caminha2017a,Caminha2017b,2019Caminha,2020Meneghetti}, and 
\citet{Bergamini2019}\footnote{We did not include the latest refined lensing model for M0416 that has recently been published in
\citet{2021Bergamini}.  Nevertheless, this does not affect our final results.}. 
For each cluster, we obtained a set of magnification maps as a function of 
the redshift of the background source.  
To visualize the magnification power in a given cluster field, we show the 
cumulative solid angle corresponding to a magnification $\mu$ higher than 
a given value in the lens plane, and to explore the redshift dependence, a 
set of curves of total solid angle with magnification $\mu$ higher than a 
given value as a function of redshift.  

In Fig. \ref{mu_angle_m0416} (left panel) we show the solid angle $A(>\mu)$ versus the
magnification $\mu$ in the case of cluster M0416.  The maximum 
reached at $\mu=1$ is set by our choice of the field of view, 
corresponding to a square of $5\times 5$ arcmin$^2$ centered on the X-ray emission peak. We note that in the high magnification regime $\mu \sim 10$, 
we have only $\sim 1$ arcmin$^2$ (at $z_{\rm s}>2$) in the lens plane, 
which approximately corresponds to $\leq 0.1$ arcmin$^2$ in the source plane.  
The dependence on the redshift of the lensed source is quite steep up to 
$z\sim 1.5$, and it finally increases only by a factor of two above 
$z\sim 2$ up to the highest redshifts.  
This is shown in the right panel of Fig. \ref{mu_angle_m0416}, where we show the 
solid angle $A(>\mu)$ in the lens plane as a function of redshift for 
$\mu = 10, 20, 50,$ and $100$ in the case of M0416.  The drop in the solid angle 
in the regime of very high magnification $\mu = 100$ is roughly an order of magnitude 
with respect to $\mu=10$, which implies a very tiny area in the source plane.  

In Fig. \ref{mu_angle} we show an overview of the entire sample.  
In the left panel we show the solid angle $A(>\mu)$ in the lens plane versus 
the magnification $\mu$ for all the clusters we used, where $\mu$ is 
conventionally computed for $z_{\rm s}=6$.  We note a large 
spread in $A(\mu)$, showing that the magnification depends on the cluster properties, 
mainly total mass and mass distribution (including member galaxies along
critical lines, ongoing mergers, etc.).  In the right panel we show the total solid 
angle obtained summing the magnification maps of the 11 clusters.  
For a magnification of $\mu=10$ we have $\sim 10 $ arcmin$^2$ in the lens plane, 
or a total of less than 1 arcmin$^2$ in the source plane.  This implies
that the probability of finding X-ray lensed sources at $\mu>10$ in our data is relatively low. 
These aspects are further explored in \S 7 when we discuss a possible extension of this work to 
current X-ray data archives and future missions.

We also explored the correlation between the virial mass of the cluster computed at
$r_{200}$ and the high-magnification area in the lens plane. 
In Fig. \ref{mu_m200} we show this relation for $\mu>10$ 
(assuming $z_{\rm s}=6$ in this case as well). 
Using a simple linear function $A(\mu>10, z_{\rm s}=6)=a\cdot ({M_{200}}/{10^{14}M_{\odot}})+b$ to fit 
this relation and considering a 20\% typical error for the solid angle, 
we find $a=0.08 \pm 0.01$ and $b=-0.17 \pm 0.02$
(see the dashed line in Fig. \ref{mu_m200}). We remark that the effect of the intrinsic scatter 
(measured to be $\sim 0.2$ dex, and probably associated with the details of the mass
distribution) and the limited size of our sample do not allow a 
meaningful investigation of this relation, which would instead be a relevant property
for the estimate of the total lensing power of the full cluster 
sample available in the {\sl Chandra} archive.  


\section{X-ray emission from lensed sources identified in the optical}

In this section we focus on the optically identified lensed sources that fall within a 
maximum field of view of $5\times 5$ arcmin$^2$ centered on the cluster X-ray center. 
Optically identified lensed sources belong to two different subsets. 
The first set includes all the sources with multiple images and is 
obtained from \citet{2019Caminha}, \citet{Bergamini2019}, \citet{2021Vanzella}, and \citet{2021Bergamini}.
These are the counterparts used in their strong-lensing analysis. 
The second set consists of all the other sources with 
spectroscopic redshift higher than the cluster redshift, and therefore
lensed at some level.  In practice, we adopted conservative cuts corresponding to 
$z>z_{\rm cl}+0.1$. In addition, we also set a lower limit 
on the expected magnification factor to avoid that it is dominated by nonlensed ($\mu \sim 1$)
sources.  Clearly, the magnification effect is maximized when we consider regions where
the magnification is significantly higher than 1.  However, the solid angle $A(\mu)$ 
for the entire sample decreases rapidly for low $\mu$, as shown in Fig. 
\ref{mu_angle} (right panel). We note that a threshold 
$\mu_{\rm max}>3$ or $\mu_{\rm max}>2$ would
reduce the useful field of view in the lens plane by a factor $\sim 10$ and $\sim 7$, 
respectively. Because the statistics in our sample is limited, we therefore preferred to adopt
a softer threshold  $\mu_{\rm max}>1.5$ that reduces the useful field of view by less than 
a factor of 3 (see dashed vertical line in Fig. \ref{mu_angle}). 
By also including the low-magnification regime down to $\mu_{\rm max}\sim 1.5$, 
we obtained a statistically significant list of lensed sources for each 
cluster on the basis of the current imaging and spectroscopic analysis 
in the optical band, as we also verified {\sl \textup{a posteriori}}.

We note that the total solid angle
in the source plane corresponding to the solid angle in the lens plane
in which we searched for X-ray emission is measured with good approximation by $\Sigma_i A_i/\mu_i,$ where 
$A_i$ is the area of a resolution element, $\mu_i$ is the maximum magnification at each position,
and the sum is performed over all the positions where $\mu_{\rm max}>1.5$.  This approximation 
does not account for the fraction of the field of view in the lens plane that is folded in the 
source plane.  However, this fraction is negligible when compared to the variation in magnification 
associated with the redshift range of the sources, as we tested by comparing this approximation with 
the source and lens-plane mapping obtained with full ray-tracing at specific redshifts.
Finally, we obtain a total of 61.9 arcmin$^2$ in the source plane.

Our strategy consisted of three steps.  
Driven by the high-resolution HST images, first we visually inspected the X-ray images 
in the soft (0.5--2 keV) and hard (2--7 keV) bands separately to search for X-ray 
emission from single sources. Then, we performed aperture photometry at the source 
position and inspected the distribution of measured values in the soft and hard bands.  
Finally, we estimated the cumulative X-ray emission from the bulk of the 
optically identified lensed sources by stacking the X-ray data at the source positions.

\subsection{Optically identified lensed sources: Visual inspection}

First, we searched for X-ray emission from single sources by visual inspection. We proceeded 
in two steps: examining the original X-ray images, and then studying the 
residual images after subtracting background emission and the ICM foreground emission.  

To obtain the residual images, we modeled the ICM component with two or more smooth ellipsoids 
in the {\sl Sherpa} software.  First, obvious unresolved 
X-ray point sources were removed from the images. The holes left in the images were filled 
with a Poissonian realization consistent with the surrounding signal. The cleaned image, 
largely dominated by the ICM, was fit with a combination of two elliptical 
$\beta$-models plus a constant background. A double $\beta$-model provides acceptable fits
for most of the clusters in our sample. However, 4 out of 11 clusters, 
namely AS1063, M0416, M1206, and R1347, are undergoing a major merger process 
of multiple subhalos and required a more complex modeling. For these clusters, 
we considered four independent $\beta$-model components. For M0416 and R1347, 
where a bright cool core coexists within a disturbed morphology due to an 
ongoing major merger, we further added 
another component for the bright emission excess in the cool core. This process was 
applied separately to the soft- and hard-band images.  Finally, the residual images were obtained by 
smoothing the original images with an FWHM of 0.5$\arcsec$ and then subtracting the 
best-fit models.  Because the signal obtained as the difference of the two images is non-Gaussian, the residual images were not used for quantitative analysis, 
but were used only as support for the visual inspection to highlight possible 
candidate X-ray sources.  Clearly, residual images are also a good resource 
to search for extended features in the ICM, such as sharp edges and cavities. 
However, we do not discuss this aspect here and refer to \citet{2016Donahue} for
a detailed ICM morphological analysis of CLASH clusters.

Before performing the visual inspection, we fixed the size of the region we used for
photometry, or extraction radius $R_{\rm ext}$ (i.e., the circle within which we measured the emission 
of an unresolved source) according 
to the PSF at the source position. For multiple observations, 
the combined PSF was obtained by weighting each PSF map by the corresponding exposure 
map and exposure time. For simplicity, we ignored the departure of the shape of the PSF 
from circular because this aspect affects our analysis only little.  We set the 
extraction radius $R_{\rm ext}$ to the size of the PSF corresponding to a 90\% enclosed energy
fraction as computed with the task {\tt mkpsfmap} within CIAO,
setting a minimum value of 1.3~arcsec\footnote{For a complete documentation of the 
{\sl Chandra} PSF see https://cxc.harvard.edu/proposer/POG/html/chap4.html.}. 
This choice was a good compromise between the need to maximize the 
encircled energy fraction and simultaneously keep 
the extraction region as small as possible to avoid the dilution of the signal 
by the high and structured foreground due to the ICM.  
This is relevant also considering that the cluster usually is 
not centered on the {\sl Chandra} aim point, so that the PSF significantly
degrades across the FOV.  In addition, we manually 
adapted the shape of the source region in the few but relevant cases in which the 
lensed sources were significantly extended as a result of strong lensing.  
A case in which X-ray emission may come
from single unresolved regions along a strong arc is expected \citep[see][]{2020Bayliss}. We 
also had to slightly move or reduce the source region to avoid the few bright X-ray 
sources (mostly foreground galaxies and AGN) in the field.  Finally, when optical 
counterparts belonged to the same family closer than the typical X-ray resolution, 
we merged the region of the nearby sources to obtain a single X-ray aperture photometry. 

Our visual inspection consisted of searching for a distribution of pixels consistent with 
a PSF-like distribution on top of the background or foreground emission, using the experience gained 
in the investigation of deep fields such as the CDFS that have been performed by our group in the past.  Although 
the identification of faint sources in extremely deep exposure is made with a high background, the
main difference here is the strong foreground with a complex small-scale structure that
makes the identification process significantly harder than in the field.
The visual inspection provides us with a preliminary list of 20 X-ray source candidates
either in the soft or in the hard band.  Most of them are barely visible against the 
overwhelming ICM emission, but they are sometimes emphasized in the residual images. 
As described above, the residual images are very sensitive to 
irregularities in the ICM distribution, which can be associated with a plethora of 
features from infalling subhalos to edges, cold fronts, and cavities in the ICM.  
While all these features are extended, which is different from the unresolved emission 
expected from background sources, it may well happen that the combination of real 
ICM features and noise makes them compatible with the signal from unresolved sources. 
Clearly, mastering this aspect goes well beyond the goal of this work.  To
measure the flux of these sources, we adopted a
simple aperture photometry. This method is certainly affected by Poissonian noise, but
it does not depend on any other assumption.  It also has the positive aspect of 
providing a simple and well-defined selection function that can be directly computed  
field by field. This aspect is crucial for the untargeted source detection 
and the measurement of the cumulative number counts, as shown in \S 5 and \S 6.

\subsection{Optically identified lensed sources: Aperture photometry}

\begin{table*}
\caption{\label{phot_src}X-ray emitting sources identified by visual inspection 
at the position of optically lensed sources. RA and DEC here correspond to the X-ray position.}
\footnotesize
\begin{center}
\begin{tabular}[width=\textwidth]{lccccccccc}
\hline\hline
Cluster & ID & RA & Dec & $z$ & $\mu_z$ & $N_{\rm soft}$ & $N_{\rm hard}$ & $f_{\rm soft}$ & $f_{\rm hard}$ \\
 &  & [h:m:s] & [d:m:s] &  &  &  &  & [$10^{-16} {\rm erg/s/cm^{2}}$] & [$10^{-16} {\rm erg/s/cm^{2}}$] \\
\hline
AS1063  &  31  &  22:48:50.43  &  -44:32:11.4 &  1.440  &  1.87  &  $144.0\pm13.3$  &  $118.0\pm11.6$  &  $107.2\pm9.6$  &  $269.7\pm28.3$  \\
AS1063  &  46 (m) &  22:48:46.11  &  -44:31:47.4  &  1.260  &  30.46  &  $17.5\pm8.6$  &  $2.3\pm7.1$  &  $14.1\pm6.7$  &  $5.8\pm18.7$  \\
AS1063  &  95  (m)  &  22:48:46.97  &  -44:32:14.7  &  3.713  &  3.75  &  $9.7\pm5.1$  &  $1.5\pm3.8$  &  $7.3\pm3.7$  &  $3.4\pm9.3$  \\
M0416  &  12  &  4:16:05.60  &  -24:04:17.2  &  0.736  &  1.53  &  $7.4\pm3.8$  &  $5.0\pm3.3$  &  $2.3\pm1.2$  &  $4.8\pm3.5$  \\
M0416  &  208 (m)$^\ast$ &  4:16:06.87  &  -24:05:09.5 &  2.218  &  5.74  &  $15.6\pm5.5$  &  $-3.4\pm3.1$  &  $4.4\pm1.5$  &  $-3.0\pm2.9$  \\
M0416  &  221  (m)  &  4:16:11.14  &  -24:04:36.2  &  4.071  &  2.17  &  $7.3\pm4.1$  &  $-0.5\pm2.6$  &  $2.0\pm1.1$  &  $-0.5\pm2.4$  \\
M0416  &  241 (m)$^\ast$    &  4:16:09.56  &  -24:04:00.9  &  5.995  &  1.73  &  $10.6\pm8.8$  &  $15.3\pm7.8$  &  $3.1\pm2.4$  &  $14.1\pm7.6$  \\
M1115  &  23  &  11:15:51.51  &  +1:30:07.5 &  3.525  &  73.86  &  $17.0\pm8.4$  &  $-4.9\pm5.2$  &  $23.0\pm11.1$  &  $-25.3\pm28.6$  \\
M1206  &  18  &  12:06:18.29  &  -8:48:39.4  &  2.427  &  2.09  &  $20.6\pm5.2$  &  $29.2\pm6.2$  &  $13.3\pm3.2$  &  $49.1\pm11.2$  \\
R2129  &  12\&13  &  21:29:39.46  &  +0:05:50.3  &  0.592  &  1.51  &  $6.2\pm2.8$  &  $0.2\pm1.3$  &  $11.6\pm5.0$  &  $1.7\pm10.4$  \\
R2129  &  73  &  21:29:39.38  &  +0:05:51.7  &  4.847  &  1.90  &  $4.4\pm2.4$  &  $-0.6\pm0.8$  &  $8.1\pm4.3$  &  $-4.6\pm6.2$  \\
\hline
\end{tabular}
\tablefoot{The source ID is an identification number associated with the optical 
sources within each field.  The flag [m] marks sources that have multiple counterparts, 
as discussed in the text and in the appendix.  
Net counts are computed with aperture photometry as described in the text. Only sources with 
photometry $S/N>2.0$ are listed (with three exceptions at slightly lower $S/N$). 
Energy fluxes in the soft and hard bands are computed as described in \S 4.2.  All the
values refer to the observed, therefore lensed, quantities.  Errors on the observed
flux include the Poissonian errors and the systematic uncertainties on the 
conversion factors. We mark with an asterisk two sources (M0416-208 and M0416-241) that
are discarded on the basis of the lack of detection of a counterpart with a higher 
magnification.  In addition, M0416-208 overlaps within $\sim 2$ arcsec with a 
foreground galaxy, thus the photometry may be contaminated.  We kept these sources in the 
list consistent with the criteria we applied, but we removed them from the following
analysis.} 
\end{center}
\end{table*}

We performed aperture photometry at the source position for the 20 source candidates 
that were visually selected. A critical aspect is the evaluation of the photometric error.
To sample the local emission from the ICM, the background region for each source was 
determined as an outer annular region with a thickness of 2$\arcsec$, with the innermost 
annulus at 1 arcsec distance from the source region.  The area occupied by 
the few bright X-ray sources (mostly foreground AGN) was masked out when 
the background was measured. However, X-ray bright spots were not masked because they might 
be associated with the intrinsic fluctuations of the ICM and must naturally be
included in the noise affecting the aperture photometry. Finally, the $1\sigma$ 
error on the net counts was conservatively computed as $\sqrt{C+2B}$, where $C$ 
is the value of the net counts and $B$ is the value of the background or foreground 
counts rescaled to the source extraction region.

\begin{figure}
\begin{center}
\includegraphics[width=0.49\textwidth, trim=15 10 15 10, clip]{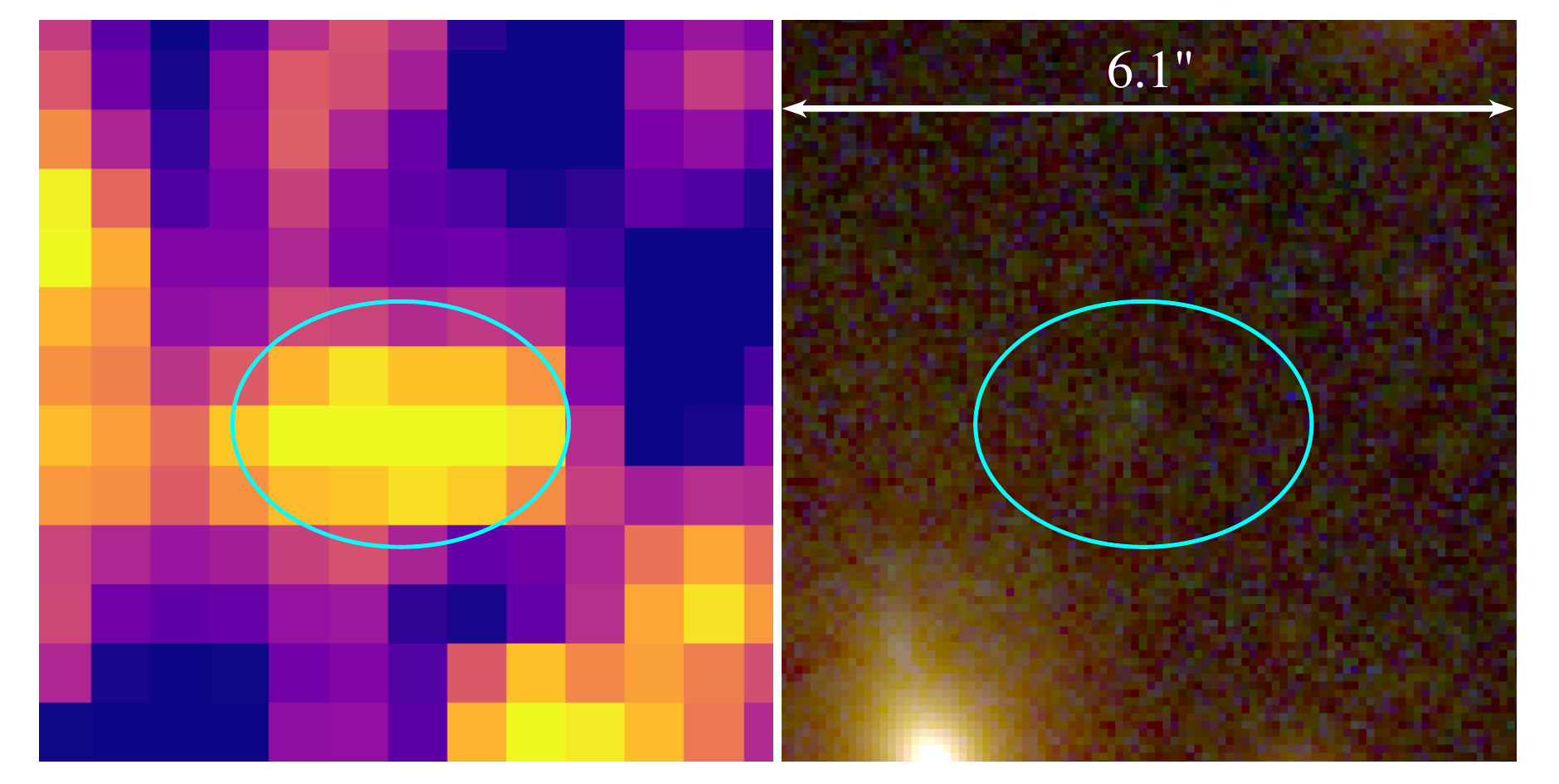}
\includegraphics[width=0.49\textwidth, trim=15 10 15 10, clip]{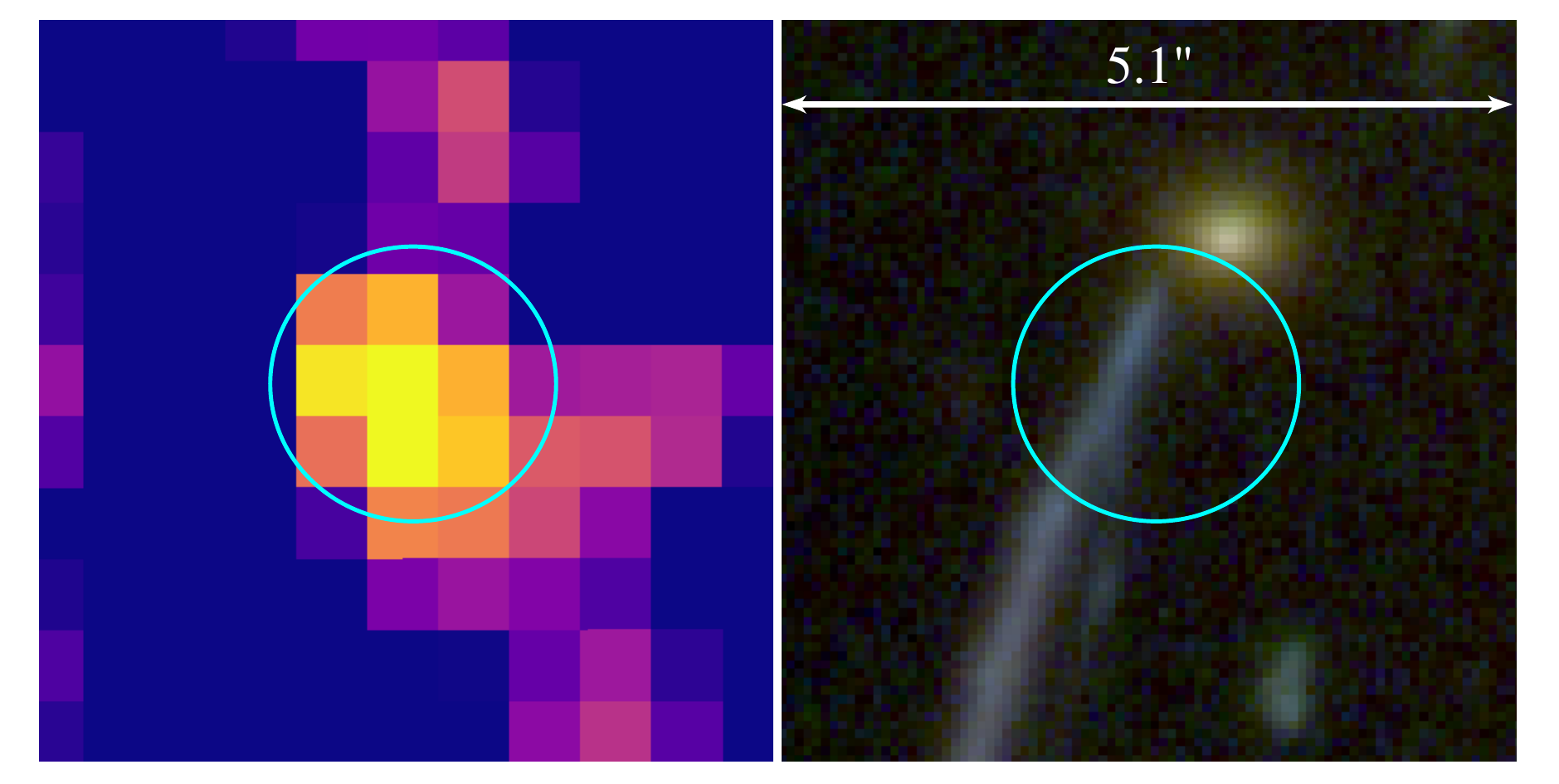}
\includegraphics[width=0.49\textwidth, trim=15 10 15 10, clip]{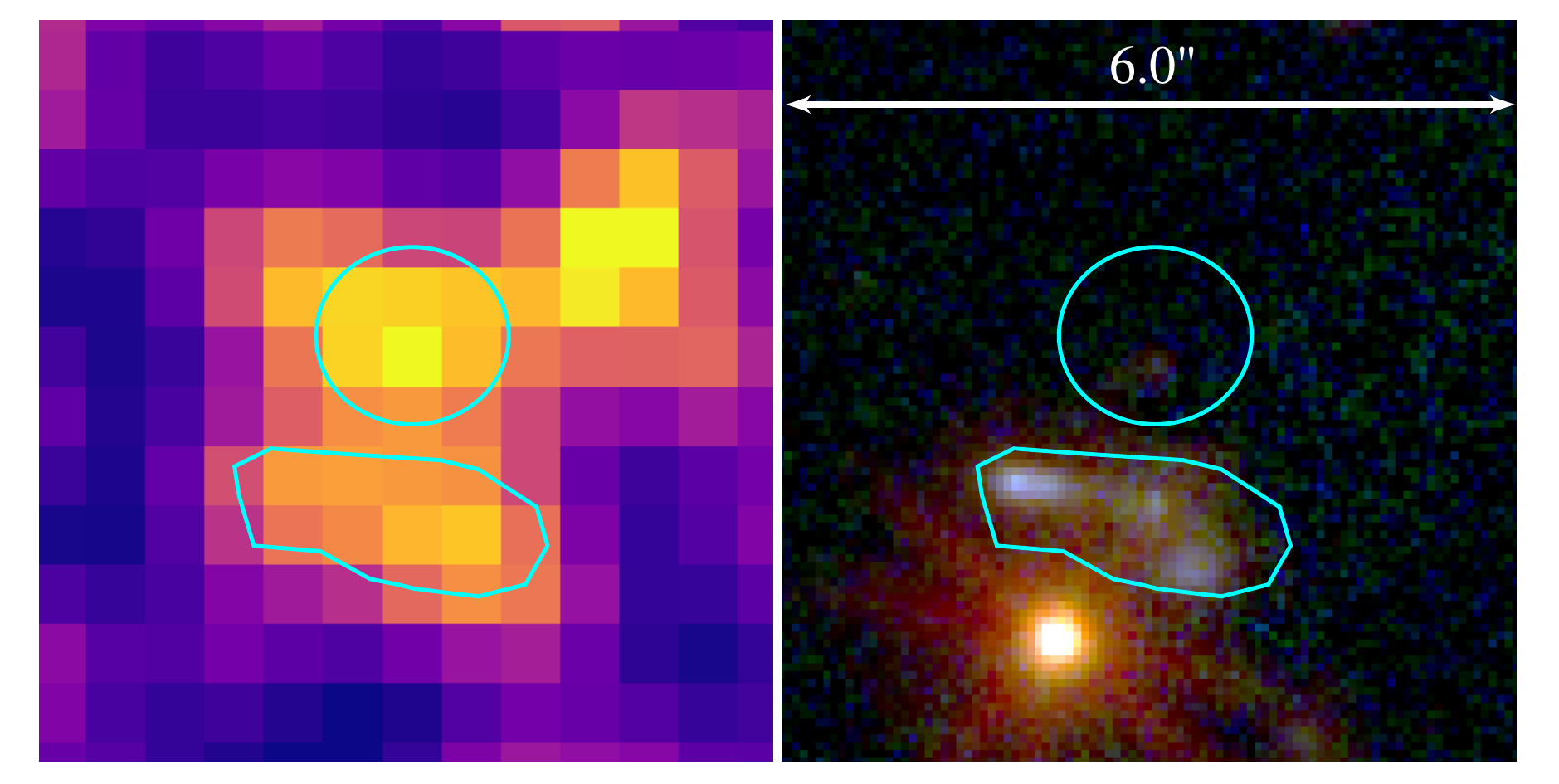}
\caption{Examples of identified X-ray counterparts of optically lensed sources. 
The left panels show the residual X-ray images (soft band), and right panels show the 
HST images. Contours are those used for aperture photometry in the X-ray bands.
Top panels: Source 23 in the FOV of M1115, with $\mu\sim 74$. Middle panels: 
Source 46 in the FOV of AS1063, with $\mu\sim30$. Bottom panels: Sources 12, 13 
(elongated region), and 73 (circular region) in the FOV of R2129, with $\mu\sim 1.5, 1.5,$ 
and 1.9, respectively. }
\label{example}
\end{center}
\end{figure}

As expected, in several cases the photometric measurement is consistent with a
null signal within one $\sigma$. Because the robustness of these sources is mostly based 
on a visible excess at the optical source position, we obtain a small but reliable list of
X-ray emitting lensed sources  when we apply a further
filter 
on the $S/N$.  The reference choice for the $S/N$ threshold
is $S/N=2$ in at least one of the two bands, and this is used generally throughout 
the paper \citep[see][]{2002RosatiCDFS}.  Only 11 sources meet this criterion, 
including a few with $1.8<S/N<2$, which were recovered on the basis of their 
convincing visual appearance. For each source we computed the  
energy flux as $f_X=S\times {\rm ecf} \times H_{\rm aimp}/H_{\rm s}$, where $S$ is the net count 
rate (the net count divided by the total exposure time in the field), 
${\rm ecf}$ is the energy conversion factor, and $H_{\rm aimp}$ and $H_{\rm s}$ are the values of 
the exposure map at the aim point and at the source position, respectively. In particular, 
$H_{\rm s}$ was averaged over the source extraction region. This was computed separately in the 
soft and hard band.  For the ${\rm ecf} $, we used the mean value of the two 
listed in table \ref{TableX}, corresponding to spectral shapes with $\Gamma=1.4$ and $1.8$. 
The full range was accounted for as a systematic error on the fluxes and was summed in quadrature to the 
dominating statistical errors from aperture photometry.  All the relevant information 
for the 11 X-ray emitter candidates are listed in Table \ref{phot_src}. 
In Fig. \ref{example} we show three examples of lensed X-ray emitting candidates, comparing the X-ray
and the HST images, and highlighting the extraction regions $R_{\rm ext}$ we used for aperture photometry. 
The first two panels show the two sources with the highest magnification values, and the third
panel shows the pair of candidates in the field of R2129; they lie at the limit of our detection
threshold.

We have five sources in two different fields that have one or more 
counterparts in addition to the counterpart that is identified as an X-ray emitter, labeled ``(m)''
in Table \ref{phot_src}.  Therefore we were immediately able to
verifiy on the X-ray images whether the photometry at the position of the other counterparts 
was consistent with our detection.  We find that in three cases the lack 
of detection in the other counterparts is still consistent within $1 \sigma$, while in the
case of M0416-241 and M0416-208, we should have detected their counterparts with a signal 
of more than $3 \sigma,$ but we obtained a null photometry at the corresponding positions.  
In addition, M0416-208 is very close to a cluster member ($<2\arcsec$) that might easily contaminate
its photometry.  Therefore these two sources were not considered in our further analysis.  The details of the confirmation of the multiple counterparts are given
in the appendix.

\begin{figure}
\begin{center}
\includegraphics[width=0.49\textwidth, trim=20 0 40 30, clip]{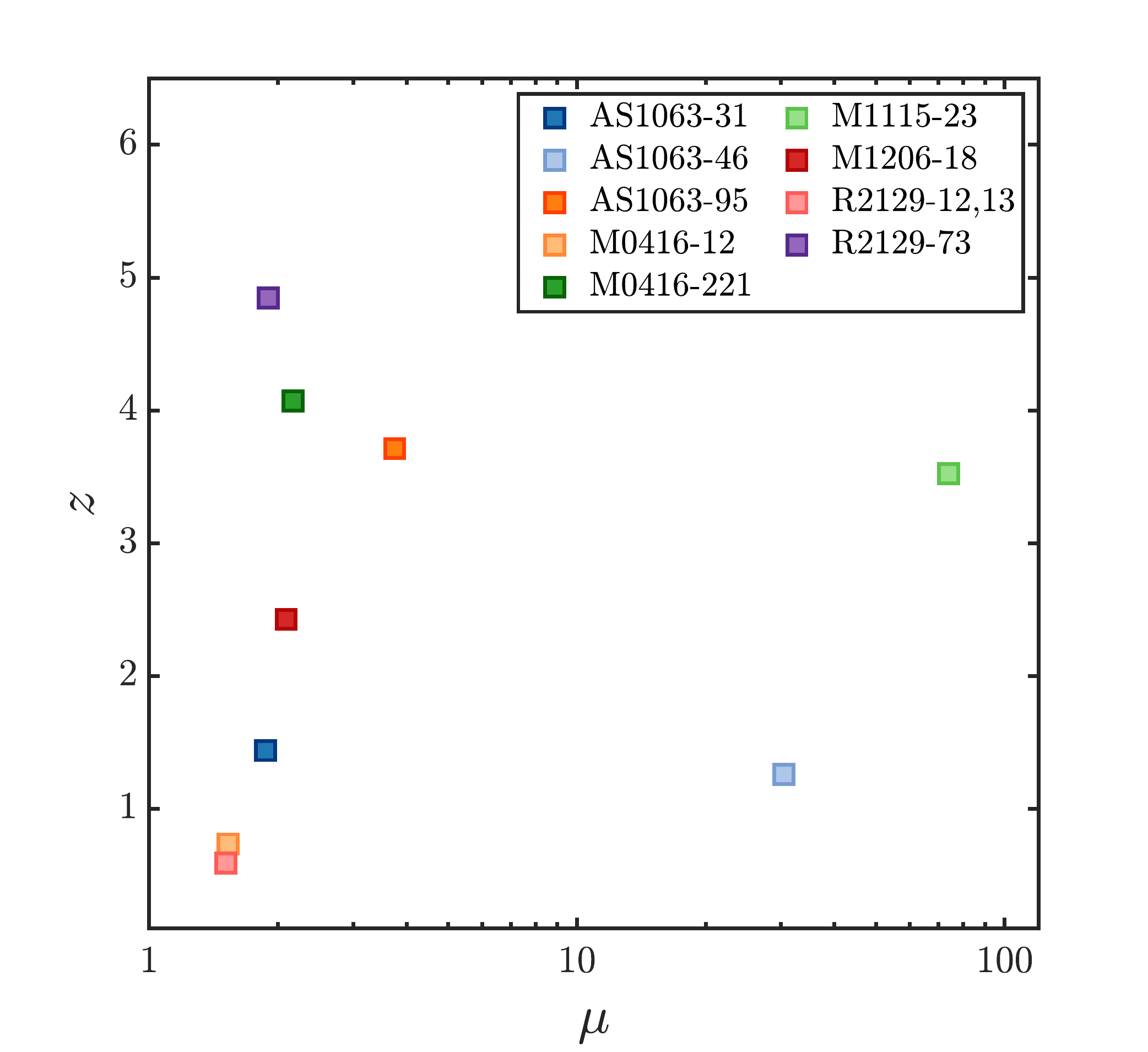}
\caption{Redshift and magnification of the sources in Table \ref{phot_src} 
identified by visual inspection after removing the two sources marked with the asterisk. 
The $1\sigma$ error on magnification is about 10\% up to $\mu=10$, and it can 
reach a factor of $2$ or higher for $\mu>20$.}
\label{muz}
\end{center}
\end{figure}

The magnification distribution peaks at low values ($\sim\!2$), and only two
sources lie in the high-magnification regime ($\mu = 30$ and $74$). 
In Fig. \ref{muz} we show the distribution in the $z-\mu$ plane for the nine sources that were 
selected as reliable X-ray emitter candidates.  
Typical errors on the magnification are modest, about 10\%, up to values
$\mu\sim 10$, but they rapidly increase above $\mu\sim 20$ \citep[see, e.g.,][]{2021Bergamini}.  
Most of the detections are found in the low-magnification
regions, and the probability of observing a highly magnified X-ray source among 
the optically lensed sources is very low.

In Table \ref{lum_X} we list the rest frame delensed X-ray luminosity of the X-ray emitting 
sources listed in Table \ref{phot_src} except for the two that have been discarded.  
Luminosities are corrected for Galactic absorption 
but not for intrinsic absorption, therefore only a flux-luminosity conversion is used
in each band, with a $K$-correction computed for an average spectral slope of $\Gamma = 1.6$.  
The uncertainty in the spectral shape is already accounted for in the error of the flux.  
In several cases, particularly in the hard band, we are able to only set a $1\sigma$ upper 
limit.  Applying a crude classification in terms of X-ray luminosity, we find three sources 
(one of which includes two optical counterparts)
consistent with being dominated by X-ray emission from star formation, with luminosities 
$L_X<10^{42}$ erg/s in the soft or $L_X<2\times 10^{42}$ erg/s in the total 0.5--10 keV band
(AS1063-46, M0416-12, R2129-12 and 13, and barely M1115-23). The luminosities of the other six sources lie in the range $8\times 10^{42} - 5\times 10^{43}$ erg/s in the soft band, 
corresponding to moderate Seyfert-like galaxies. The only two sources with hard-band 
detection also appear to have significant absorption.  The two sources in the high -magnification regime are instead consistent with being powered in the X-ray band by 
star formation. This is expected because at their delensed flux level ($\sim\! 3 \times 
10^{-17}$ erg/s/cm$^{2}$), the number density of star-forming galaxies is smaller by 
a factor of 3 than the number density of AGN. Overall, the few sources 
we identified so far show the typical mix of sources expected in this flux range.
We did not attempt to further characterize these sources because the quality of the X-ray spectra is poor.

\begin{table*}
\caption{\label{lum_X}Rest frame delensed X-ray luminosity of the X-ray emitting sources 
identified by visual inspection at the position of optically lensed source and 
listed in Table \ref{phot_src}.  Upper limits refer to the $1\sigma$ confidence level.}
\begin{center}
\begin{tabular}[width=\textwidth]{lcccc}
\hline\hline
Cluster & ID  & $z$ & $L_{X, \rm soft}$ & $L_{X, \rm hard}$ \\
 &  & & [${\rm erg/s}$] & [${\rm erg/s}$] \\
\hline
AS1063  &  31  & 1.440 &    $(5.3 \pm 0.5)\times 10^{43}$ &  $(1.3\pm 0.2) \times 10^{44}$  \\
AS1063  &  46 (m) & 1.260 &  $(3.2 \pm 1.5) \times 10^{41}$ &  $<5.5\times 10^{41}$ \\
AS1063  &  95  (m) & 3.713  &   $(1.4\pm 0.7)\times 10^{43}$ &  $<2.5\times 10^{43}$ \\
M0416  &  12  &  0.736   &  $ (3.0\pm 1.6)\times 10^{41}$ & $<1.1\times 10^{42}$   \\
M0416  &  221 (m) & 4.071  & $(8.1\pm 4.5)\times 10^{42}$   &  $<7.7\times 10^{42}$  \\
M1115  &  23  &    3.525  &  $(2.0\pm 1.0)\times 10^{42}$  &   $<3.0 \times 10^{41}$ \\
M1206  &  18  &  2.427  &  $(1.9 \pm 0.5) \times 10^{43}$  &  $(6.9\pm 1.6) \times 10^{43} $    \\
R2129  &  12\&13  &  0.592  &  $(9.3 \pm 4.0) \times 10^{41}$  &  $<1.0 \times 10^{42} $    \\
R2129  &  73  &  4.847  &  $(5.4 \pm 2.8) \times 10^{43}$  &  $<1.1 \times 10^{43} $    \\
\hline
\end{tabular}
\end{center}
\end{table*}

\subsection{Optically identified lensed sources: Cumulative emission}

\begin{figure}
\begin{center}
\includegraphics[width=0.49\textwidth, trim=0 0 20 15, clip]{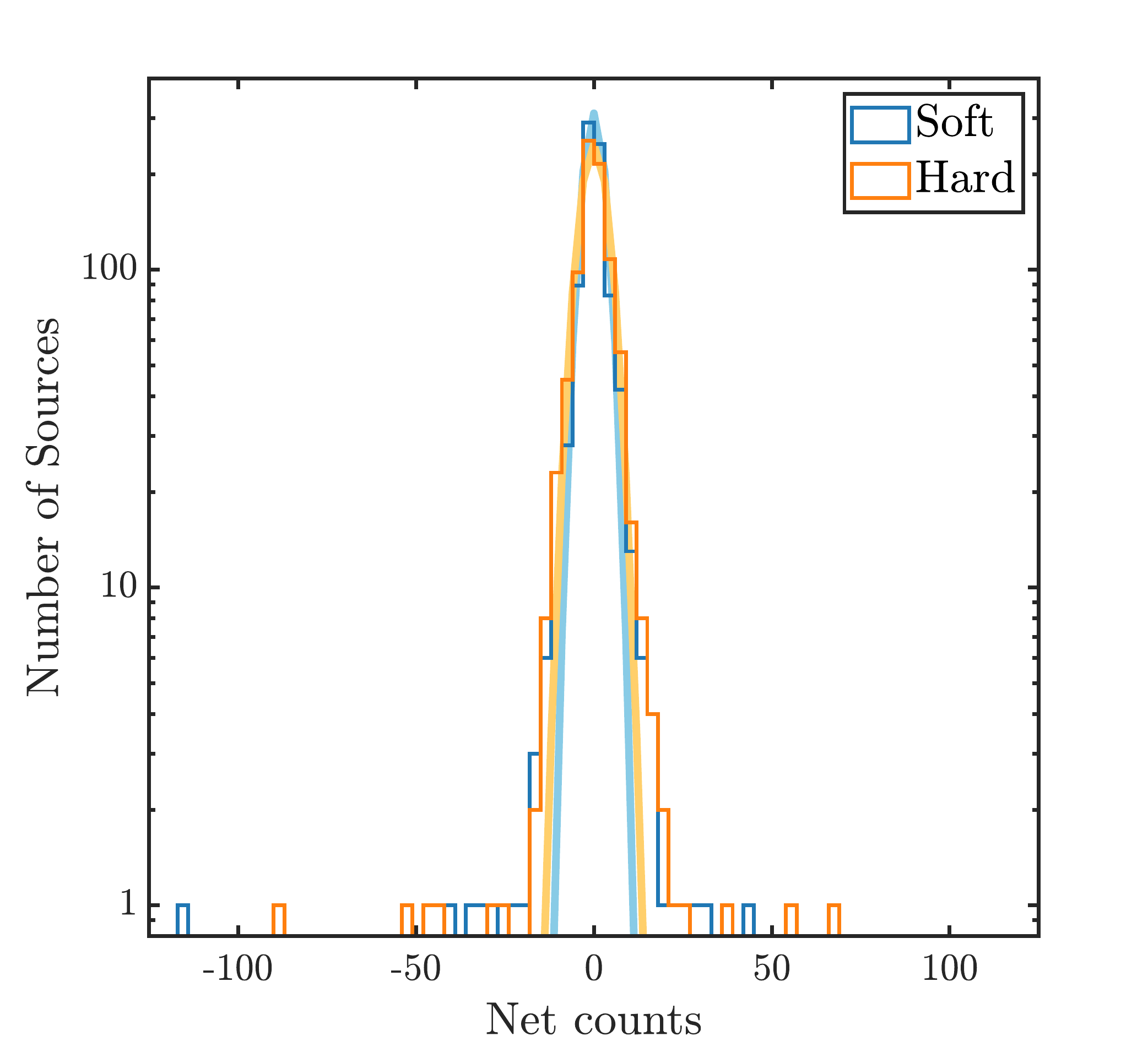}
\caption{Histogram of aperture-photometry values in the soft and hard bands (cyan and orange
histogram, respectively) for the 
837 optically lensed sources in the 11 cluster fields after removing 
the visually identified sources listed in Table \ref{phot_src}. The thick solid lines show 
the normalized best-fit Gaussian functions. } 
\label{phot}
\end{center}
\end{figure}

\begin{figure}
\begin{center}
\includegraphics[width=0.49\textwidth, trim=20 40 10 40, clip]{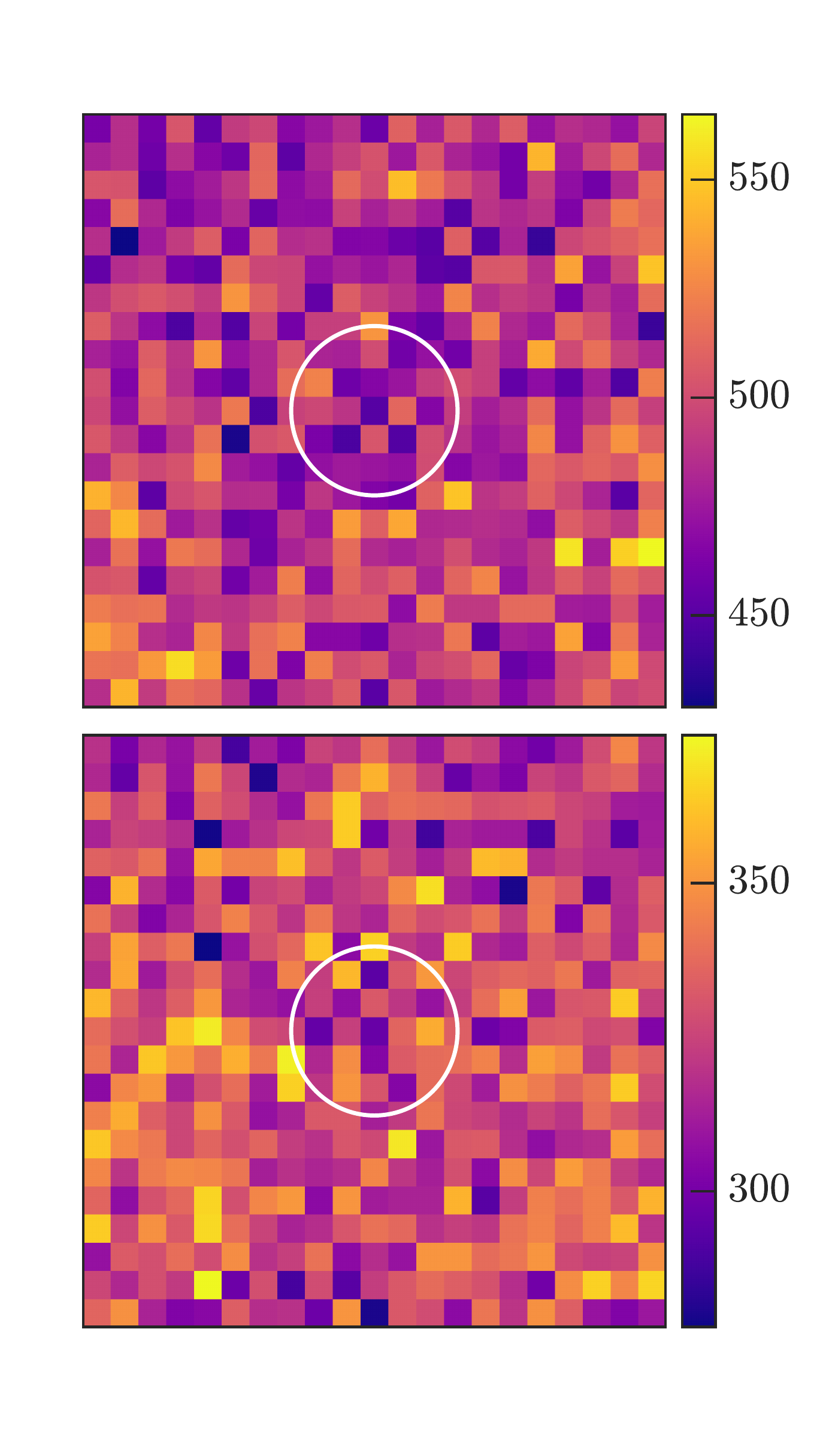}
\caption{Stacked X-ray image of all the optically identified lensed sources 
after removing the 11 X-ray detected sources and clipping the extreme photometry 
values as described in the text.  Images are 10$\arcsec$ across, and the
color scale indicates net counts.
Upper and lower panels show soft and hard bands, respectively. The white circle marks the $3$ arcsec aperture we used for photometry in the stacked images. }
\label{cutout}
\end{center}
\end{figure}

We searched for additional signal from the bulk of the optically identified lensed sources.
The distribution of aperture photometry for the entire sample of 837 source counterparts (after 
removing the 11 sources with identified X-ray emission in the previous step) 
is shown in Fig. \ref{phot}.  We note that the values on the negative and 
positive side of the photometry distribution are occasionally very high.  These events are associated with a region of 
extremely high ICM emission, and the corresponding values have an extremely large error as 
well.  However, some signal may be present close to $\sim 10$ net counts
in both bands, possibly associated with a subthreshold signal from the source population. 
Therefore we removed 47 anomalously high photometric values by performing a 
$3\sigma$ clipping and computed the stacked photometry (the sum of all the photometric values).
We found no positive signal, and we are able to place only upper limits on the average 
observed flux
of $4\times 10^{-16}$ erg/s/cm$^{2}$ and $1.8\times 10^{-15}$ erg/s/cm$^{2}$ in the soft and hard bands, respectively, at a $3\sigma$ confidence level.
To better appreciate this result, in Fig. \ref{cutout} we show the stacked X-ray images of all the 
sources in the soft and hard band. We also computed the cumulative photometry on the
stacked image in an aperture of 3 arcsec, finding $-218\pm 155$ and $-90\pm 128$ 
net counts in the soft and hard bands, respectively.
These values are consistent with what we previously found by summing the aperture photometry 
at each source position, and we confirm that the cumulative X-ray emission from the bulk of the
optically identified lensed sources is well below our detection limits.

\section{Untargeted search for X-ray sources in the high-magnification region}

The targeted search for X-ray emission from lensed sources in the field of 
our CLASH clusters provided us with only 11 candidates drawn from the lensed sources 
previously identified in the optical band.  Clearly, we still have 
X-ray lensed sources that were not previously identified in the optical.  Some may 
be associated with optical counterparts that have no redshift, and in some other cases, with sources with high $f_X/f_{\rm opt}$; in both cases they are potential X-ray lensed source candidates.  
Therefore we proceeded with the untargeted detection 
of sources in the X-ray images.  Although the ICM makes the X-ray 
detection process quite different and much harder than in the field, we did 
not develop an {\sl \textup{ad hoc}} algorithm for the ICM 
emission and its spatial variations. A proper and self-consistent modeling 
of the ICM would require an effort well beyond the possibilities of 
current X-ray imaging analyses.  Instead, we adopted a standard procedure 
that was validated and calibrated {\sl \textup{a posteriori}} with imaging simulations. 

\subsection{X-ray lensed source candidates identified with {\tt wavdetect}}

The X-ray source detection was performed using the {\tt wavdetect} tool \citep{2002Freeman} 
on the soft (0.5--2 keV), hard (2--7 keV), and total (0.5--7 keV) band
images. The wavelet scales were set as 1, 2, 4, 8, 16, and 32, and the false-positive probability 
threshold was set to $10^{-5}$.  This high threshold (compared to the 
$10^{-6}$--$10^{-7}$ range that is commonly used for robust detection) does not correspond to any 
predictable contamination fraction because of the strong and rapidly varying foreground, 
and it is expected to provide a generous number of source candidates. A source list obtained with this strategy therefore needs to be further refined with much tighter criteria. 
We merged the source lists obtained independently from the soft, hard, 
and full bands to include all the source candidates that were detected in at least one of the 
three bands. The threshold for merging sources is a distance smaller than $R_{\rm ext}$. 
A few cases of double detections that were not resolved by our matching criteria\footnote{This often happens
when the PSF becomes strongly asymmetric, in the outskirts of the field of view.} were fixed
after visual inspection.  The direct output of the {\tt wavdetect} algorithm 
provides us with 388 source candidates in the 11 cluster fields.  Then, we 
carefully filtered this list by applying a series of steps.

First, we verified whether we recovered the 11  previously identified sources by 
matching the {\tt wavdetect} X-ray position to the optical position by applying 
a criterion based on the distance.  Matched sources are those for which the 
distance is smaller than $R_{\rm ext}$.  Only 3 
out of the 11 sources (AS1063-31, M0416-208, and M1206-18) were recovered. This 
implies that 8 sources are below the {\tt wavdetect} detection threshold
and were identified only by the combination of a positive aperture 
photometry and the spatial overlapping with a magnified source.  This indicates that
the combination of a high {\tt wavdetect} selection algorithm with
threshold parameter $10^{-5}$ and the $S/N>2$ condition
may not give completeness $\sim 1$, particularly close to the $S/N>2$ threshold 
and in the regions that are deeply embedded in the strong 
ICM emission of the cluster core.  This aspect was considered with 
simulations as explained in the next section.

In the second step, we searched in the list of published redshifts from the 
CLASH-VLT survey
\citep{2013Biviano,2017Monna,Caminha2016,2016Balestra,2017Karman,2019Caminha,Bergamini2019}
for optical counterparts with a redshift lower than the cluster redshift
(or, more accurately, $z_{\rm s}<z_{\rm cl}+0.1$).  In this way, we 
identified 24 foreground sources with X-ray emission that were removed from 
the list of X-ray lensed source candidates.  

Because we chose from the beginning to focus on the solid angle corresponding to 
$\mu_{\rm max}>1.5$ to avoid a large contribution from 
the much larger field of view with low magnification, in the third step we also removed the 
source candidates that were found at $\mu_{\rm max}<1.5$. We remark that the choice
of using the maximum magnification is just a convention, and another choice
(e.g., using the magnification at $z_{\rm s}=6$) would not affect
our results because this step is needed uniquely to restrict our 
field of view to the most relevant regions for our purposes.  
We find that 140 source candidates correspond to positions with $\mu_{\rm max}<1.5$. 
After this step, 221 sources were left.

In the fourth step, we performed aperture photometry.  Before this, 
we visually inspected the extraction regions of source and background, defined 
as described in Sect. 4. In some cases we manually adjusted $R_{\rm ext}$ 
, and more often, the size of the background regions, particularly when these extended
across areas in which the ICM emission varies significantly.  This process mostly affects the 
few sources that are deeply embedded in the core of the cluster, where 
the small-scale fluctuations of the ICM inevitably introduce large uncertainties.  
Therefore we also considered the background and foreground estimated from our modeling 
of the ICM emission in the soft and hard band, which allowed us to measure it directly
in the source region, under the assumption that the ICM is well represented by the smooth
surface-brightness fit. We identified several sources for which the two photometry methods
provide significantly different results, showing that ICM fluctuations heavily contaminate
the sources candidates obtained with {\tt wavdetect}.  All these cases are located
in the core or slightly outside the core.  Therefore we conservatively included only 
the sources for which the two methods provided consistent photometry. We inspected 
the removed sources to verify whether some reliable source candidate might have been discarded 
in this step, but we found none.  Finally, we selected all the sources with $S/N>2$ in the soft 
or in the hard band.  

\begin{figure}
\begin{center}
\includegraphics[width=0.49\textwidth, trim=0 0 0 10, clip]{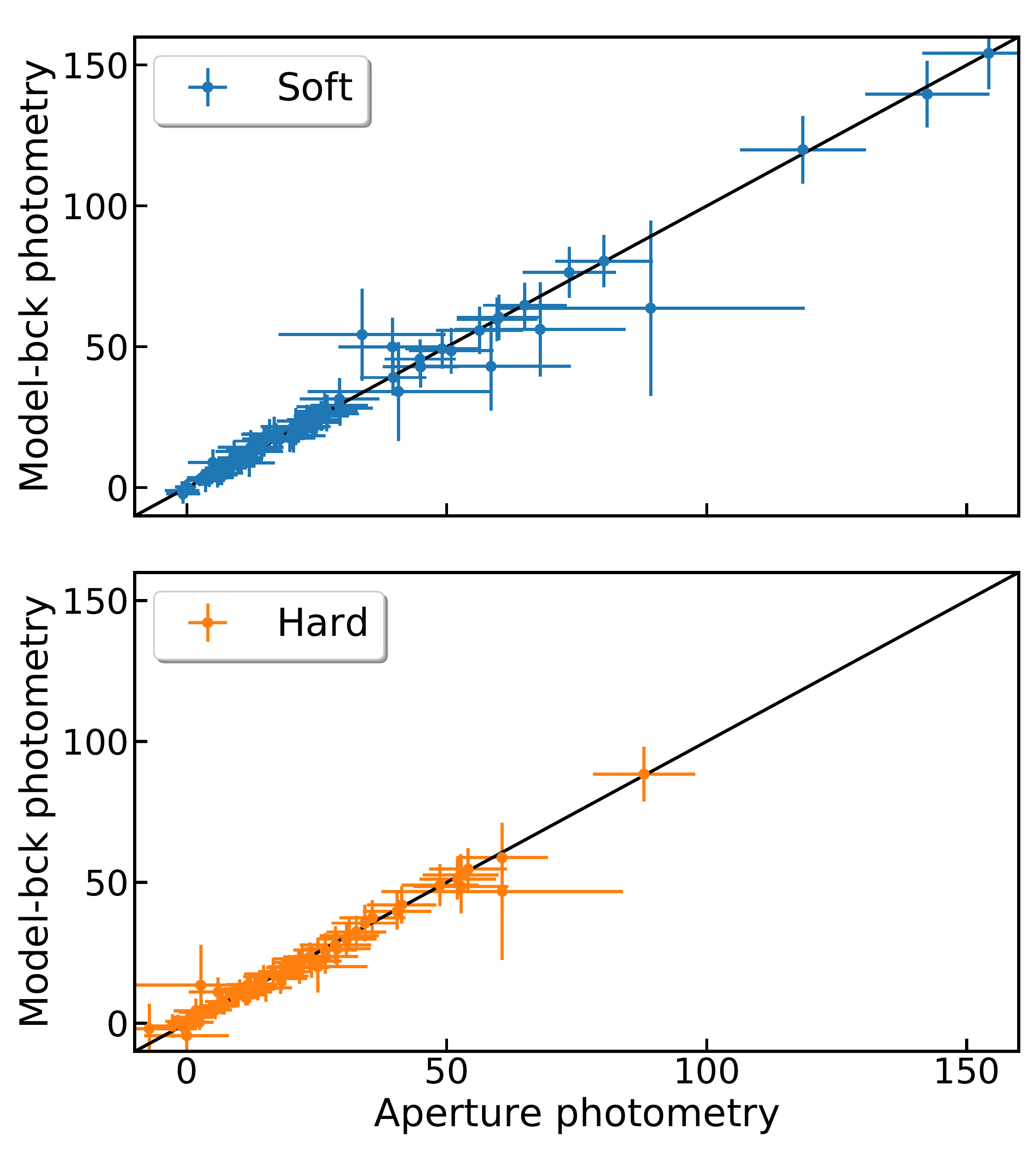}
\caption{Comparison of the aperture photometry values obtained with modeled or directly sampled
background and foreground for the soft band (upper panel) and hard band (lower panel).  The solid line shows the
$x=y$ relation.}
\label{photometry}
\end{center}
\end{figure}

This series of steps provided us with a list of 80 candidate X-ray lensed sources with 
a well-defined selection criterion in the X-ray band.  The two 
methods for performing aperture photometry (sampling the background or foreground from the surrounding 
regions, or modeling with a global fit of the surface brightness) 
are shown to be consistent with each other within the statistical errors 
for the list of 80 source candidates 
in Fig. \ref{photometry}.   At this point, we applied a fifth and last selection step 
by investigating the source candidates with visual inspection of the
corresponding HST and Subaru images.  This step allowed us to identify the cases in which the
X-ray emission is associated with an obvious foreground object at low 
redshift or with a cluster member.  This step is complementary to the removal of the 
sources with measured redshift $z_{\rm s}<z_{\rm cl}+0.1$ because not all the foreground sources 
have a measured redshift.  We removed 14 sources in this last step.  
All the X-ray sources without redshift that we kept in our sample are consistent with 
being in the background of the cluster.  However, we acknowledge that some of them may still be in the
foreground, and therefore our final source list is an upper limit to the number of 
X-ray sources that are lensed by the clusters with $\mu>1.5$.

In addition, we took advantage of the comparison with HST and 
Subaru images to refine the source position, particularly for the sources that lie very close to the 
critical lines. Only one source (R1347-7) has a rapidly changing 
magnification within the X-ray position error, therefore we used the position of the HST 
counterpart to improve the accuracy of the magnification.  All the other sources lie at positions far from the critical lines and therefore depend smoothly on the 
magnification of the centroid position. In the low-magnification regime 
($\mu<6$) an uncertainty of 1$\arcsec$ in position corresponds to an uncertainty of  
$\sim 5$\% on $\mu$, and therefore we assumed this value as an upper limit 
of the uncertainty on the magnification associated with the position.
However, the lack of a redshift for most of the
X-ray lensed source candidates implies an additional uncertainty on the magnification. 
Because it is not feasible to extract the redshift from the sparse optical information we have for
our X-ray lensed source candidates, we chose an approximation based on the actual 
redshift distribution of X-ray sources found in the CDFS \citep[see Figure 9 in][]{2017Luo}. 
We considered star-forming galaxies and the AGNs together because we lack a source 
classification in the redshift interval $0.5<z<4$ in which most of the lensed X-ray sources are expected in our fields.  
The probability for a source in the CDFS to be in a given redshift bin is
approximated by the probability distribution shown in Table \ref{zdist}.  When we
consider only sources in a flux range comparable to our source 
list ($10^{-16}$ -- $10^{-14}$ erg/s/cm$^{2}$ and 
$10^{-15}$ -- $10^{-13.5}$ erg/s/cm$^{2}$ in the soft and hard bands, respectively), 
we find a very similar distribution, with
differences smaller than 10\% in every bin. Therefore the probability 
of any source in these flux intervals to have a given redshift is approximately described by
the stepwise normalized function $p(z)$ shown in Table \ref{zdist}.
By weighting the $\mu(z)$ at each position in the FOV by the average redshift distribution
of CDFS sources, we obtain the best guess for the magnification associated with each source
simply as $\mu_{\rm w}\equiv \Sigma_i \mu(z_i)\times p(z_i)$ (where $z_i$ are the mid
points of the redshift bins in the first column of Table \ref{zdist}).

\begin{table}
\caption{Fractional distribution in bins of redshift for the sources of the CDFS
(normalized total number of sources in the range $0.5<z<4$).}
\label{zdist}
\begin{center}
\begin{tabular}[width=0.5\textwidth]{cc}
\hline
\hline
redshift bin & source fraction  \\
\hline
0.5--1   &     0.34 \\
1--1.5    &     0.21 \\
1.5--2    &     0.18 \\
2--2.5     &    0.13 \\
2.5--3     &    0.07 \\
3--3.5     &    0.05 \\
3.5--4    &    0.02  \\
\hline
\end{tabular}
\end{center}
\end{table}

\begin{figure}
\begin{center}
\includegraphics[width=0.49\textwidth, trim=0 0 0 0, clip]{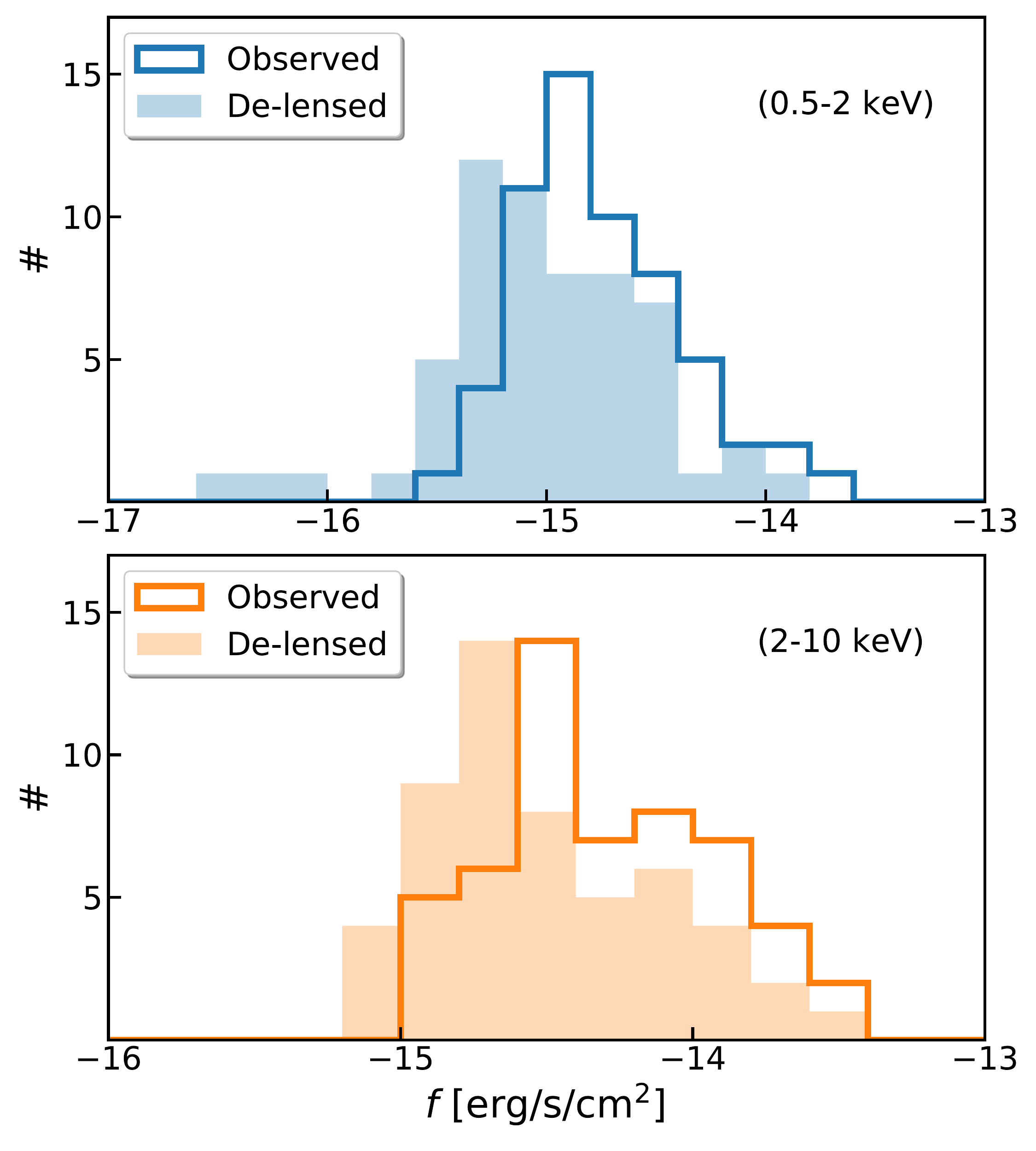}
\caption{Distribution of observed and delensed 
fluxes in the soft (upper panel) and hard 
(lower panel) bands for the sources detected with our algorithm (Table \ref{x_src}) plus sources
with optical counterpart (as in Table 3). Only sources with $S/N\geq 2$ 
in each band are shown.}
\label{flux_histo}
\end{center}
\end{figure}

\begin{figure}
\begin{center}
\includegraphics[width=0.49\textwidth, trim=0 0 0 0, clip]{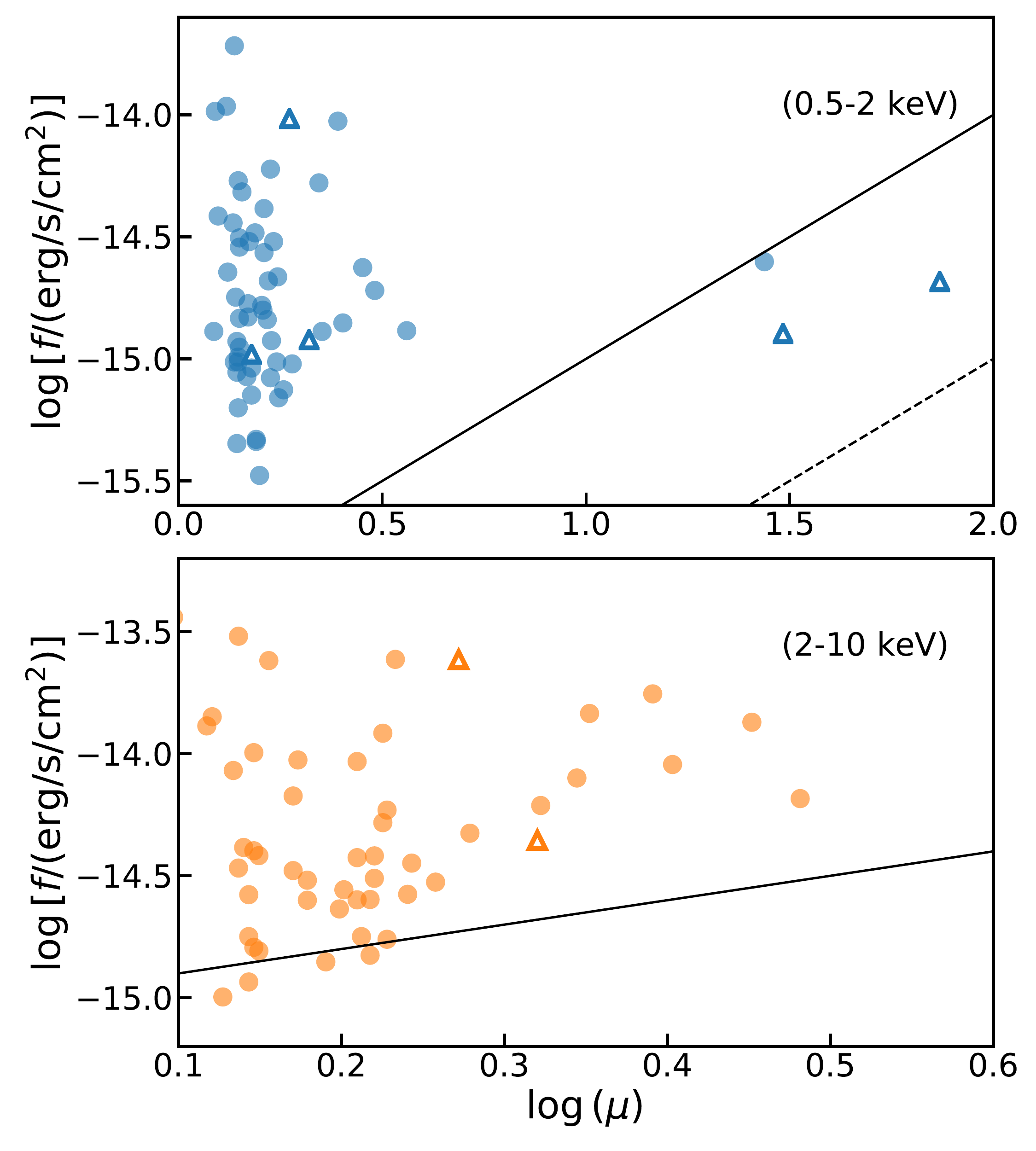}
\caption{Scatter plot of the soft (top panel) and hard (bottom panel) observed flux vs. the 
weighted magnification.  In the top panel the region below the solid and dashed black lines 
corresponds to delensed fluxes below $10^{-16}$ and $10^{-17}$ erg/s/cm$^{2}$, respectively. 
In the bottom panel, the region below the black line corresponds to delensed fluxes below 
$10^{-15}$ erg/s/cm$^{2}$. Sources marked with circles are X-ray selected, 
and triangles correspond to sources identified as X-ray counterparts of 
optically lensed sources (listed in Table \ref{phot_src}). Only sources with $S/N\geq 2$ 
in each band are shown.}
\label{flux_vs_mu}
\end{center}
\end{figure}

The properties of the final sample of 66 X-ray selected sources, including 
observed flux and weighted magnification, are shown in Table \ref{x_src}.  In Fig. \ref{flux_histo}
we show the histogram distribution of the lensed and delensed fluxes in the soft and hard band.  
The lensing effect consists of a shift of about 
a factor of 2 in sensitivity for the bulk of the sources in both bands. 
However, the sensitivity of each field is lower on average than the sensitivity of a blank field with the same {\sl Chandra} exposure because of the foreground ICM emission.  
A similar result can be seen in the scatter plots of observed flux versus magnification 
in Fig. \ref{flux_vs_mu}. Only two sources in the soft band occupy the 
lower right part of the plot, where the combination of low flux and magnification brings the 
delensed flux range below the $10^{-16}$ erg/s/cm$^{2}$ limit, marked by the solid black line.  
None of our sources is identified below $10^{-17}$ erg/s/cm$^{2}$ (dashed blackline), which
is about three times higher than the flux limit of the CDFS in the soft band. 
In the hard band (lower panel) the solid line marks the delensed flux limit $10^{-15}$ erg/s/cm$^{2}$
, which is 1.5 orders of magnitude higher than the hard-band flux limit in the CDFS.  
Only two sources (with low magnifications) are observed below this limit. 
Clearly, these detection limits are significantly affected by the ICM 
emission, an important effect that we better quantify in the next two subsections. 

\begin{table*}
\caption{\label{x_src} X-ray lensed source candidates identified 
with untargeted X-ray detection (complementary to those in Table \ref{phot_src}). Columns 3 and 4 show the X-ray position.  In columns 5 and 6 we list the 
detected net counts in the soft and hard band, respectively.  Soft- and 
hard-band fluxes (columns 7 and 8) are
observed values, therefore the intrinsic fluxes can be obtained by dividing by the weighted magnification
$\mu_{\rm w}$ (column 9).}
\footnotesize
\begin{center}
\begin{tabular}[width=\textwidth]{lcccccccc}
\hline\hline
Cluster & ID & RA & Dec & $N_{\rm soft}$ & $N_{\rm hard}$ & $f_{\rm soft}$ & $f_{\rm hard}$ & $\mu_{\rm w}$ \\
 &  & [h:m:s] & [d:m:s] &  &  & [$10^{-16}$erg/s/cm$^2$] & [$10^{-16}$erg/s/cm$^2$] &  \\
\hline
AS1063 & 1 & 22:48:37.46 & -44:34:07.6 & $42.9\pm7.4$ & $17.5\pm5.5$ & $34.8\pm5.8$ & $42.5\pm14.7$ & 1.41 \\
AS1063 & 2 & 22:48:51.50 & -44:32:59.1 & $80.4\pm9.4$ & $49.0\pm7.4$ & $59.7\pm6.7$ & $112.3\pm18.7$ & 1.40 \\
AS1063 & 3 & 22:48:37.69 & -44:32:42.4 & $31.5\pm7.5$ & $58.8\pm8.9$ & $26.3\pm6.1$ & $149.6\pm24.9$ & 2.83 \\
AS1063 & 4 & 22:48:45.03 & -44:33:46.9 & $26.2\pm6.2$ & $19.8\pm5.5$ & $19.9\pm4.5$ & $45.9\pm14.0$ & 1.38 \\
AS1063 & 5 & 22:48:48.49 & -44:33:25.7 & $12.9\pm4.8$ & $12.7\pm4.6$ & $9.8\pm3.5$ & $29.4\pm11.8$ & 1.39 \\
AS1063 & 6 & 22:48:34.18 & -44:33:21.5 & $-2.2\pm3.5$ & $23.7\pm6.2$ & $-1.8\pm2.8$ & $58.0\pm16.6$ & 1.68 \\
AS1063 & 7 & 22:48:32.98 & -44:33:18.3 & $5.3\pm4.3$ & $15.8\pm5.5$ & $4.7\pm3.7$ & $41.7\pm15.9$ & 1.62 \\
AS1063 & 8 & 22:48:33.21 & -44:32:54.7 & $2.6\pm4.2$ & $26.5\pm6.5$ & $2.2\pm3.4$ & $65.3\pm17.5$ & 1.69 \\
AS1063 & 9 & 22:48:54.55 & -44:30:24.4 & $3.8\pm2.5$ & $11.9\pm3.8$ & $2.9\pm1.9$ & $28.1\pm9.8$ & 1.65 \\
AS1063 & 10 & 22:48:52.74 & -44:29:31.0 & $14.3\pm4.0$ & $3.2\pm2.2$ & $12.4\pm3.3$ & $8.8\pm6.5$ & 1.41 \\
AS1063 & 11 & 22:48:32.88 & -44:32:01.4 & $8.9\pm4.6$ & $12.6\pm5.0$ & $7.9\pm3.9$ & $33.7\pm14.8$ & 1.51 \\
M0329 & 1 & 03:29:37.00 & -02:12:19.2 & $24.2\pm5.1$ & $11.4\pm3.5$ & $23.2\pm4.7$ & $42.4\pm14.5$ & 1.66 \\
M0329 & 2 & 03:29:41.76 & -02:10:57.3 & $60.5\pm8.1$ & $23.6\pm5.1$ & $58.5\pm7.6$ & $88.4\pm21.1$ & 2.21 \\
M0329 & 3 & 03:29:35.82 & -02:10:30.1 & $28.8\pm5.6$ & $6.9\pm3.0$ & $30.3\pm5.7$ & $28.0\pm13.5$ & 1.62 \\
M0329 & 4 & 03:29:38.08 & -02:10:17.7 & $8.6\pm3.4$ & $8.9\pm3.3$ & $8.3\pm3.1$ & $33.1\pm13.6$ & 1.81 \\
M0416 & 1 & 04:16:08.57 & -24:05:22.1 & $120.0\pm12.0$ & $301.7\pm17.9$ & $33.6\pm3.2$ & $270.8\pm17.6$ & 1.71 \\
M0416 & 2 & 04:16:10.14 & -24:05:10.2 & $18.4\pm6.0$ & $4.5\pm4.2$ & $5.1\pm1.6$ & $3.9\pm4.1$ & 1.55 \\
M0416 & 3 & 04:16:10.82 & -24:04:47.7 & $48.6\pm8.2$ & $22.1\pm6.0$ & $13.2\pm2.2$ & $19.3\pm5.8$ & 1.69 \\
M0416 & 4 & 04:15:59.88 & -24:05:18.2 & $4.6\pm3.7$ & $12.1\pm5.1$ & $1.3\pm1.0$ & $11.2\pm5.2$ & 1.34 \\
M0416 & 5 & 04:16:17.29 & -24:02:55.2 & $17.4\pm4.5$ & $14.0\pm4.2$ & $5.0\pm1.3$ & $12.9\pm4.3$ & 1.39 \\
M0416 & 6 & 04:16:13.58 & -24:02:43.3 & $17.6\pm4.9$ & $16.7\pm4.8$ & $5.2\pm1.4$ & $15.6\pm4.9$ & 1.55 \\
M0416 & 7 & 04:16:03.68 & -24:04:29.1 & $12.8\pm5.8$ & $27.7\pm6.7$ & $3.7\pm1.6$ & $25.7\pm6.8$ & 1.58 \\
M0429 & 1 & 04:29:34.42 & -02:52:08.3 & $6.2\pm2.8$ & $-0.4\pm0.7$ & $18.4\pm8.1$ & $-5.4\pm8.8$ & 1.60 \\
M1115 & 1 & 11:15:54.79 & +01:27:31.6 & $29.2\pm5.5$ & $18.1\pm4.5$ & $40.1\pm7.4$ & $94.9\pm25.7$ & 1.36 \\
M1115 & 2 & 11:15:53.66 & +01:27:48.6 & $39.1\pm6.4$ & $51.2\pm7.3$ & $53.7\pm8.5$ & $267.7\pm41.8$ & 1.43 \\
M1115 & 3 & 11:15:53.48 & +01:28:59.8 & $76.4\pm9.0$ & $37.4\pm6.4$ & $104.7\pm12.0$ & $195.5\pm36.5$ & 2.46 \\
M1115 & 4 & 11:15:54.95 & +01:30:21.8 & $10.6\pm3.7$ & $31.1\pm5.7$ & $14.4\pm4.8$ & $162.5\pm33.0$ & 2.25 \\
M1115 & 5 & 11:15:52.61 & +01:29:13.9 & $10.6\pm4.4$ & $3.5\pm2.9$ & $14.5\pm5.8$ & $18.2\pm16.8$ & 3.63 \\
M1115 & 6 & 11:15:56.01 & +01:29:25.3 & $2.9\pm2.3$ & $13.2\pm3.8$ & $3.8\pm3.0$ & $68.2\pm21.7$ & 2.10 \\
M1206 & 1 & 12:06:13.54 & -08:49:22.9 & $27.1\pm5.9$ & $22.8\pm5.6$ & $16.5\pm3.5$ & $36.9\pm9.9$ & 1.48 \\
M1206 & 2 & 12:06:06.82 & -08:49:14.8 & $12.8\pm3.9$ & $12.4\pm4.0$ & $7.0\pm2.0$ & $17.9\pm6.3$ & 1.40 \\
M1206 & 3 & 12:06:08.54 & -08:48:36.5 & $18.7\pm5.0$ & $35.5\pm6.5$ & $10.6\pm2.8$ & $52.5\pm10.6$ & 1.90 \\
M1206 & 4 & 12:06:12.98 & -08:49:08.6 & $28.3\pm6.3$ & $11.1\pm4.8$ & $16.1\pm3.5$ & $16.6\pm7.8$ & 1.65 \\
M1206 & 5 & 12:06:14.69 & -08:49:05.1 & $16.6\pm5.6$ & $3.8\pm4.3$ & $9.3\pm3.0$ & $5.7\pm6.9$ & 1.68 \\
M1206 & 6 & 12:06:16.81 & -08:47:32.5 & $19.1\pm6.1$ & $19.9\pm6.0$ & $10.8\pm3.3$ & $29.5\pm9.8$ & 1.74 \\
M1206 & 7 & 12:06:15.55 & -08:47:10.3 & $8.8\pm5.0$ & $11.1\pm5.2$ & $5.8\pm3.2$ & $19.0\pm9.8$ & 1.62 \\
M1206 & 8 & 12:06:13.26 & -08:46:30.1 & $19.0\pm5.4$ & $2.1\pm3.7$ & $10.8\pm3.0$ & $3.1\pm6.0$ & 1.40 \\
M1206 & 9 & 12:06:06.86 & -08:49:29.5 & $0.0\pm1.4$ & $6.6\pm3.1$ & $0.0\pm0.7$ & $9.6\pm4.9$ & 1.35 \\
M1206 & 10 & 12:06:17.53 & -08:48:57.7 & $-1.0\pm3.3$ & $12.3\pm5.1$ & $-0.6\pm1.9$ & $19.8\pm9.0$ & 1.63 \\
M1311 & 1 & 13:11:03.78 & -03:11:33.2 & $139.7\pm11.8$ & $54.7\pm7.4$ & $213.2\pm17.5$ & $336.8\pm50.3$ & 1.37 \\
M1311 & 2 & 13:10:58.70 & -03:10:45.2 & $6.5\pm2.7$ & $5.7\pm2.5$ & $10.8\pm4.4$ & $37.8\pm18.4$ & 1.37 \\
M1931 & 1 & 19:31:48.08 & -26:36:53.5 & $49.3\pm7.1$ & $32.4\pm5.8$ & $45.9\pm6.4$ & $103.3\pm20.3$ & 1.62 \\
M1931 & 2 & 19:31:47.54 & -26:35:47.6 & $21.0\pm5.0$ & $39.7\pm6.5$ & $15.6\pm3.6$ & $100.4\pm18.0$ & 2.53 \\
M1931 & 3 & 19:31:55.13 & -26:35:22.8 & $19.9\pm4.7$ & $6.2\pm2.8$ & $16.3\pm3.7$ & $17.3\pm8.5$ & 1.41 \\
M1931 & 4 & 19:31:45.81 & -26:35:14.9 & $3.0\pm2.6$ & $13.7\pm4.0$ & $2.2\pm1.9$ & $34.3\pm11.1$ & 1.66 \\
M1931 & 5 & 19:31:44.72 & -26:35:00.5 & $45.6\pm7.0$ & $42.0\pm6.6$ & $33.6\pm4.9$ & $104.8\pm18.1$ & 1.49 \\
M1931 & 6 & 19:31:44.67 & -26:34:35.5 & $25.4\pm5.3$ & $29.9\pm5.7$ & $18.7\pm3.8$ & $74.6\pm15.5$ & 1.48 \\
M1931 & 7 & 19:31:45.47 & -26:34:34.5 & $3.6\pm2.9$ & $12.4\pm4.0$ & $2.6\pm2.1$ & $30.8\pm10.8$ & 1.59 \\
M1931 & 8 & 19:31:48.05 & -26:32:19.9 & $17.2\pm4.8$ & $7.7\pm3.5$ & $13.1\pm3.5$ & $19.8\pm9.8$ & 1.39 \\
M1931 & 9 & 19:31:45.90 & -26:33:25.9 & $0.3\pm2.4$ & $6.1\pm3.1$ & $0.3\pm1.7$ & $15.7\pm8.8$ & 1.55 \\
M2129 & 1 & 21:29:22.03 & -07:42:44.1 & $7.8\pm2.9$ & $10.8\pm3.3$ & $14.4\pm5.1$ & $78.5\pm26.6$ & 1.22 \\
M2129 & 2 & 21:29:29.36 & -07:42:25.2 & $64.8\pm8.1$ & $19.8\pm4.5$ & $120.5\pm14.5$ & $144.5\pm35.9$ & 1.31 \\
M2129 & 3 & 21:29:21.54 & -07:41:54.7 & $9.3\pm3.3$ & $1.6\pm1.6$ & $17.6\pm6.0$ & $11.6\pm12.7$ & 1.61 \\
M2129 & 4 & 21:29:31.33 & -07:41:36.1 & $17.8\pm4.3$ & $3.8\pm2.0$ & $36.5\pm8.5$ & $30.7\pm18.1$ & 1.54 \\
M2129 & 5 & 21:29:33.88 & -07:40:41.4 & $21.7\pm4.7$ & $52.7\pm7.3$ & $42.8\pm9.0$ & $403.3\pm61.4$ & 1.25 \\
M2129 & 6 & 21:29:34.43 & -07:40:39.0 & $59.7\pm7.8$ & $21.6\pm4.7$ & $114.9\pm14.4$ & $161.9\pm38.9$ & 1.23 \\
M2129 & 7 & 21:29:18.65 & -07:41:59.6 & $13.6\pm3.8$ & $21.6\pm4.7$ & $25.2\pm6.8$ & $157.6\pm37.9$ & 1.32 \\
M2129 & 8 & 21:29:31.84 & -07:41:47.7 & $5.8\pm2.5$ & $5.9\pm2.5$ & $11.3\pm4.6$ & $44.5\pm20.7$ & 1.40 \\
R1347 & 1 & 13:47:33.16 & -11:45:40.3 & $50.0\pm10.4$ & $48.5\pm9.5$ & $21.2\pm4.3$ & $72.8\pm15.6$ & 3.03 \\
R1347 & 2 & 13:47:36.34 & -11:44:42.1 & $55.8\pm8.4$ & $25.9\pm6.0$ & $24.1\pm3.5$ & $39.6\pm10.1$ & 1.75 \\
R1347 & 3 & 13:47:36.97 & -11:44:08.6 & $154.2\pm12.8$ & $88.4\pm9.8$ & $66.6\pm5.3$ & $134.9\pm16.4$ & 1.68 \\
R1347 & 4 & 13:47:25.77 & -11:46:38.4 & $17.8\pm5.1$ & $0.4\pm2.8$ & $7.7\pm2.2$ & $0.6\pm4.6$ & 1.76 \\
R1347 & 5 & 13:47:38.15 & -11:44:44.3 & $21.5\pm5.3$ & $4.7\pm3.4$ & $9.4\pm2.2$ & $7.2\pm5.6$ & 1.47 \\
R1347 & 6 & 13:47:31.07 & -11:43:07.3 & $23.7\pm5.7$ & $18.4\pm5.1$ & $10.2\pm2.4$ & $27.9\pm8.5$ & 1.51 \\
R1347 & 7 & 13:47:31.08 & -11:45:24.0 & $63.7\pm31.2$ & $46.8\pm24.4$ & $27.8\pm13.2$ & $72.5\pm41.4$ & 27.41 \\
R2129 & 1 & 21:29:33.54 & +00:04:22.7 & $17.3\pm4.5$ & $6.0\pm2.8$ & $31.9\pm8.1$ & $43.6\pm22.6$ & 1.41 \\
\hline
\end{tabular}
\end{center}
\end{table*}

\subsection{Completeness and contamination maps with imaging simulations}

\begin{figure*}
\begin{center}
\includegraphics[width=0.245\textwidth, trim=50 20 50 35, clip]{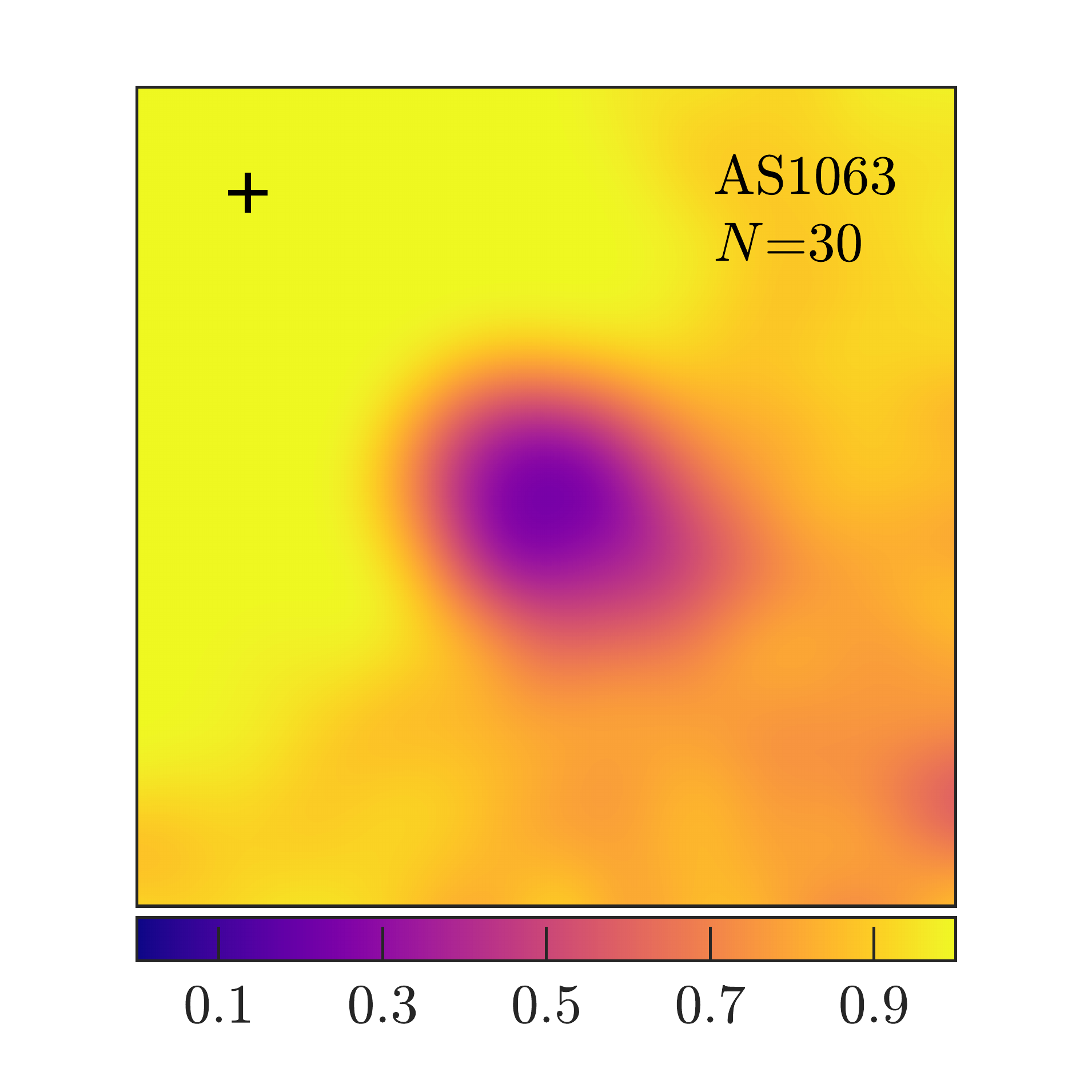}
\includegraphics[width=0.245\textwidth, trim=50 20 50 35, clip]{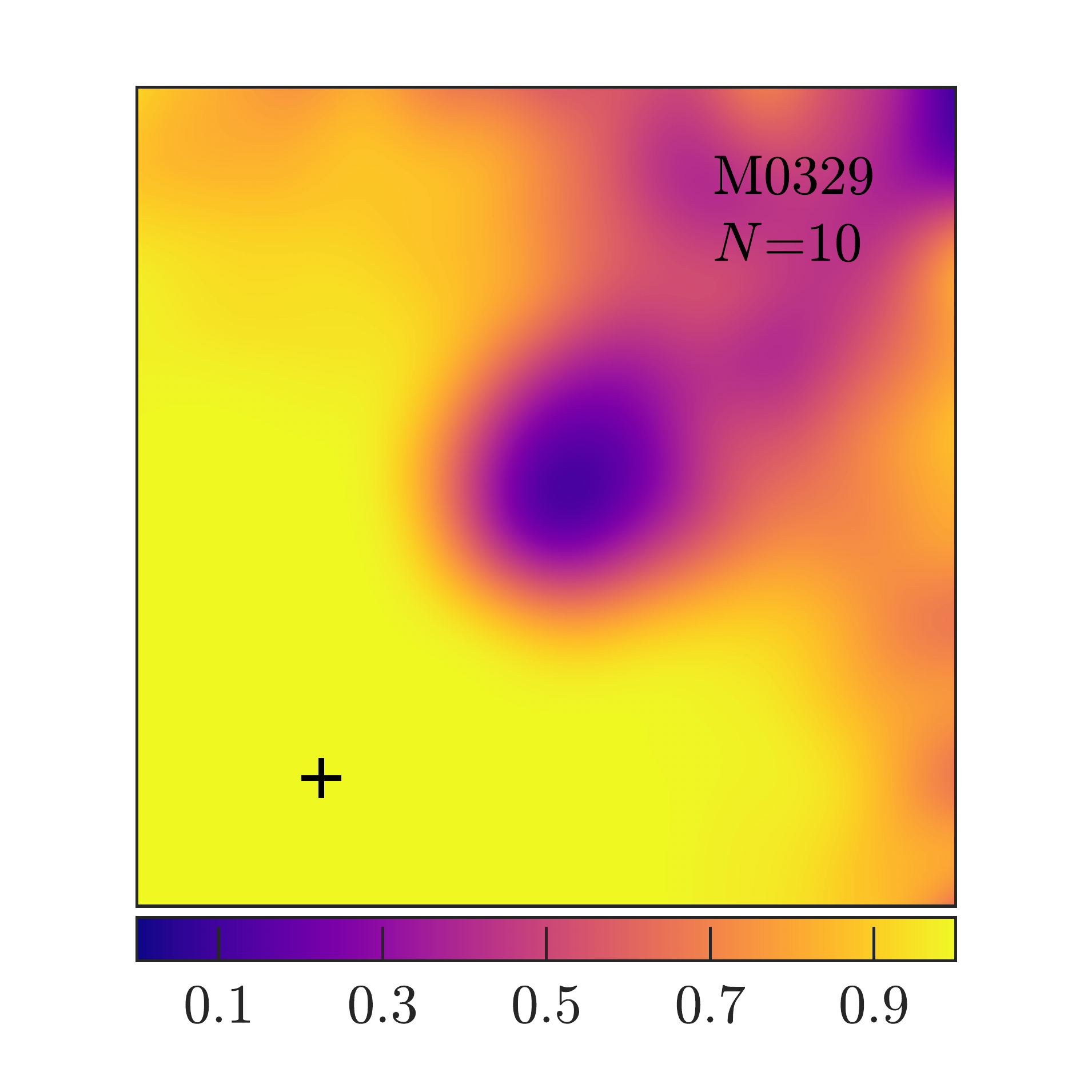}
\includegraphics[width=0.245\textwidth, trim=50 20 50 35, clip]{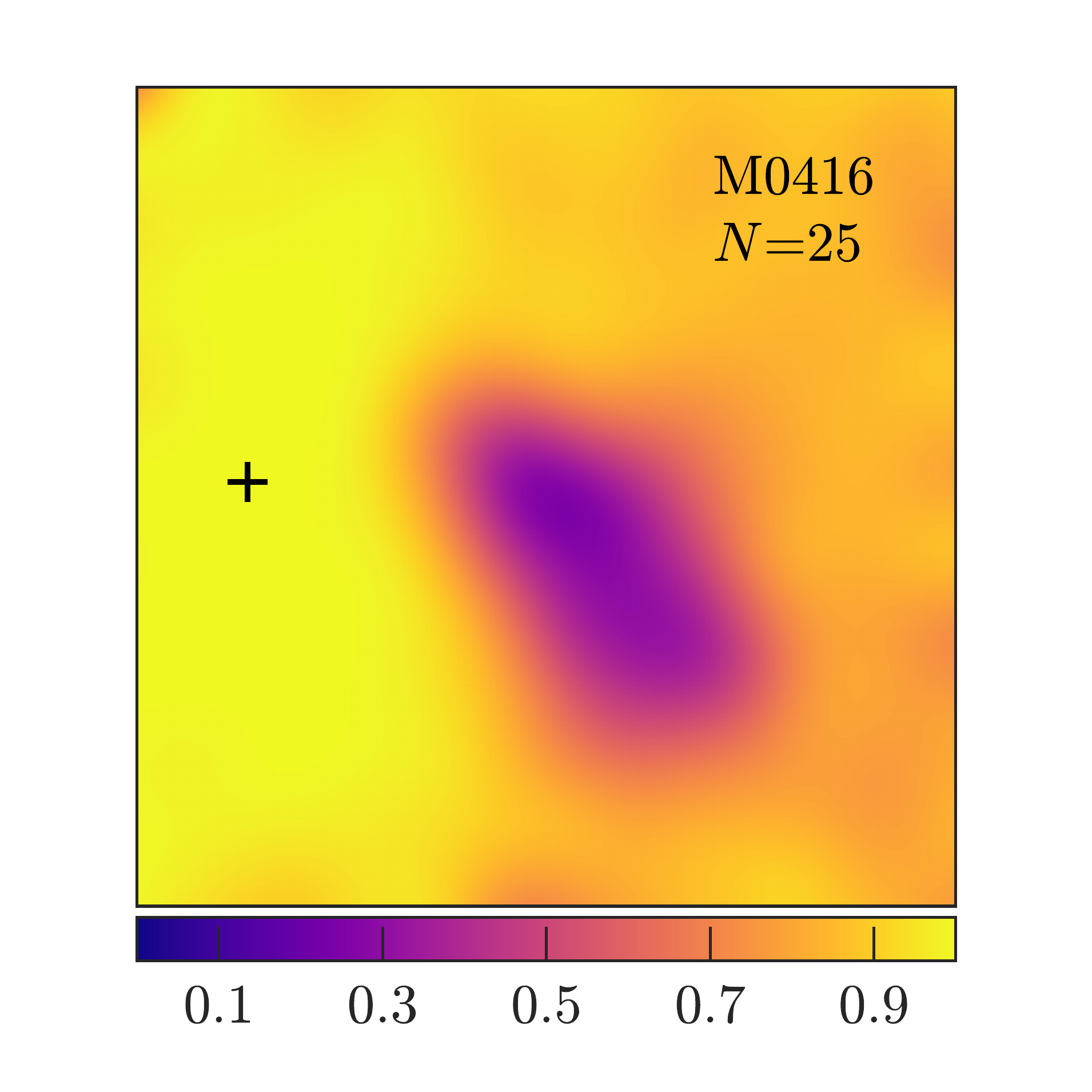}
\includegraphics[width=0.245\textwidth, trim=50 20 50 35, clip]{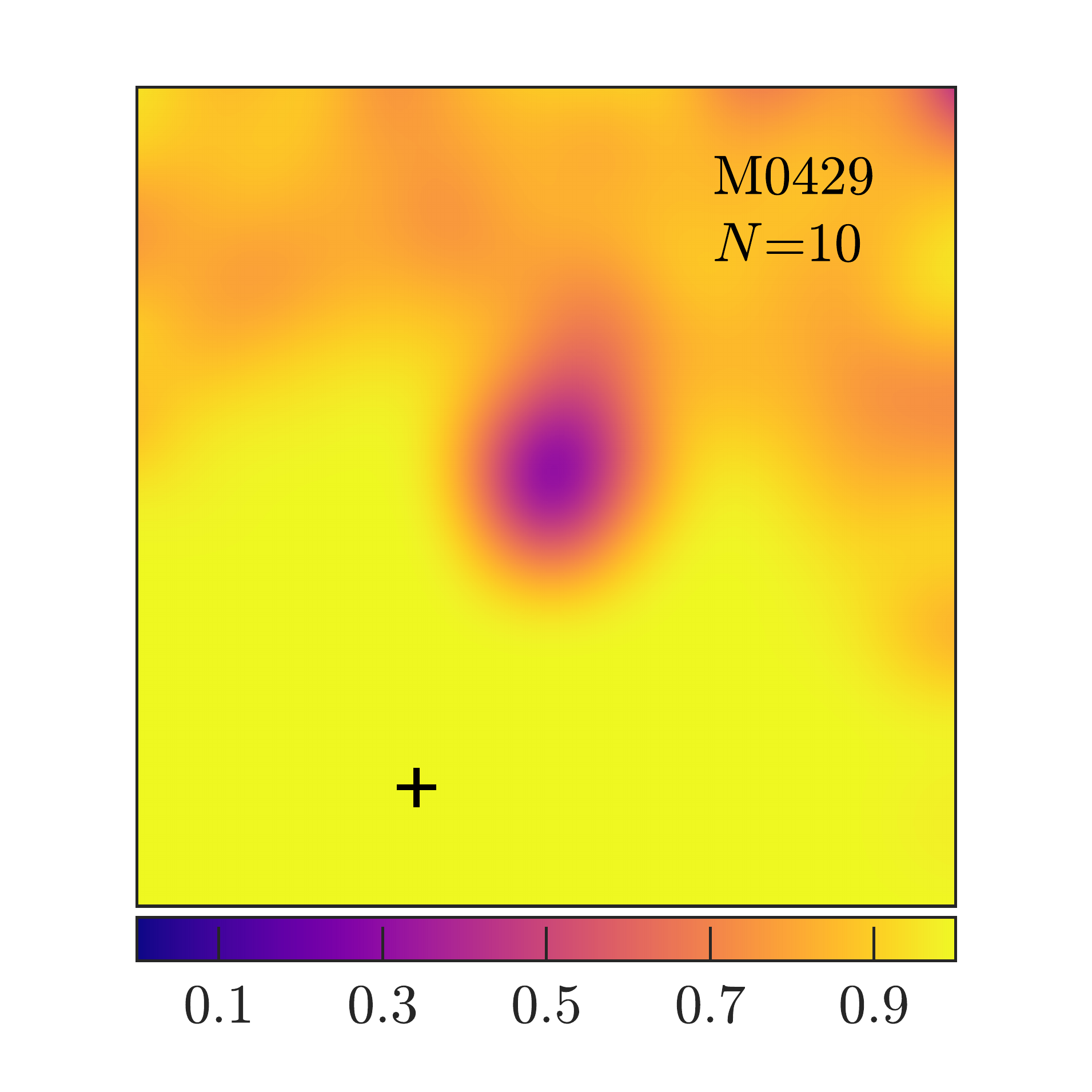}
\includegraphics[width=0.245\textwidth, trim=50 20 50 35, clip]{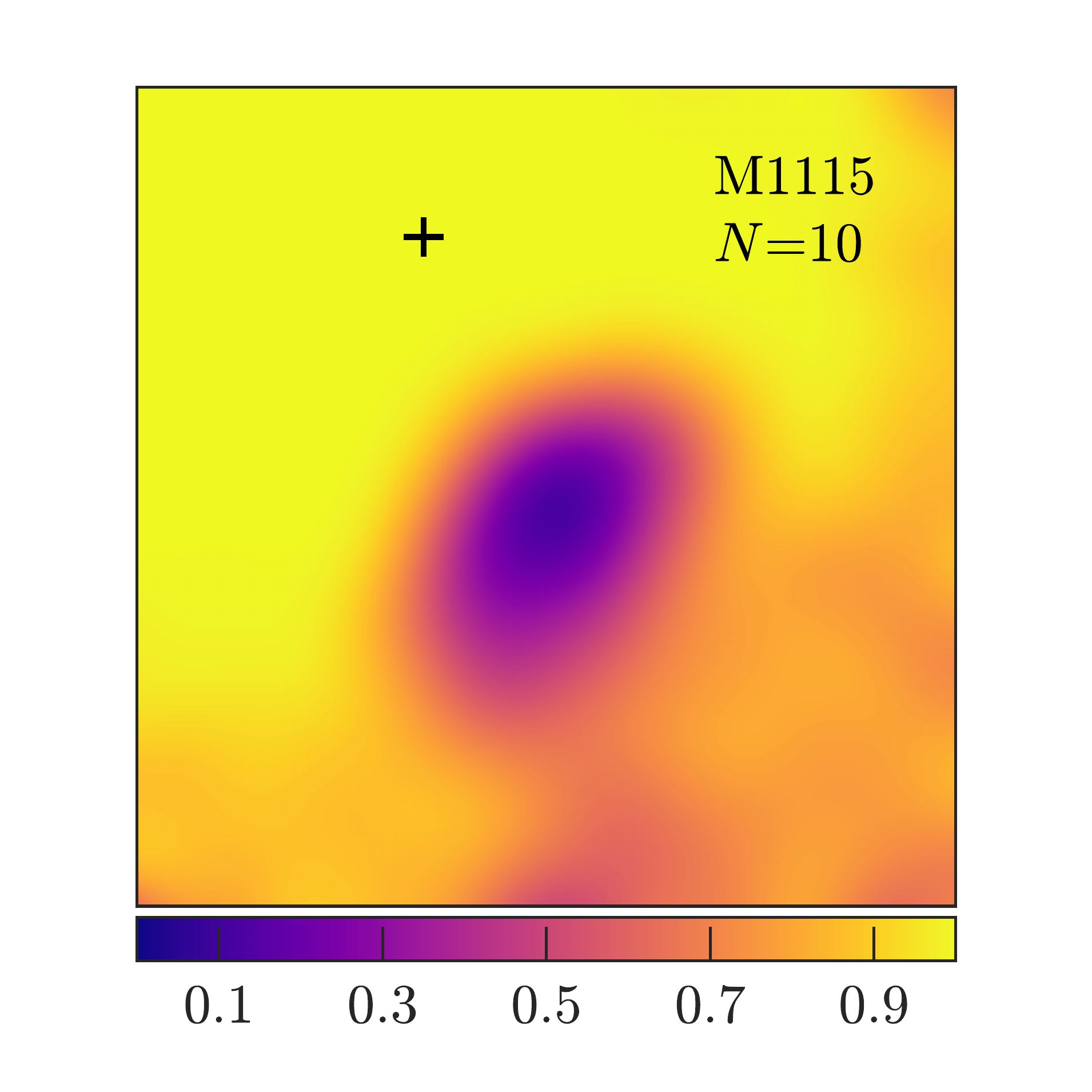}
\includegraphics[width=0.245\textwidth, trim=50 20 50 35, clip]{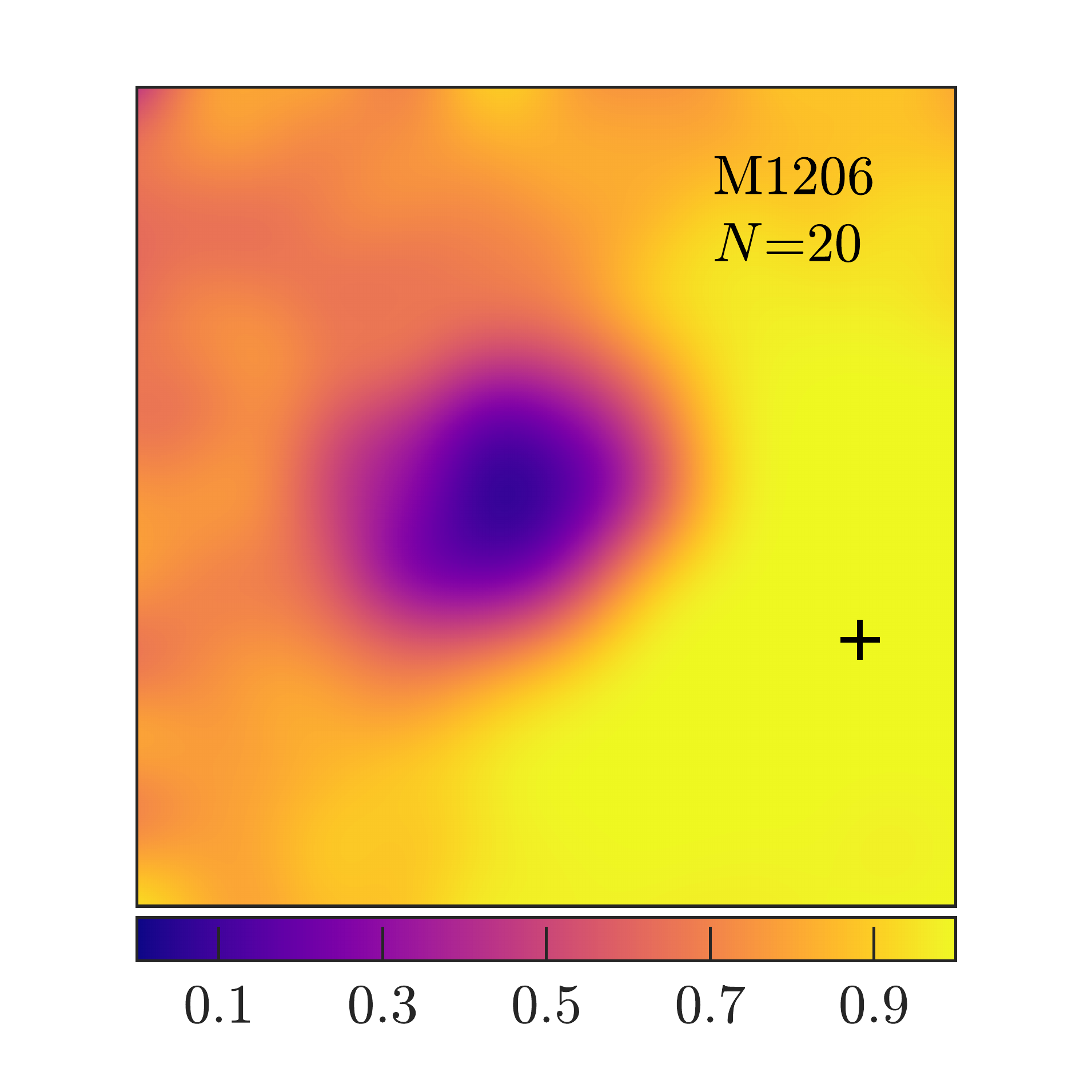}
\includegraphics[width=0.245\textwidth, trim=50 20 50 35, clip]{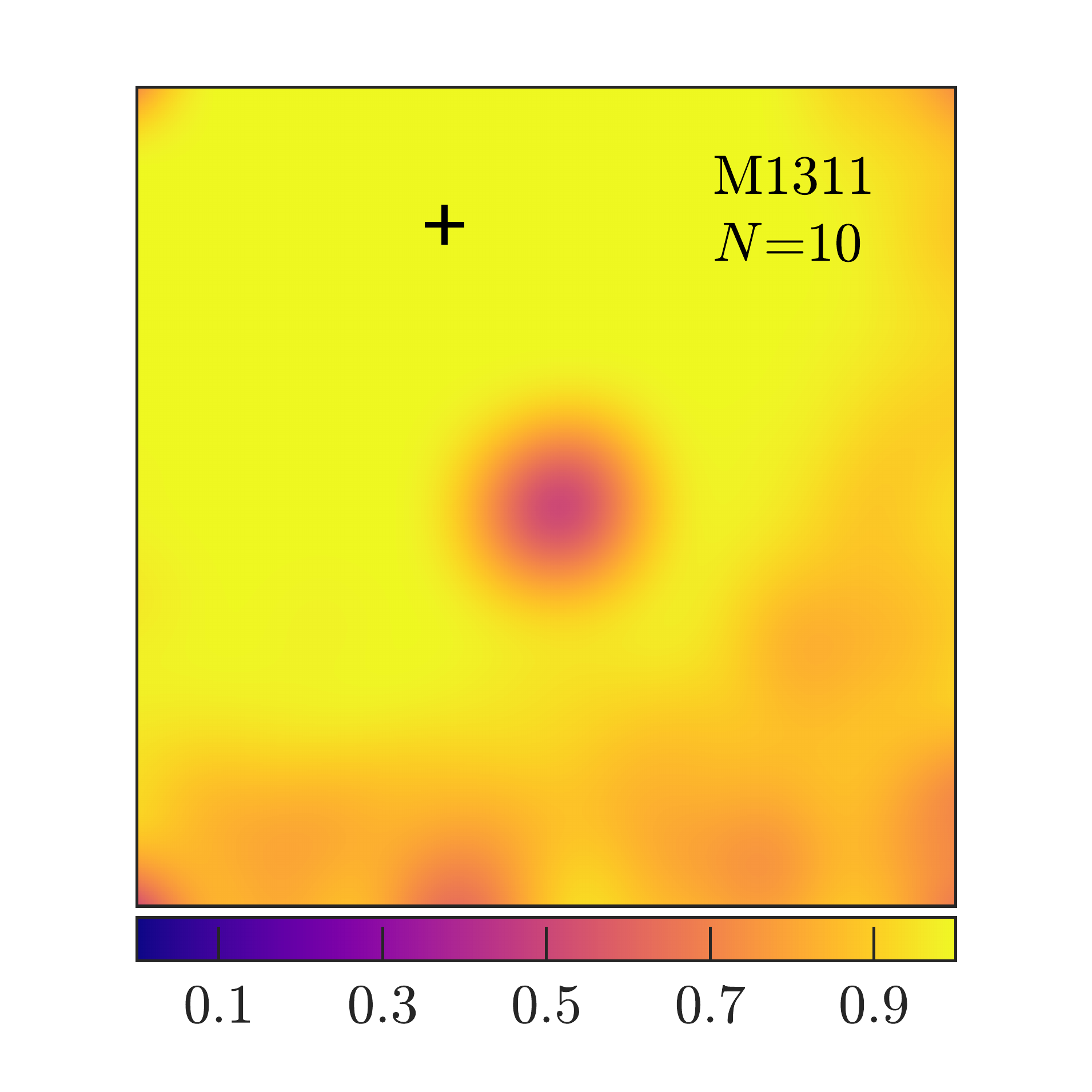}
\includegraphics[width=0.245\textwidth, trim=50 20 50 35, clip]{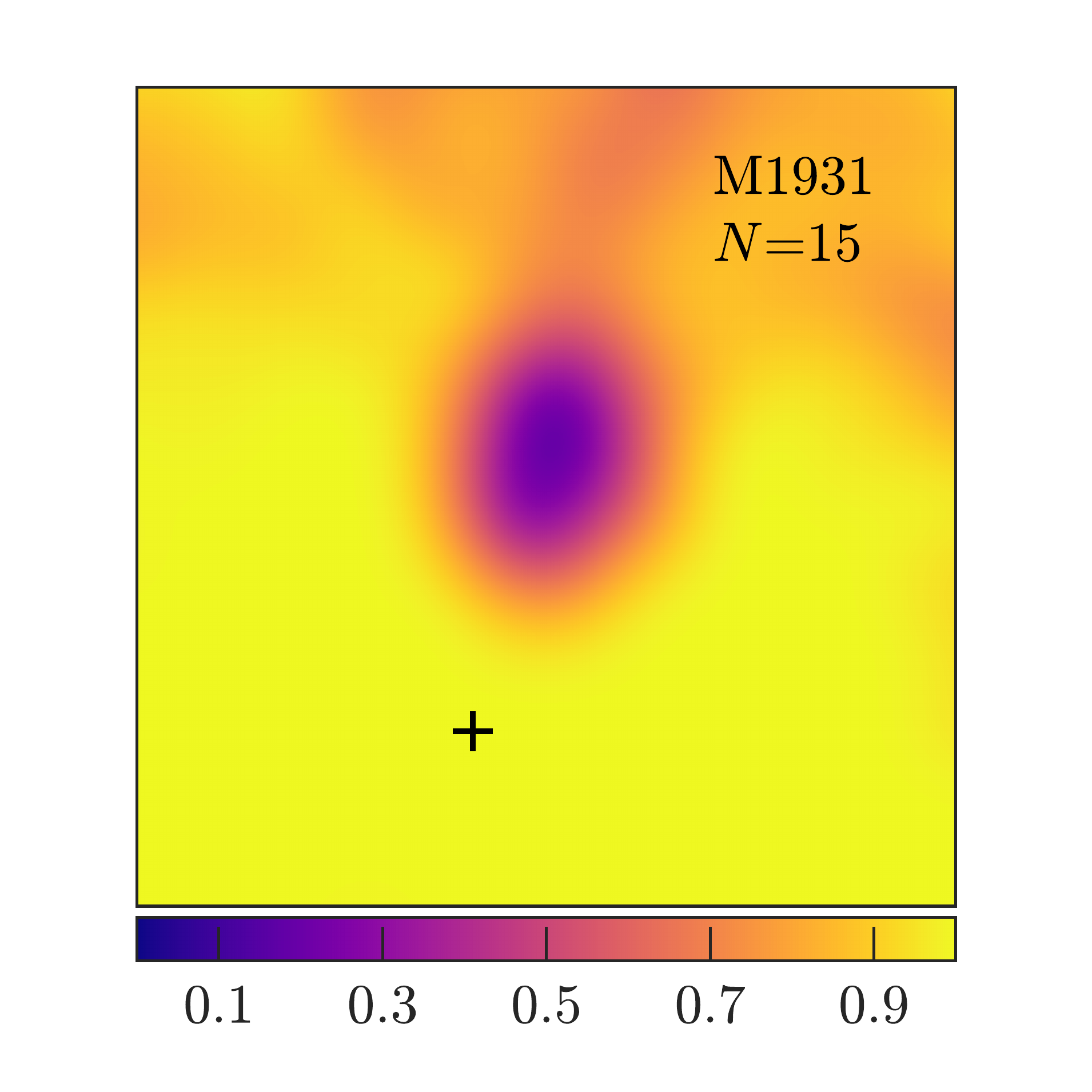}
\includegraphics[width=0.245\textwidth, trim=50 20 50 35, clip]{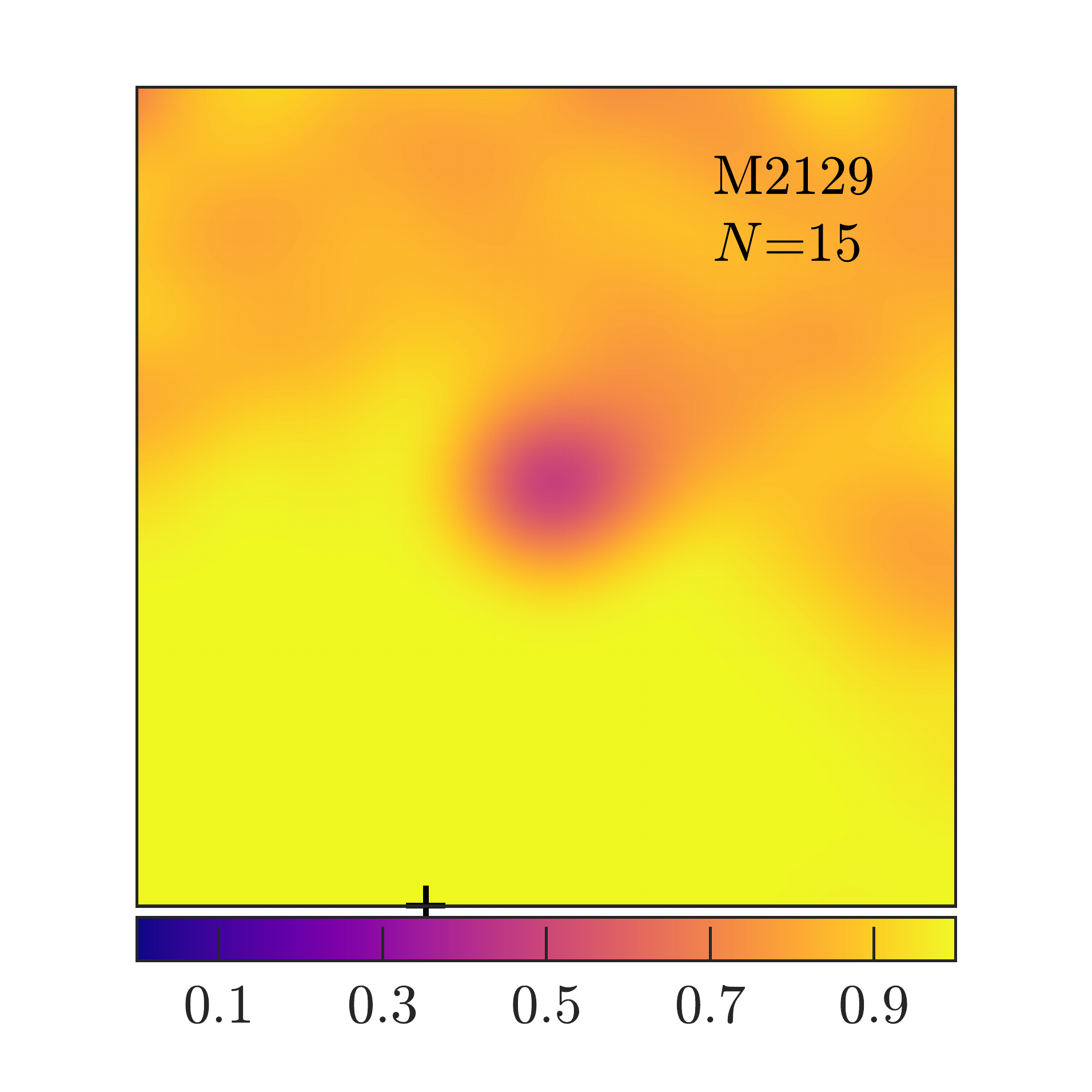}
\includegraphics[width=0.245\textwidth, trim=50 20 50 35, clip]{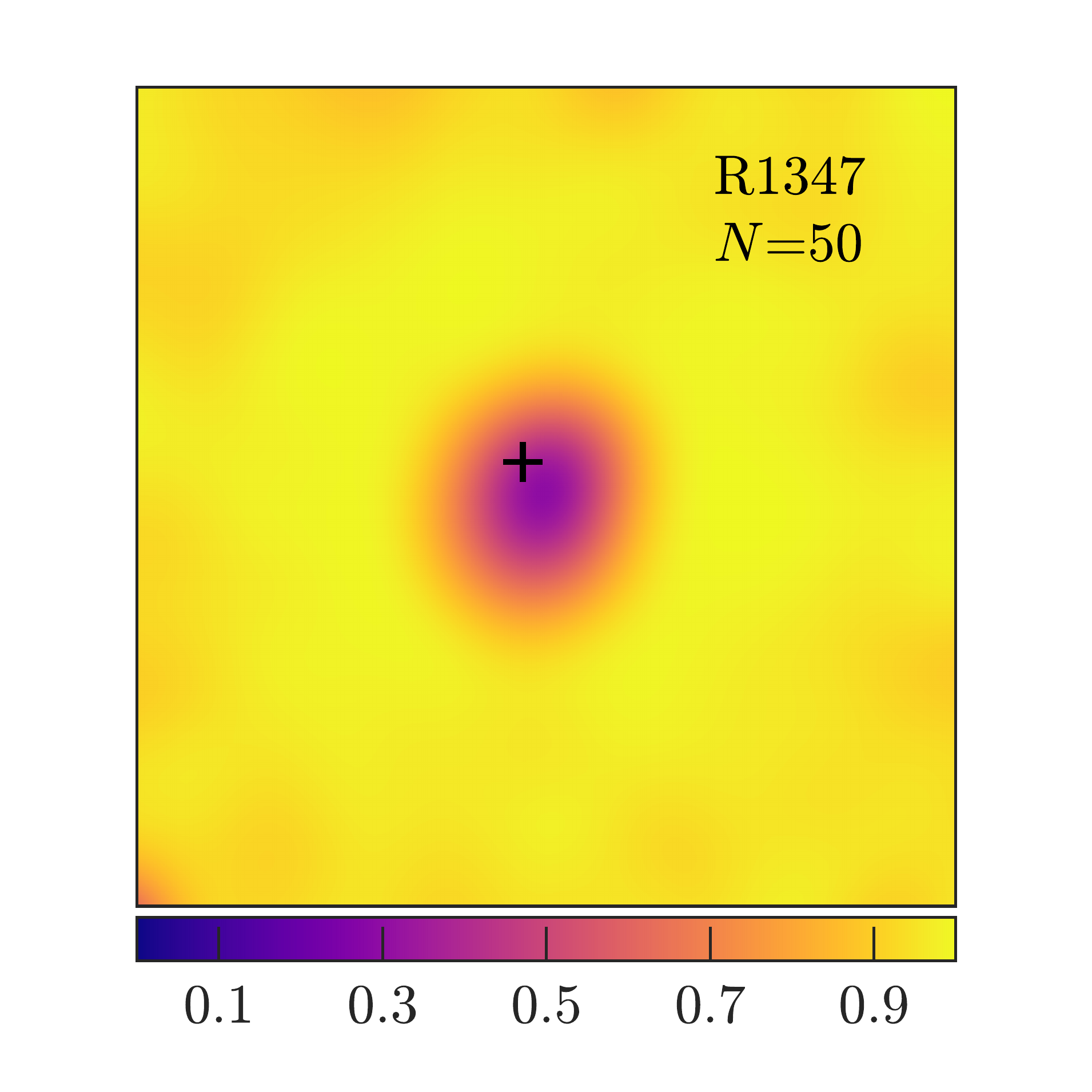}
\includegraphics[width=0.245\textwidth, trim=50 20 50 35, clip]{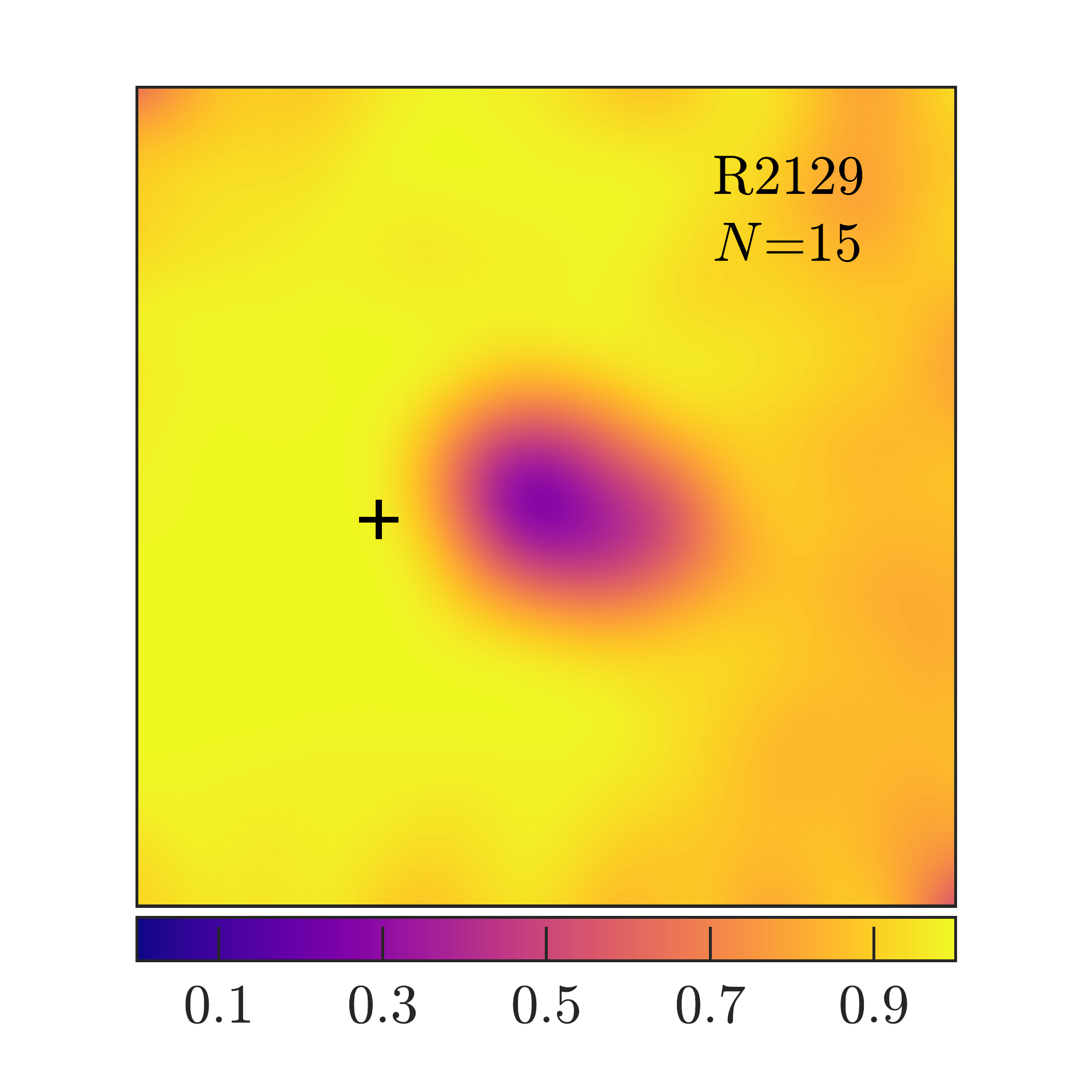}
\caption{Completeness maps in the 0.5--2 keV band for all
the sources with $S/N>2$ detected with {\tt wavdetect} in the 11 
cluster fields. The color code indicates the completeness or recovery rate at a given 
input net counts $N$, as shown in the color bar, in the 0.5--2 keV band. The maps are 5$\arcmin$ across and are smoothed with a Gaussian FWHM of 
$10\arcsec$. The black crosses mark the position of minimum PSF in each field, 
resulting from the overlapping of several exposures with different aim points.
The low recovery rates in cluster centers and in the corners of the field of view
are due to the strong ICM emission and the increase in PSF size, respectively. }
\label{map}
\end{center}
\end{figure*}

\begin{figure}
\begin{center}
\includegraphics[width=0.49\textwidth, trim=5 80 10 140, clip]{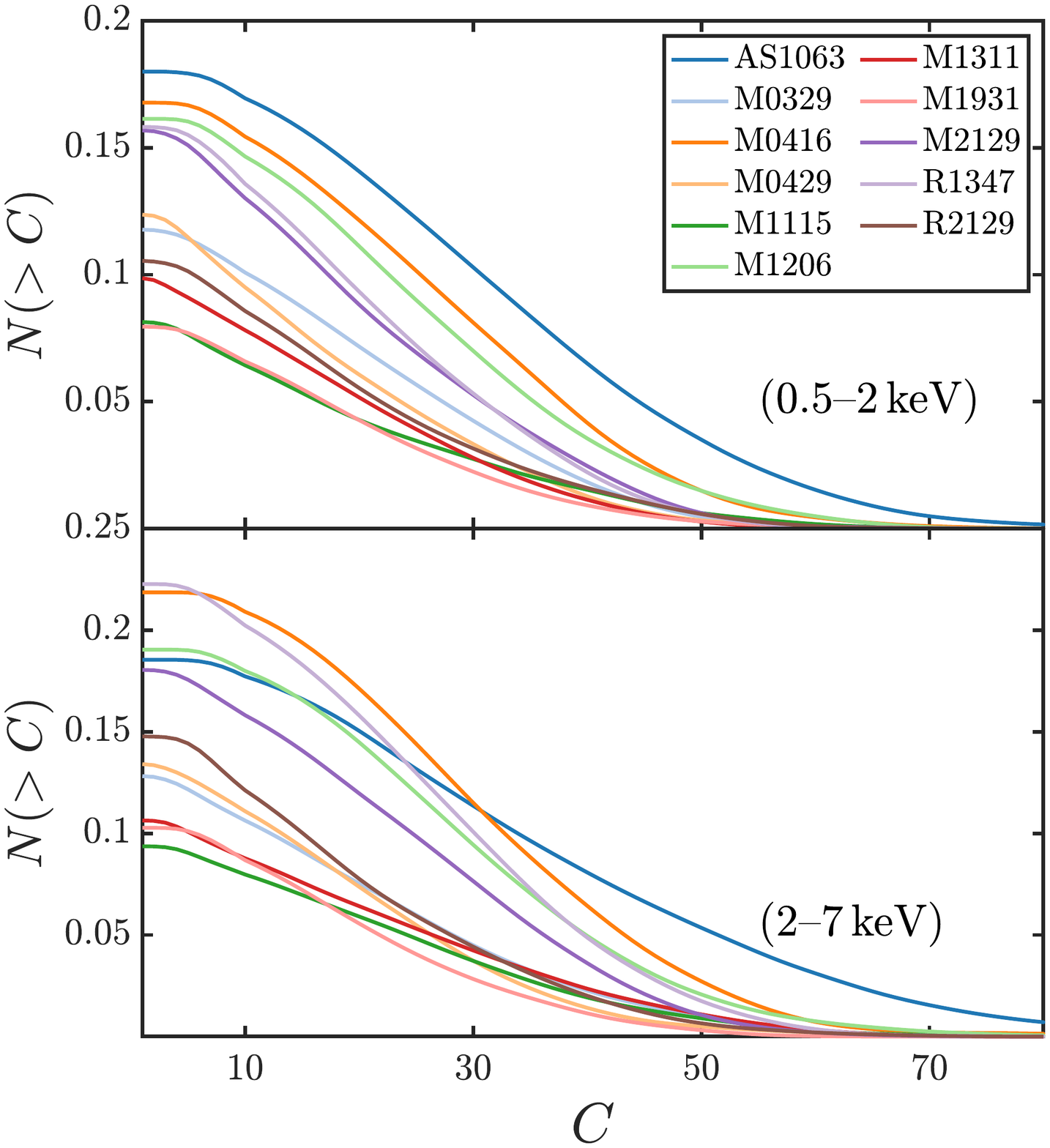}
\caption{Cumulative number of spurious sources $N$ expected per field with measured net counts 
below $C$ for each cluster in the soft (upper panel) and hard (lower panel) bands.}
\label{spur_simu}
\end{center}
\end{figure}

Before computing the flux limit corresponding to our selection threshold 
$S/N>2$, we evaluated the completeness, defined as the ratio of the 
recovered over the total sources above the detection threshold.  This step would not
be necessary if the detection algorithm, in our case {\tt wavdetect}, were able to 
identify all the sources down to $S/N=2$ or below.  Clearly, a solution is to 
raise the selection threshold up to a level above which the detection 
algorithm is complete, but this would result in a significantly lower
number of selected sources and therefore a poor statistics.  Another strategy 
is to refine the detection algorithm at the point of reaching completeness at the 
desired selection threshold, but this would imply a significant amount of work, 
and, unavoidably, an increasing contamination from spurious sources. 
Clearly, the best solution is a compromise, with a reasonably low threshold 
\citep[$S/N>2$ is the same used in the CDFS, see][]{2002RosatiCDFS}, and an
accurate estimate of the completeness (or recovery rate) in order to be
able to apply this correction {\sl a posteriori}. 

To compute the completeness correction, we used a direct approach using imaging simulations.
As a first step, we obtained background- and foreground-only images in the 
soft and hard bands by removing a circle defined by $R_{\rm ext}$ around each source 
identified in the first step with {\tt wavdetect},  
and filling it with a Poisson realization consistent with the surrounding background and foreground. 
Then we simulated point sources distributed randomly in our $5\times 5$ arcmin$^2$
field of view centered on the cluster center, distributing them across the pixels
according to a PSF consistent with that expected at each position in the real data. 
For simplicity, instead of generating the 2D PSF by ray-tracing for each position 
in the image, we used a 1D Gaussian PSF, which is a good approximation to the real composite
PSF except in the outskirts of the field, which are excluded by our selection.
For each field of view and 
a given value of input net counts, we simulated $\sim 1000$ images with only $\text{ten}$ sources 
each to avoid overcrowding.  Because of the low density of X-ray sources in our fields and 
the high angular resolution of {\sl Chandra}, we can ignore confusion effects and correlation.  
The input count for 
each source ranged from 5 to 100, with a step of 5 counts. Simulated images were then analyzed by 
running {\tt wavdetect} with the same threshold parameters of $10^{-5}$ as we used on the real data. 
Then we matched the position of the detected sources with the input positions, adopting a matching 
radius proportional to $R_{\rm ext}$. We 
considered only sources with a measured aperture photometry corresponding to $S/N>2$, following 
the same selection strategy we used for the real data.  To derive the net counts, 
we applied aperture photometry using the background and foreground as modeled by our fit of the 
soft and hard surface brightness. 

Finally, for each simulation, we collected three types of sources: Recovered, 
lost, and spurious sources as a function of the input 
net counts and the position in the image.  The completeness, or recovery rate at $S/N>2$, is 
defined as the ratio of recovered to simulated sources for a given range of input 
net counts.  The completeness maps were obtained in the following way.  
For each pixel in the image, we computed the average recovery rate within 
a circle of $r = 15$ pixels as a function of the input net counts $C$.  
This fraction depends in a nontrivial way on the 
PSF shape and the background and foreground 
level in a given region of the image.  In particular, we note that in most of the cases, 
the aim point of a single observation does not coincide with the cluster center, a choice 
that is often adopted by the observers to avoid the CCD gaps in the core regions.  
While this choice is optimal for the ICM science, it is not optimal for the search 
for unresolved sources close to the critical lines because the PSF quality rapidly 
decreases at off-axis angles larger than 3 arcmin.  As a result, the best PSF of the 
combined image is not obtained at the center of the cluster.  This
aspect, as we discuss below, significantly limits the search for X-ray lensed sources in 
{\sl Chandra} images, as opposed to future X-ray missions that aim at $\sim 1 $ arcsec 
resolution on a large field of view.  

Based on the imaging simulations, we obtained 20 completeness maps, corresponding 
to 20 different values for the input counts, for each cluster in the soft and hard bands
(for a total of 440 completeness maps). Some of them are shown in Fig. \ref{map} 
for some representative values of net input counts.  We note that completeness is 
rather high ($>90$\%, corresponding to bright yellow) for a large part of the 
FOV considered here (5 arcmin by side centered on the cluster).  A significant fraction of the 
field of view has a completeness of $\sim 0.5$, while only the central regions, swamped by the 
core emission, are highly incomplete (up to a factor of 10).  As previously discussed, 
the completeness maps are not symmetrical with respect to the cluster center because of the 
offset of the minimum PSF (marked in Fig. \ref{map} with crosses) with
respect to the cluster, and this significantly affects the capability
of detecting faint sources with our algorithm.
Most importantly, these maps highlight a strong 
difference between the sensitivity and the magnification in the most interesting regions close
to the critical lines. This aspect is further discussed in \S 7. The 
completeness maps were used to weight each source by the probability 
of detecting it according to its position in the field and its measured 
net counts. This weight is assumed to be the inverse of the completeness value.   

\begin{figure}
\begin{center}
\includegraphics[width=0.49\textwidth, trim=5 80 10 140, clip]{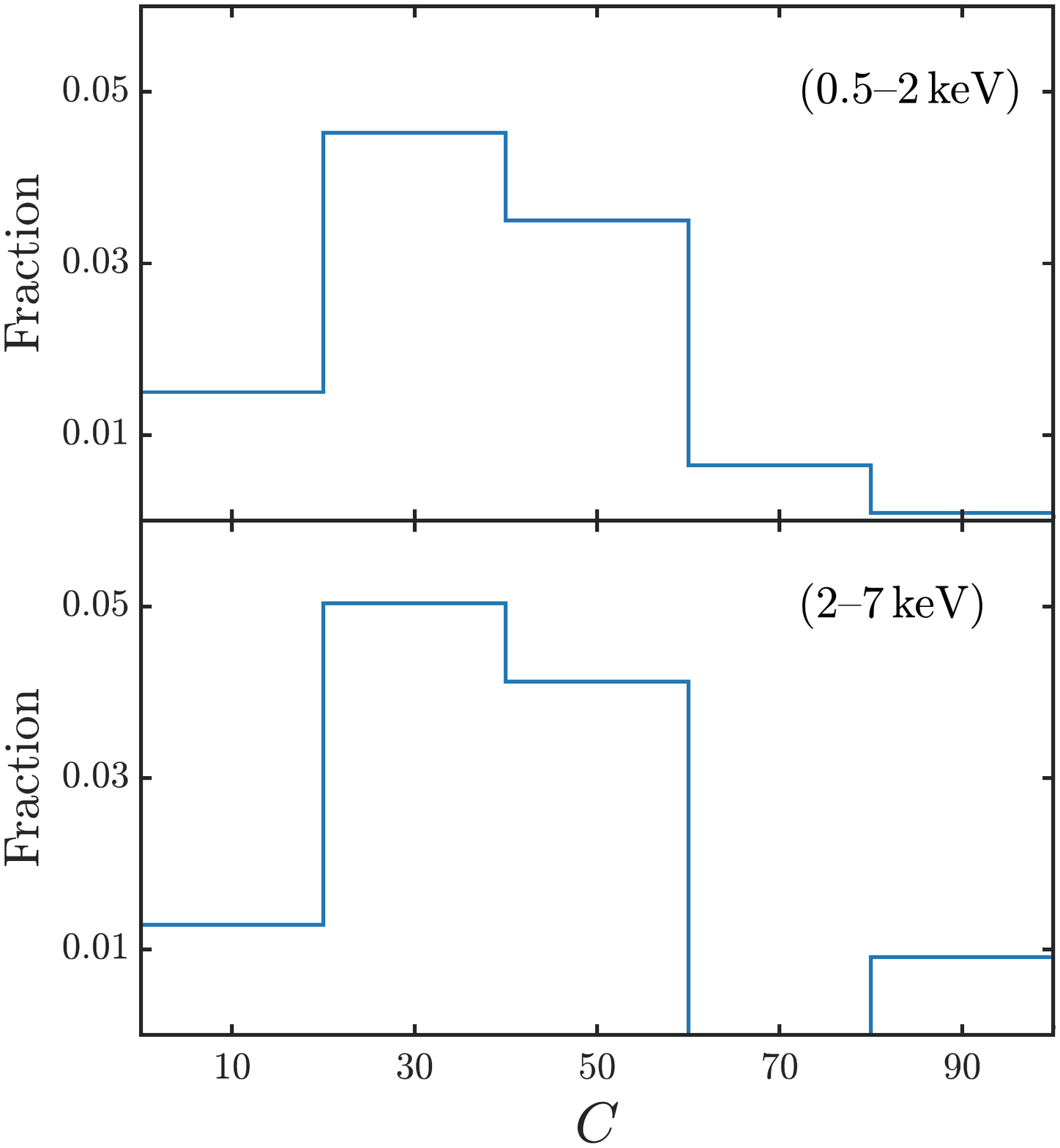}
\caption{Ratio of spurious sources over detected sources in the entire survey in bins of 
observed net counts in the soft (top panel) and hard (bottom panel) bands.}
\label{spur_histo}
\end{center}
\end{figure}

Focusing on the spurious sources, we directly obtained their distribution in each field
in the soft and hard bands by computing their average number at each position for a given
value of the observed net counts.  The expected number of spurious
sources with apparent $S/N>2$ per field is about 0.10--0.15 in the soft and
0.10--0.20 in the hard band, most of which 
are concentrated in the cluster core, as expected.  The cumulative number of spurious
sources expected per field with measured net counts above a given value 
in the soft and hard bands
is shown in Fig. \ref{spur_simu}.  We immediately note that the 
total expected number of spurious sources at $S/N>2$ for the entire survey (the sum of the 
11 fields) is about 2--3 summed in the soft and hard bands.  
Considering that we also applied an accurate visual inspection that led to the removal 
of several unreliable sources in the core region, we can reasonably expect a low 
fraction of spurious sources, distributed in counts according to our simulation as shown 
in Fig. \ref{spur_histo}. The corresponding 
fraction is computed as the ratio of the average number of spurious
sources found with simulation in the entire survey, divided by the detected sources in each count bin.
From this distribution, we can associate with each source that is detected with a given number of counts
in each band the probability of being a real source equal to one minus the spurious fraction.  
The fraction 
of spurious sources drops in the lowest count bin instead of being maximum, as expected. The reason is that the spurious sources with observed counts between 5 and 20 are excluded by the 
$S/N>2$ criterion.  
  
\subsection{Flux-limit maps and sky coverage}

\begin{figure*}
\begin{center}
\includegraphics[width=0.99\textwidth, trim=70 0 70 30, clip]{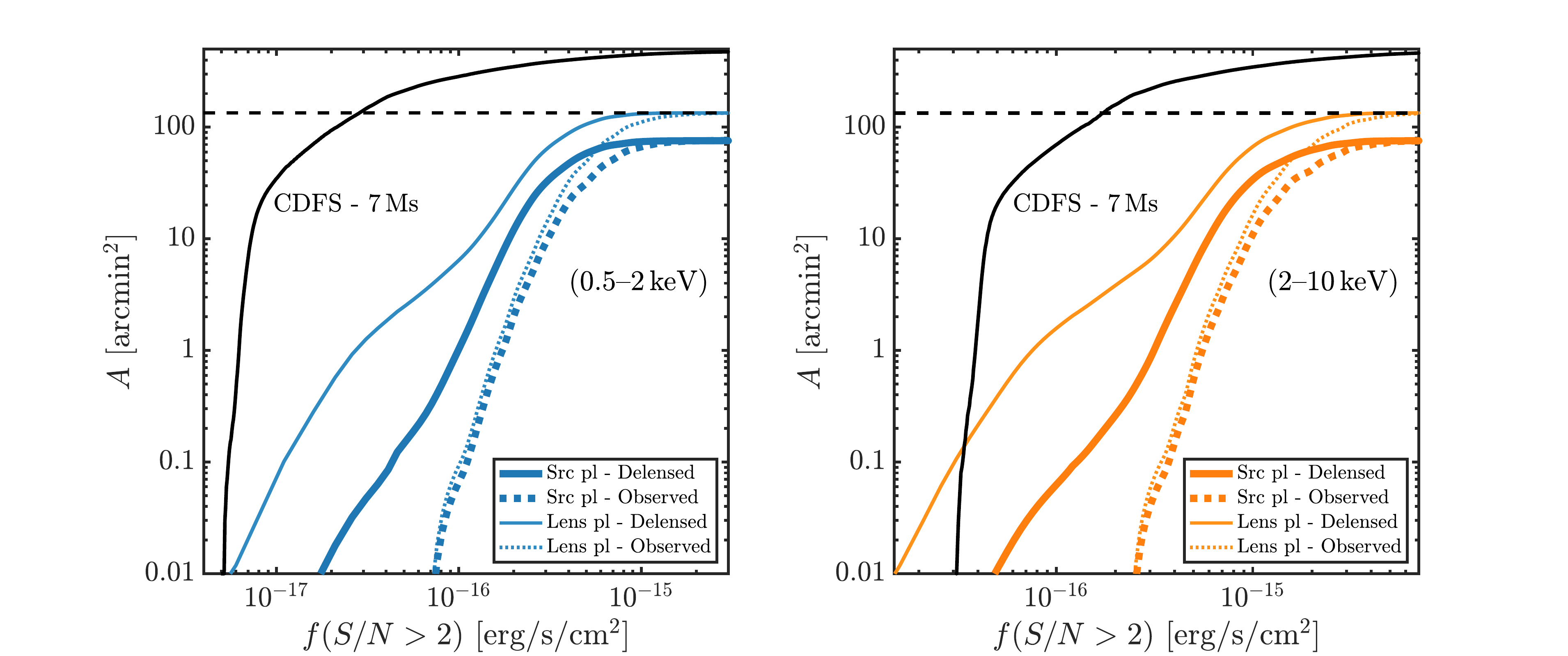}
\caption{Sky coverage as a function of flux for the 11 CLASH cluster fields. The thick solid curve shows the sky coverage in the source plane 
as a function of the de-lensed flux in the soft (left panel) 
and hard (right panel) bands. 
The thick dotted curve instead shows the same 
sky coverage in the source plane, but as a function of the observed flux.
The sky coverage in the source plane was used to compute the delensed
log$N$-log$S$. 
The sky coverage in the lens plane is also shown as a function of the delensed 
flux (continuous curve) and of the observed flux (thin dotted curve). 
The horizontal dashed line indicates the total solid angle, 
by the criterion $\mu_{\rm max}>1.5$. The black curve shows the sky coverage of the 7\,Ms CDFS survey in the corresponding energy ranges \citep{2017Luo}. The hard-band flux limits in \citet{2017Luo} given in 2--7 keV are converted into 2--10 keV by assuming $\Gamma=1.8$.  } 
\label{sky_cov}
\end{center}
\end{figure*}

Finally, we computed the sky coverage as a function of the source flux in the soft 
and hard band separately.  The sky coverage is defined as the total solid angle in 
our survey where a source of a given flux can be detected with our selection 
criteria.  We obtained a flux-limit map by computing the flux corresponding
to our selection threshold $S/N>2$ in each point of our image. 
Technically, the flux limit is computed on the basis of the X-ray images
cleaned from the unresolved sources that we also used for the imaging simulations. 
From the condition $S/N \equiv C/\sqrt{C+2B}>2$, we computed at each position 
the minimum net counts $C_{\rm min}$ that satisfied our selection criterion. We
recall that $B$ represents the counts expected within $R_{\rm ext}$ contributed
by the background and the foreground (ICM). Then 
we considered the net count rate corrected for vignetting and multiplied it by the
average conversion factor, as explained in \S 4.2, to obtain the flux limit at 
each position.  The flux limit therefore depends
on the background and foreground and on the size of the extraction region in which the 
aperture photometry is performed, both of which strongly depend on  the
position of the source in the image.  The total solid angle of our 
survey as a function of the flux $f_X$ 
was obtained by computing the total solid angle whose flux limit $f_{\rm lim}$ is below $f_X$. 
The sky coverage corresponding to a given flux is therefore simply the
solid angle satisfying both $\mu_{\rm max}>1.5$ and $f_{\rm lim}<f_X$.

We can plot the sky coverage measured on the sky (lens plane) or delensed, that is, obtained
by mapping the solid angle in the lens plane back to the source plane and weighting the source-plane
maps by the average redshift distribution. Here we approximated this step by
weighting each pixel by $\mu_{\rm w}^{-1}$.  As described above, weighting the solid angle 
by the inverse of the magnification value in each position does not take into account the
regions in the lens plane that correspond to the same region in the 
source plane. We evaluated the effect of this approximation by comparing the 
average field magnification obtained with the full ray-tracing 
for a specific redshift \citep[see, e.g.,][]{2008Meneghetti,2010Meneghetti,2019Plazas} to that obtained with the simple inversion, and found that 
the differences are in the range 1\%--3\%, with a maximum of 6\% in the case of R1347. 
Because the uncertainties associated with the unknown redshift of the sources are 
significantly larger, we conclude that our strategy to compute the solid angle in the source plane 
by dividing the lens plane value by $\mu_{\rm w}$ is acceptable. 

The two sky coverage functions for the 
entire survey (11 CLASH fields) as a function of the observed (lensed) flux are 
shown in Fig. \ref{sky_cov} with dashed lines in the soft and hard band. The maximum solid 
angle in the lens plane, defined by the condition $\mu_{\rm max}>1.5$ (marked with a 
horizontal dashed line), is reached 
at $\sim 10^{-15}$ and $\sim 4 \times 10^{-15}$ erg/s/cm$^2$ in the soft and hard bands, respectively.  
The survey limits, conventionally defined by a sky coverage as small as $0.1$ arcmin$^2$, is reached at 
$10^{-16}$ and $3.5 \times 10^{-16}$ erg/s/cm$^2$ in the soft and hard bands, respectively.

In Fig. \ref{sky_cov} we also plot the lens- and source-plane sky coverage as a function of the
delensed flux (solid lines).  The sky coverage in the lens-plane as 
a function of the
delensed flux is significantly deeper.  However, a more useful quantity is the sky coverage in the
source plane versus the delensed flux.  When we compare this to the same quantity plotted 
versus the observed flux, the effect resulting from the combination of magnification 
and source dilution of gravitational lensing is immediately clear.  The net result is that the 
lens allows us to reach deeper on average by only a factor of two, as we already noted when 
we compared the distribution of observed and delensed fluxes.  
The same behavior is observed in the soft and hard bands.  Unfortunately, 
when combined with the flattening of the soft and hard bands log$N$-log$S$ at the faint end, 
this implies that an increase of a factor of $\sim 2$ in depth corresponds to a limited
increase in the number of faint sources.

To compute the log$N$-log$S$, we used the source-plane sky coverage either matching the de-lensed 
flux or the observed flux to the corresponding flux of each source, and we plot the cumulative
source number
as a function of the delensed flux for a direct comparison 
with the log$N$-log$S$ measured in blank fields or modeled.  


\section{log$N$-log$S$ of X-ray lensed sources in CLASH fields}

\subsection{Expected log$N$-log$S$}

The deepest X-ray field to date (and in the future, until
a new X-ray mission with arcsec resolution will be operational) is provided by the 
CDFS \citep[][]{2001Giacconi,2002RosatiCDFS,2017Luo}, which has
reached the unparalleled depth of 7 Ms.  
From the number counts published in \citet[][see their Figure 31]{2017Luo}, we 
note that the density of X-ray sources in the sky is rather low. We considered
the delensed flux of $\sim 10^{-16}$ and  $\sim 3\times 10^{-16}$  erg/s/cm$^{2}$ in the
soft and hard bands, respectively, that in our survey both corresponds to $\sim 1$ arcmin$^2$.
In both cases the expected number of sources is slightly higher than one, implying that 
the CLASH survey runs out of sources below these fluxes.  

However, we cannot directly compare our results with the full number counts as measured 
in the fields of deep or medium-deep surveys such as CDFS or COSMOS. 
We have reported
the number counts of lensed sources only, which implies that we ignored all the nonlensed
foreground sources with $z_{\rm s}<z_{\rm cl}+0.1$.  This choice was adopted to focus on lensed 
sources and avoid being dominated by unlensed sources as in blank fields. 
For a proper comparison, we therefore need to correct the expected
log$N$-log$S$ by removing the sources with $z_{\rm s}<z_{\rm cl}+0.1$. 
To this aim, we used mock catalogs and tools\footnote{http://cxb.oas.inaf.it/} based on the XRB model 
by \citet{2007Gilli} \citep[see also][]{2005Hasinger}, which includes an exponential decline 
in the AGN space density at redshifts above $z = 2.7$ to cope with the results from wide-area surveys 
\citep[e.g., SDSS][]{2006Richards,2006Fan}.
Because the \citet{2007Gilli} model includes only AGN, we added the steep galaxy
number counts as measured by \citet{2017Luo} in the CDFS.  However, normal
(star-forming) galaxies are relevant only at fluxes well below $10^{-16}$ erg/s/cm$^2$. 
This model provides cumulative number counts that agree with those measured
in the SDSSJ1030 \citep{2020Nanni} or the XBootes \citep{2020Masini} fields 
and are slightly higher than those in the COSMOS
field \citep{2016Civano}. The predictions are brought into good agreement with the 
AGN log$N$-log$S$ in the CDFS when the normalization is reduced by a factor 
$\sim 0.8$ in both bands. This is expected because of the already noted 
lower normalization of the cumulative number counts in the CDFS with respect to 
other deep fields \citep[see, e.g., Figure 7 in][ ]{tengliu2020}.  
Considering the uncertainties on the log$N$-log$S$, also associated with the
actual source flux (for which an average spectral shape is assumed that is not representative
of the wide distribution in intrinsic absorption), we decided to focus on the
predictions from \citet{2007Gilli}.  

\begin{figure}
\begin{center}
\includegraphics[width=0.49\textwidth, trim=0 0 0 0, clip]{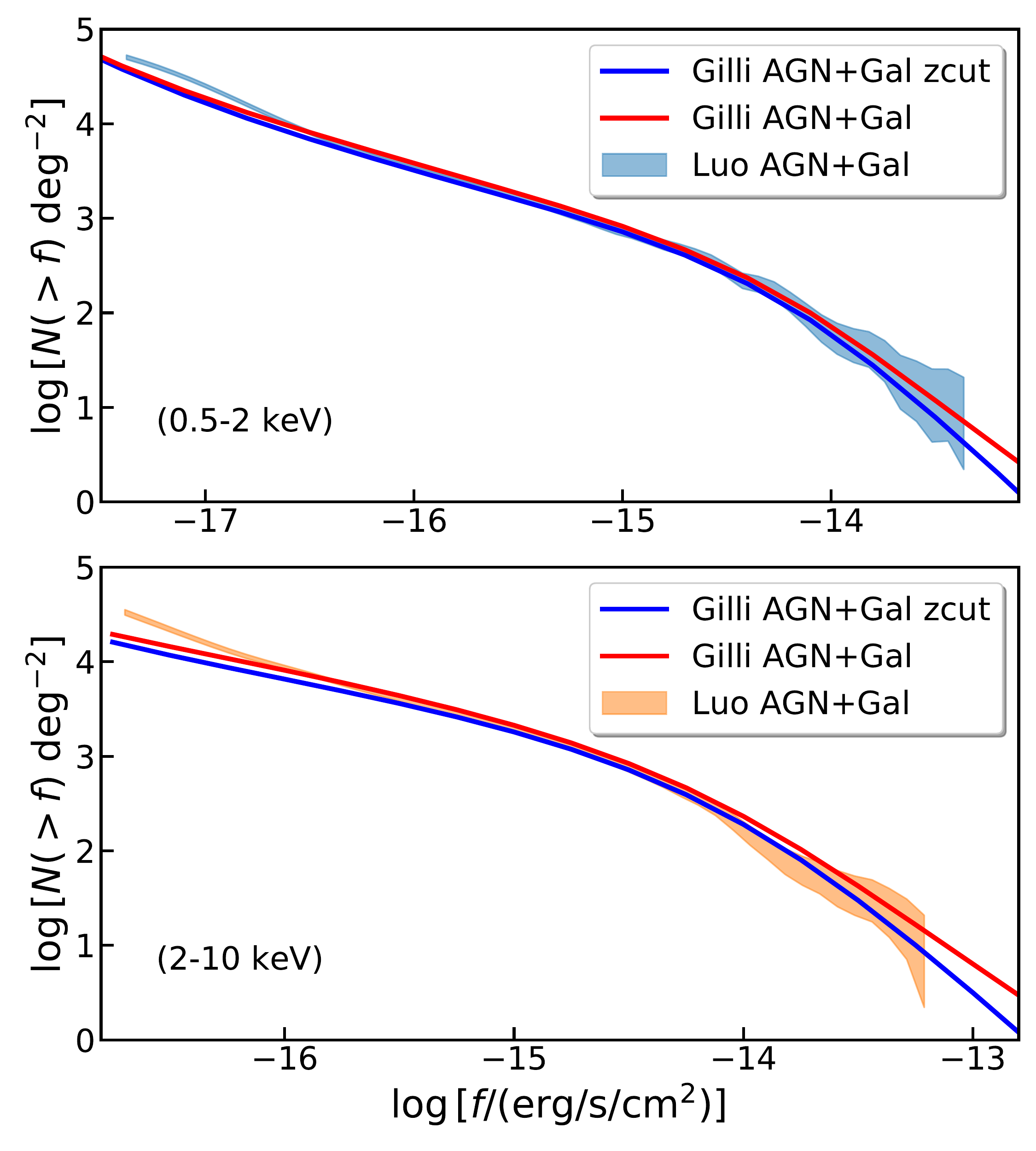}
\caption{Effects of the cuts in redshift for $z_{\rm s}<z_{\rm cl}+0.1$ on the expected log$N$-log$S$ of lensed
X-ray sources in the soft (upper panel) and hard (lower panel) bands. The upper red lines show
the total number counts from the model by \citet{2007Gilli}, and the lower 
blue lines show the number counts expected 
for the lensed sources ($z_{\rm s}<z_{\rm cl}+0.1$) on the basis of \citet{2007Gilli}. The shaded areas are the CDFS number counts from \citet{2017Luo}. The hard-band flux limits in \citet{2017Luo} given in 2--7 keV are converted into 2--10 keV by assuming $\Gamma=1.8$.   The predictions from \citet{2007Gilli} are multiplied by a factor of 0.8 to 
be consistent with the CDFS results.}
\label{lognlogs_zcut}
\end{center}
\end{figure}

In the model by \citet{2007Gilli}, the dependence on  redshift is built in on the basis of the 
luminosity function of AGN as implemented in the model. This allows us to immediately compute  the
expected log$N$-log$S$ only for sources with $z_{\rm s}>z_{\rm cl}+0.1$ in each field. Because the redshift cut
is different in each field, we summed the expected log$N$-log$S$ in each field using a weight 
proportional to the total solid angle considered in our study (defined by the condition 
$\mu_{\rm max}>1.5$.  
The expected cumulative number counts after removing the foreground 
X-ray sources are plotted in Fig. \ref{lognlogs_zcut}. 
As expected, the decrease has a modest effect at the low-flux end, but it starts 
to be more evident at fluxes $S>10^{-14}$ erg/s/cm$^{2}$.

\subsection{Cumulative number counts}

The final list of sources we considered was obtained by adding the sources in Table 
\ref{phot_src} (excluding the two flagged with the asterisk) 
to the X-ray sources detected with {\tt wavdetect} listed
in Table \ref{x_src}. Therefore we considered a total of 68 sources.
Among these we selected for each band those that satisfied $S/N>2$, finding 55 sources in the 
soft band and 56 in the hard band. The cumulative number 
counts per solid angle were obtained by summing all the sources above a given flux, 
where each source was multiplied by three weights: by the inverse of sky coverage 
(computed at the corresponding source flux), 
the inverse of the completeness (which depends on the source position and source net counts), 
and the probability of not being spurious (which depends on the net counts). 
In a blank field (i.e., in absence of lensing), observed and intrinsic fluxes are the same, 
and therefore we have

\begin{equation}
N(>f) = \Sigma_{f_i>f} {{1 \, (1-P_s(C_i))}\over {\Omega(f_i)\, I(x_i,y_i,C_i) }}
,\end{equation}

\noindent
where $\Omega(f_i) = \Sigma_j A_j(f_{\rm lim}<f_i,\mu_{\rm max}>1.5)$ is the sky coverage corresponding 
to the flux of the $i^{}$ th source, 
$I(x_i,y_i,C_i)$ is the completeness at the source position and for the net counts $C_i$ 
measured for that source, and $P_s(C_i)$ is the probability of being a spurious source.  
As previously discussed, the factor $1/\Omega(f_i)$ is the 
weight to obtain the contribution of a given source to the source number density in the 
assumption of perfect detection (all the sources with intrinsic $S/N>2$ are detected), 
while the factor $1/I(x_i,y_i,C_i)$ provides the correction due to the probability of detecting 
this specific source with a given detection algorithm.  The factor $(1-P_s(C_i))$ instead 
provides the small correction due to the probability of including spurious sources in the sample, 
and it only depends on the measured net counts $C_i$.

The additional complication in our survey is that in order to measure the true number counts
(i.e., those that would be measured without the lensing cluster),
we need to delens the observed fluxes and delens
the sky coverage corresponding to a given {\sl \textup{observed}} flux.  
Therefore we define $f_{i,dl}=f_i/\mu$, and 
$\Omega_{dl}(f_{i,dl}) = \Sigma_j A_j(f_{{\rm lim},dl}<f_{i,dl},\mu_{\rm max}>1.5)/\mu_j$, where $f_{{\rm lim},dl}$ is the 
delensed flux limit.  An equivalent quantity can be obtained using the observed flux
and matching it to the observed flux of each source, given that the log$N$-log$S$ is plotted
against the delensed flux.  The two approaches are not identical, but are expected to provide 
consistent results.  We verified that the two log$N$-log$S$ obtained by matching the 
delensed and the observed fluxes have negligible differences. 
Therefore we can obtain the delensed (or source plane) number counts as

\begin{equation}
N_{dl}(>f_{dl}) = \Sigma_{f_{i,dl}>S_{dl}} {{1 \, (1-P_s(C_i))}\over {\Omega_{dl}(f_{i,dl})\, 
I(x_i,y_i,C_i) }}  \, .
\end{equation}

\begin{figure*}
\begin{center}
\includegraphics[width=0.49\textwidth, trim=0 0 0 0, clip]{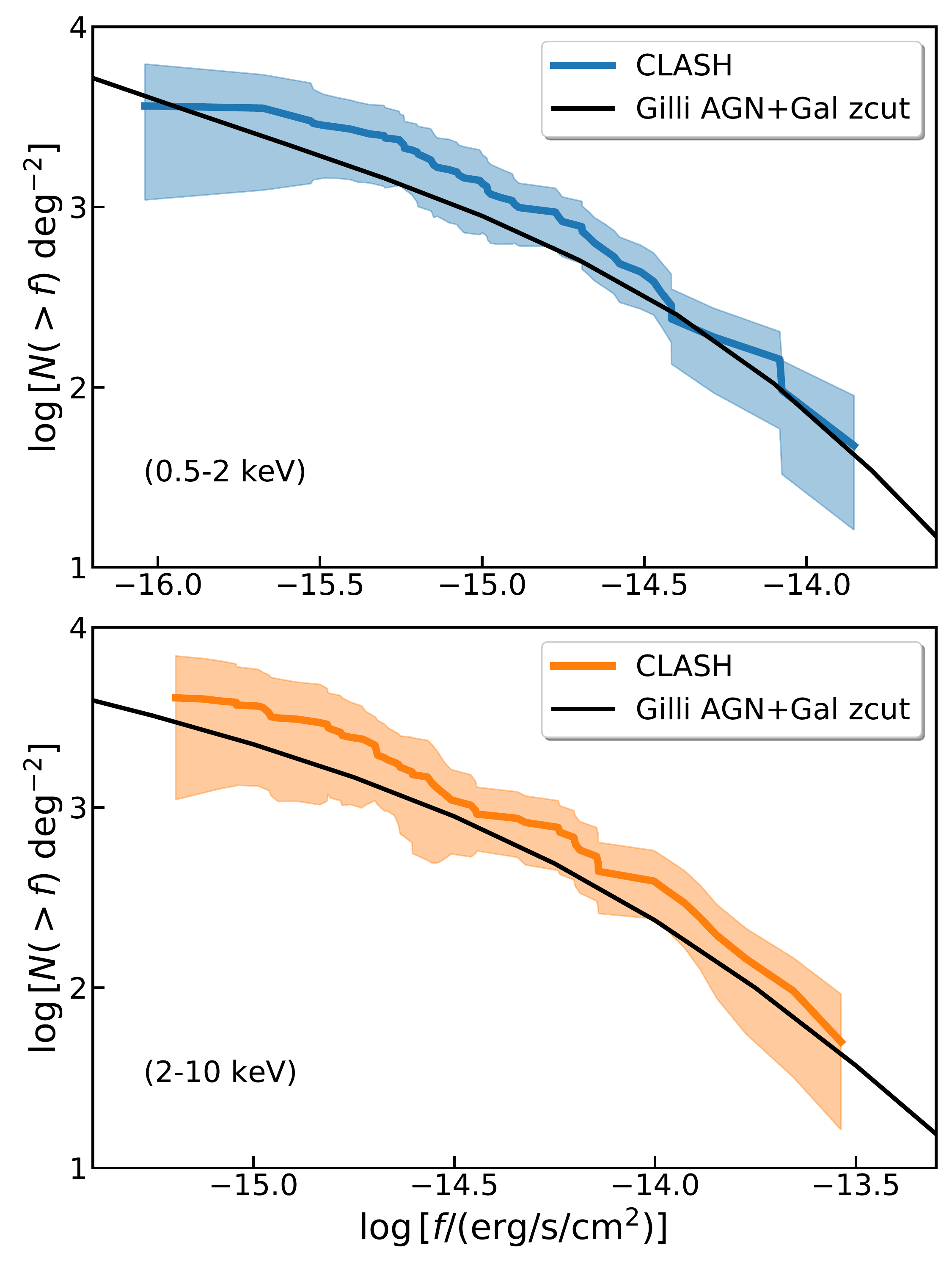}
\includegraphics[width=0.49\textwidth, trim=0 0 0 0, clip]{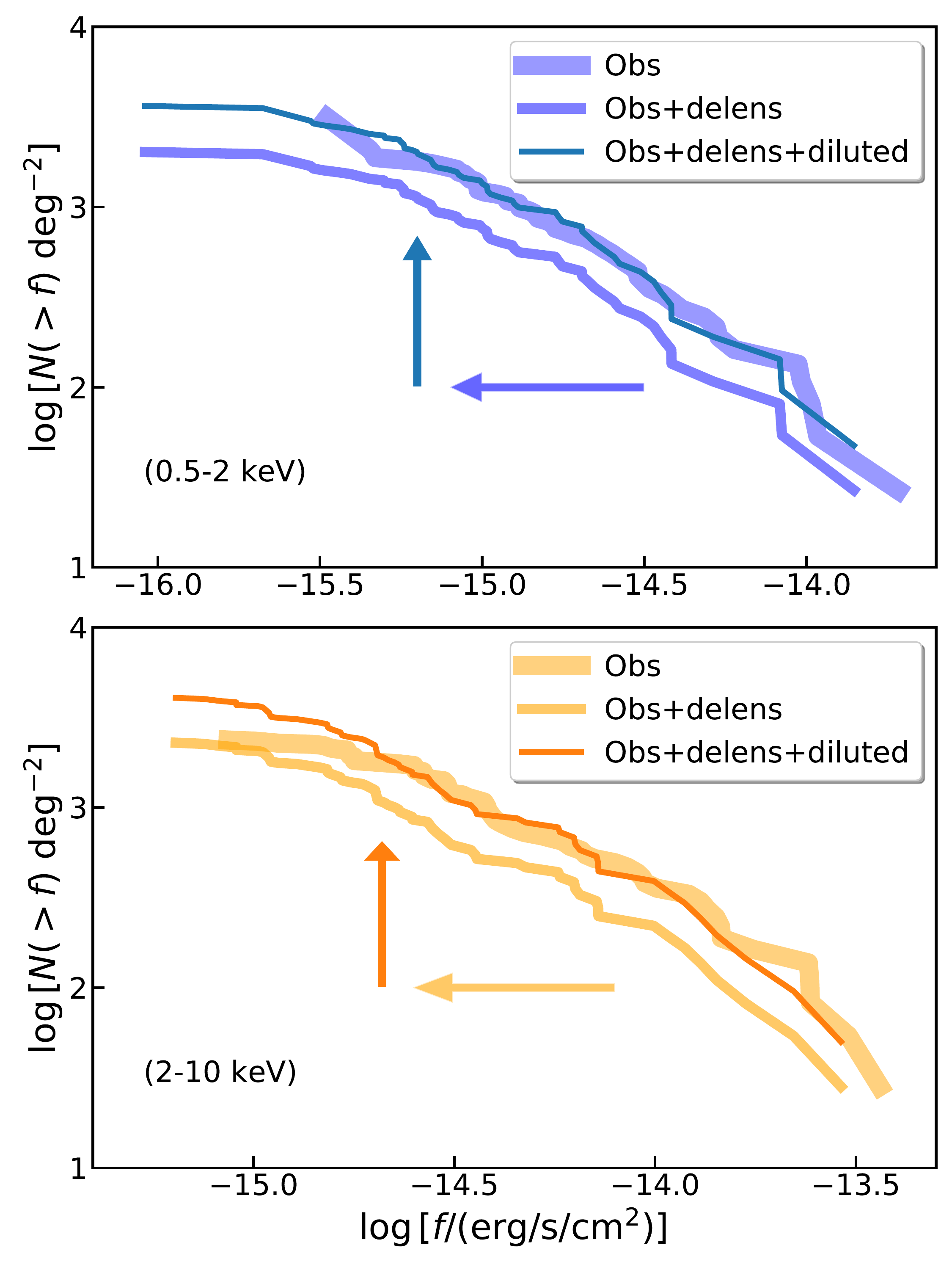}
\caption{Number counts of sources detected in the CLASH fields as a function of flux. {\sl Left panels}: Cumulative number counts of the lensed sources 
in the soft band (upper left) and hard band (lower left). The number counts are
compared with the prediction by the model of \citet{2007Gilli} after removing the 
foreground sources. {\sl Right panels}: Effects of the flux magnification and source dilution 
shown separately in the soft band (upper right) and hard band (lower right).}
\label{lognlogs_clash}
\end{center}
\end{figure*}

The delensed number counts in the soft and hard bands are shown in the left panels 
of Fig. \ref{lognlogs_clash}. The 1 $\sigma$ uncertainty includes the Poissonian 
error on the source number and the error on the flux, obtained as the quadratic 
sum of the Poissonian error on the net counts and the uncertainty in the conversion 
factor.  We note that the uncertainty in the cumulative number counts 
(shown as the shaded area) is dominated by the source statistics for bright sources, 
and as a consequence, it decreases with flux.  On the other hand, at the faint 
end it is dominated by the error on the fluxes, and therefore it increases with 
decreasing flux.  The uncertainty on the magnification, which would add another 
10\% to be summed in quadrature, has not been included here. 
When we compare our measurements to the expected log$N$-log$S$ from 
\citet{2007Gilli} plus the contribution from star forming galaxies as in \citet{2017Luo} 
(shown with a solid thin line), we find a good agreement within 
the 1$\sigma$ uncertainty, although the CLASH fields log$N$-log$S$ 
shows an excess of approximately 20\% where the discrepancy is 
maximum, at about $3\times 10^{-16}$ and $10^{-15}$ erg/s/cm$^2$ in the soft and hard bands, 
respectively, and an average excess of $\sim 10$\% in both bands over the entire flux range.
We note that differences of $\sim$ 10\%--20\% associated with cosmic variance have been 
observed in the deepest X-ray surveys such as CDFS, CDFN, and COSMOS. In a recent 
paper, the results from the XMM-RM survey have been compared with the CDFS and XMM-COSMOS
number counts, and a substantial agreement was found except for a discrepancy of a factor 1.5
between the CDFS and the other surveys \citep[see Figure 7 in][]{tengliu2020} in 
a relatively bright flux range.  In general, a thorough discussion of the 
fluctuations in the normalization of the log$N$-log$S$ due to cosmic variance
has not been performed in the literature so far, 
and it is expected to be superseded by a detailed account of the
distribution of AGN and star forming galaxies in redshift based on the increasing 
availability of complete multiwavelength surveys. Considering that the mild excess 
found in our measurements is well within the 1 $\sigma$ uncertainty and that
our source selection may have included a few foreground sources because we lack redshift 
information, we do not regard this excess as significant.
As a final point, we recall that
the delensed fluxes (for single sources and for the flux-limit maps) were obtained assuming a
representative redshift distribution derived from the complete CDFS survey.  The actual 
redshift distribution of our selected sources may be  different from the assumed one, 
causing a miscalibration of the average magnification effect.  To summarize, considering the assumptions 
we adopted, we find a reasonable agreement of our results with the expectations.

To better appreciate our results, we separately show the effect of flux magnification and 
source dilution as discussed in the optical band 
in \citet{1995Broadhurst} and in the X-ray band in \citet{1996Refregier}. 
In a very simplified but useful 
scheme, we can assume a power-law behavior, with $dN/df\propto S^{-\alpha-1}$, 
and therefore $N(>f) \propto f^{-\alpha}$.  We assume that we observe a reasonably large
region with constant magnification $\mu$, then we can express the combined effect of the flux magnification
and source dilution by writing the real cumulative number counts as a function of the observed ones as
$N(>f)={1\over \mu} \Big( {f_{\rm obs}\over \mu}\Big)^{-\alpha}= \mu^{\alpha-1}\times N_{\rm obs}(>f_{\rm obs})$. 
The delensed (intrinsic) and observed number counts would look similar 
when $\alpha \sim 1$.  This is the approximated value for the slope
in the considered flux range.  We show in the right panels of Fig. \ref{lognlogs_clash} 
the two effects at play 
in the soft and hard bands. The thickest line is the directly observed log$N$-log$S$ , obtained by 
plotting the cumulative number counts as a function of the observed flux and weighting by the 
corresponding solid angle in the lens plane. The medium-thick line is obtained by 
delensing the observed fluxes. This step is clearly roughly equivalent to a rigid shift toward 
the faint end.  Finally, the thin line is the actual log$N$-log$S$, obtained by plotting the observed 
cumulative number counts as a function of the delensed fluxes weighting by the corresponding 
solid angle in the source plane.  Again, this step is roughly equivalent to a rigid shift in 
the vertical direction.  As previously shown, because the slope in both bands is close to 
$\alpha =1$, the two effects almost cancel out, and the net results is a shift along the 
relation toward the faint end, therefore a higher depth than in the observed flux.  

The increased depth isthe most relevant effect we searched for because a significant 
shift toward the faint end would allow us to explore the deep X-ray sky through the 
lensing effect of the clusters.  Unfortunately, we note that the increase in depth is limited, 
and it is dominated by very few sources with high magnification or with low intrinsic flux.
The faintest delensed flux in our survey is 1.4 and 1.6 dex 
higher than the faint end of the CDFS in the soft and hard band, respectively. 
Therefore we find that the gain in sensitivity caused by the intervening clusters  
due to the combined effect of the small solid angle at high magnification and of the strong 
diffuse X-ray emission from the ICM is a factor of $\sim 2$ compared to blank fields 
with the same exposure.  The implications of our study for the future use of 
massive clusters as cosmic telescopes in X-ray extragalactic surveys is discussed 
in the next section.


\section{Perspective for future studies}

Our results show that the combination of low number statistics for X-ray sources, the 
small solid angle associated with high magnification, and the
overwhelming foreground due to the diffuse ICM emission significantly weaken 
the power of cosmic telescopes in the X-ray band.  In particular, out of 11 massive clusters
with exquisite optical data, we were able to identify only 3 X-ray sources that are
magnified by more than a factor of 10.  As a result, the delensed cumulative number counts 
of the X-ray lensed sources is far from approaching the flux limit of the deepest X-ray 
surveys: it reaches only $\sim 10^{-16}$ erg/s/cm$^{2}$ in the soft band and 
$\sim 5 \times  10^{-16}$ erg/s/cm$^{2}$ in the hard band.
The question now is whether we can develop a strategy on the basis of the
current results to probe the 
X-ray sky population at fainter fluxes using massive clusters as cosmic telescopes. 

On one hand, it is always convenient to determine whether highly magnified sources 
identified in the optical band are also X-ray emitters. 
The discovery of a bright, rare, high-redshift, highly magnified X-ray source, possibly 
with multiple images (as in the few known 
Einstein crosses associated with multiply lensed bright quasars) would be relevant not only 
to studying the spectral properties of a source that would otherwise
have been weak, but also for the possibility of monitoring its flux variations, and with an accurate
mass model of the lens, performing cosmological test as in the case of SN Refsdal \citep{2015Kelly,2020Grillo}. Intrinsically bright X-ray sources offer the advantage of being variable on 
short timescales, potentially offering a very efficient cosmological test.  
For this reason, it is always convenient to investigate the X-ray images of massive 
clusters that have been observed with deep optical data (the combination of HST and MUSE 
has been shown to be optimal for these studies) in search of such an object.  
However, this occurrence is expected to be very rare.  

\begin{figure}
\begin{center}
\includegraphics[width=0.49\textwidth, trim=55 25 55 30, clip]{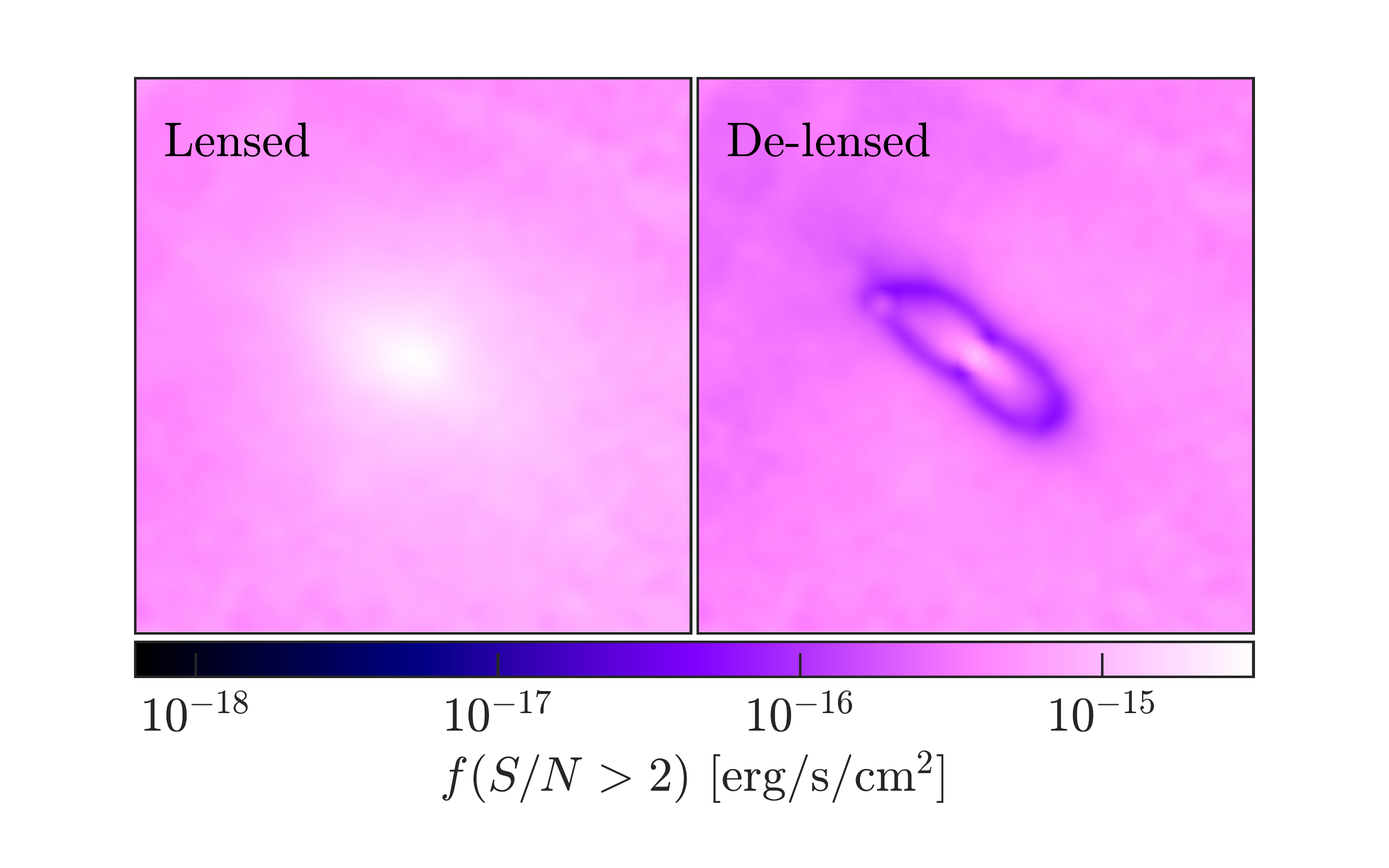}
\caption{Flux limit corresponding to the selection criterion $S/N>2$ in the soft band for the 
AS1063 field.  The observed flux limit is shown in the left panel, and the delensed flux
limit is shown in the right panel.  The images are 5$\arcmin$ across and are smoothed with an FWHM of 3$\arcsec$. }
\label{as1063_flux}
\end{center}
\end{figure}

\begin{figure}
\begin{center}
\includegraphics[width=0.49\textwidth, trim=50 10 50 0, clip]{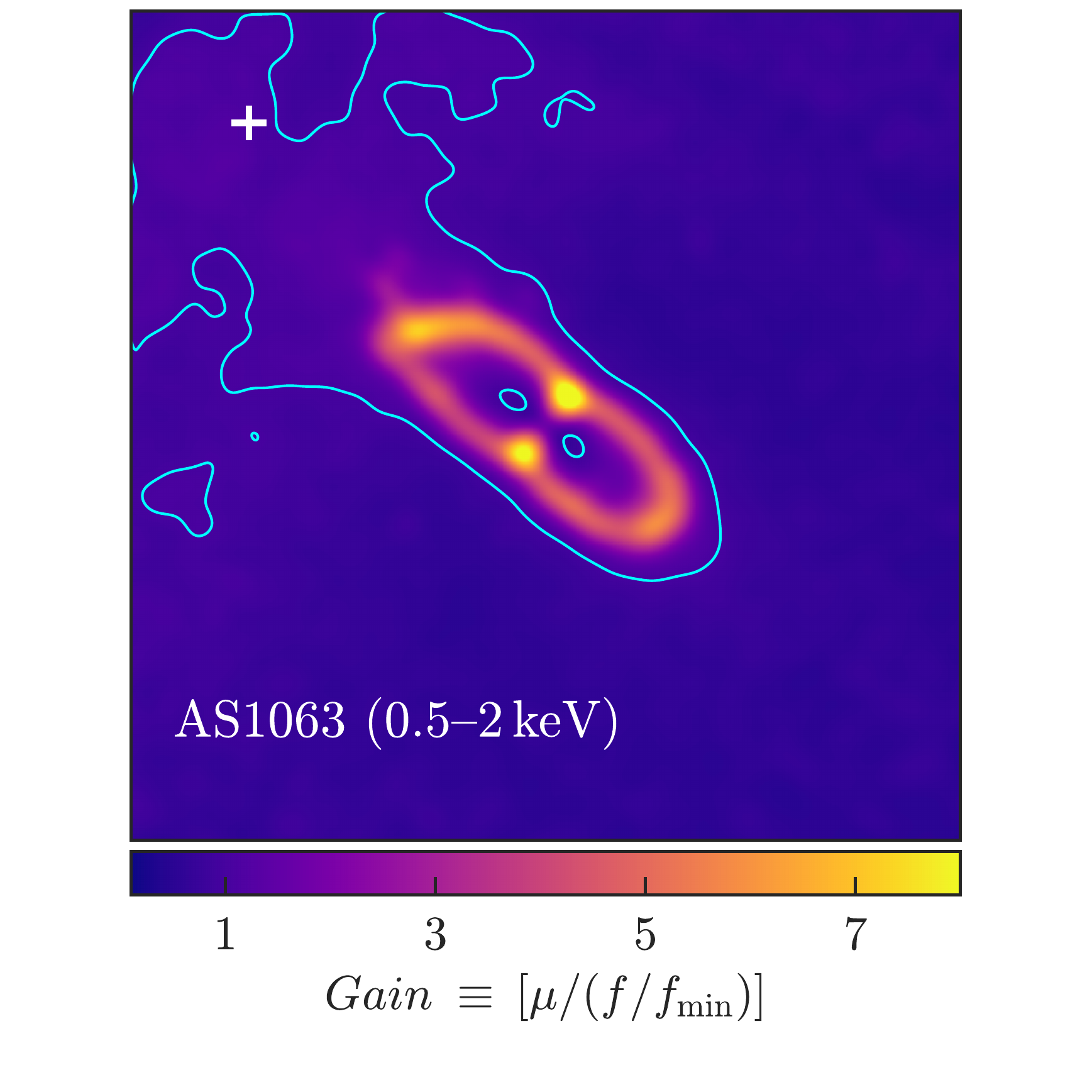}
\caption{$Gain$ map for the field of AS1063 in the soft band. The contours mark the transition
from the region with $Gain<1$ to that with $Gain >1$, where the magnification due to strong lensing 
exceeds the sensitivity loss due to the ICM emission and the PSF smearing. The white cross marks the position of the smallest PSF size. }
\label{as1063_gain}
\end{center}
\end{figure}

On the other hand, the systematic search for lensed sources to reach fainter 
X-ray sources has been shown to be difficult.  Clusters of galaxies are
not transparent in the X-ray, and the solid angle associated with high magnification 
typically is low.  Together with the low number density per solid angle of X-ray background 
sources, this implies that the trade-off between the 
probability of finding X-ray sources is very close to a critical line, and the capability of identifying 
unresolved sources in the midst of the bright ICM emission is quite unfavorable.  
This discrepance is visualized
in Fig. \ref{as1063_flux}, where we show the flux-limit maps in the field of AS1063
for the observed and the delensed fluxes.  Most the field 
reaches a delensed flux level slightly below the observed flux limit (about a factor of two), 
while in only a tiny region corresponding to the cluster critical lines does the 
magnification have a significant effect and overcomes the ICM bright emission. 

\begin{figure}
\begin{center}
\includegraphics[width=0.49\textwidth, trim=12 100 15 100,clip]{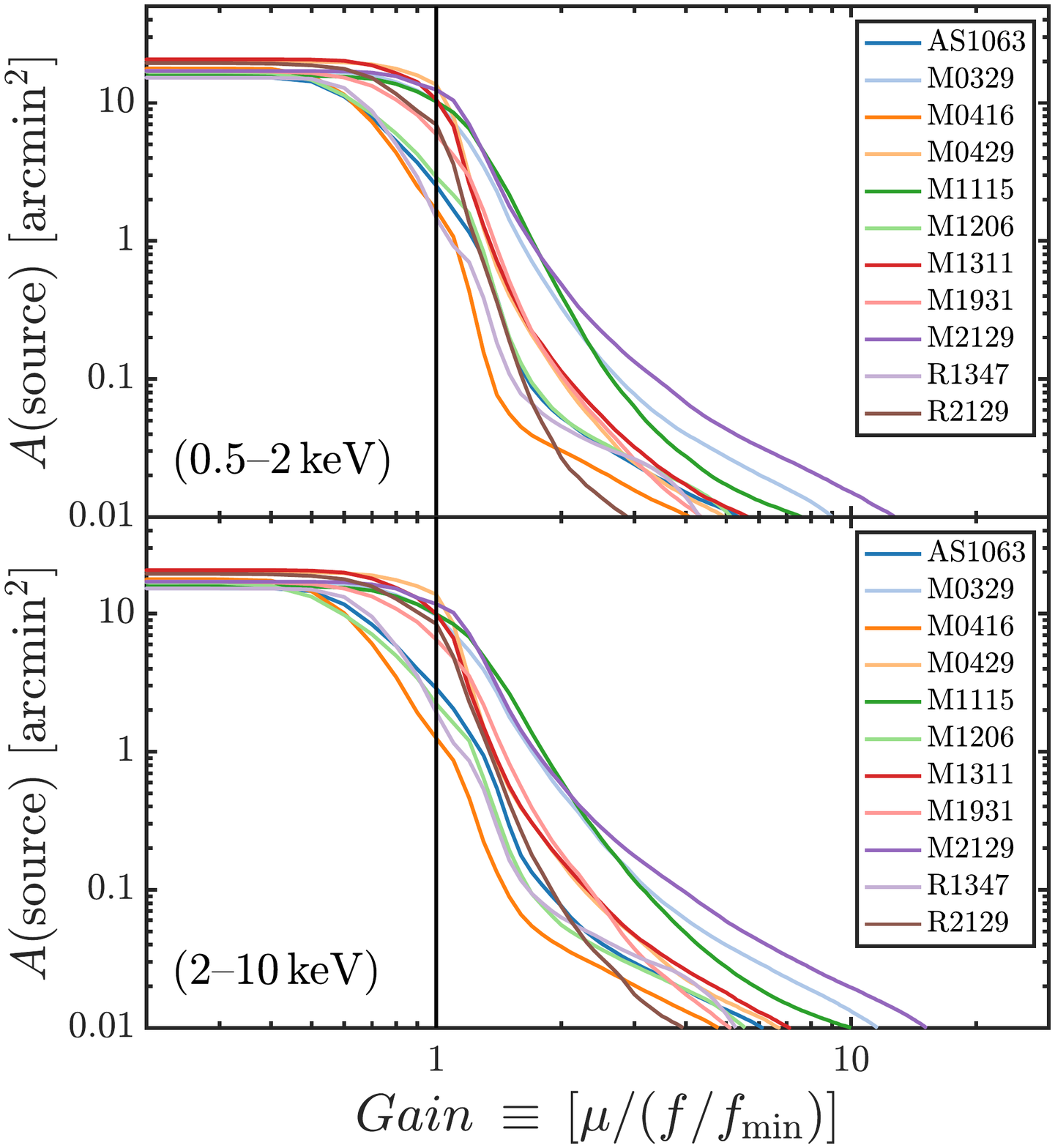}
\caption{Cumulative solid angle as a function of the $Gain$ in the soft
(upper panel) and hard (lower panel) bands.}
\label{gain}
\end{center}
\end{figure}

We can better illustrate and quantify this trade-off by computing the maximum magnification at each position in our images and the decrease in sensitivity with respect to the field.
The loss in sensitivity here is computed, consistently with our method, as the ratio of flux 
$S_{\rm lim}$ corresponding to a $S/N>2$ in a given position, divided by the minimum detectable flux in 
the same X-ray field $S_{\rm field}$, 
far from the ICM emission.  The latter value is a reasonable proxy for the 
flux limit that we would have in the same observation if the ICM emission were not present
(i.e., the sensitivity corresponding to the best PSF and the lowest 
background and foreground for a given observation).  We also convolved the map by the completeness
(this only applies to the {\tt wavdetect} detection algorithm and is therefore valid 
for a standard approach of source detection). The two effects compete directly, so that the 
comparison of the values of  $\mu_{\rm max}$ and the ratio $S_{\rm lim}/S_{\rm field}$ indicates the 
regime in which the magnification exceeds the loss in sensitivity.  
In Fig. \ref{as1063_gain} we show the $Gain$, defined as the ratio 
$\mu_{\rm max} \, / \, (S_{\rm lim}/S_{\rm field})$, in the soft band in the field of AS1063.
We find that the solid angle where the magnification exceeds the 
loss in sensitivity ($Gain>1$) is limited to a very small area, 
is dominated by values slightly higher than unity, and clearly peaks 
where the critical lines are.  This visualizes the modest efficiency in using massive clusters as
cosmic telescopes in the X-ray band.
To be more quantitative, in Fig. \ref{gain} we plot the cumulative solid angle in each field 
where the gain is higher than a given value. The solid angle drops rapidly between 
$Gain=1$ and $Gain=2$. The solid angle associated with a high $Gain$ amounts to a few percent of 
the entire field of view.

The next question then is how many {\sl Chandra} fields with massive clusters, modeled on 
the CLASH sample, are required in order to investigate the X-ray sky down to the CDFS limits.
We can estimate this by requiring a 
number of sources below a given delensed flux limit comparable to the 
same flux range at the faint end of the CDFS log$N$-log$S$.
For example, the CDFS includes about 400 and 160 sources below $5\times 10^{-17}$ and 
$2\times 10^{-16}$ erg/s/cm$^{2}$ in the soft and hard band, respectively
(see https://heasarc.gsfc.nasa.gov/W3Browse/all/chandfs7ms.html).  
We considered three of the fileds with the longest exposure times, 
 M0416, M1206 and AS1063, with 320, 200 and 125 ks, respectively.
Then we computed the total number of sources expected from the best fit of the 
CDFS number counts, using Equation (3) and Table 8 in \citet{2017Luo}.
The total number of sources expected in a survey made of $N_{\rm f}$ fields is
\begin{equation}
n_{\rm s} = N_{\rm f}\times \int^{S_{\rm up}}_{S_{\rm low}} \Big({{dN}\over{dS}}^{\rm AGN} +  {{dN}\over{dS}}^{\rm Gal} \Big)
\times \Omega(S) dS \, ,
\end{equation}
where $\Omega(S)$ is the sky coverage in the source plane we computed in each field.  Here 
the contributions from AGN and star-forming galaxies are summed to include all the unresolved sources. 
We also considered a 20\% higher normalization to be consistent with the predictions by the model of 
\citet{2007Gilli}.  To match the CDFS in the soft band, we need
about 9, 28, and 57 $\times 10^3$ fields comparable to  M0416, M1206, and AS1063, respectively.
In the hard band, this number reduces to 3, 7, and 16 $\times 10^3$ fields.  
We are aware that this estimate is adequate for the {\sl Chandra} performances comparable to 
the actual data, and therefore should not be applied to future {\sl Chandra} observations, 
for which the sensitivity in the soft band is unfortunately dramatically lower than 
in the data we analyzed.
Therefore it is not feasible to reach a depth comparable to the CDFS by adding all 
the fields of massive clusters, which in the {\sl Chandra} archive and for 
the considered exposure times are $\sim 100-200$. 
To be fully successful, this program should also rely on 
optical and spectroscopic data in order to build the magnification maps of the clusters, 
or on a massive dataset in the optical data. In this respect, the Euclid mission
\footnote{See https://www.esa.int/Science\_Exploration/Space\_Science/Euclid\_overview.}
will provide a key contribution.

\begin{figure}
\begin{center}
\includegraphics[width=0.49\textwidth, trim=12 100 15 100,clip]{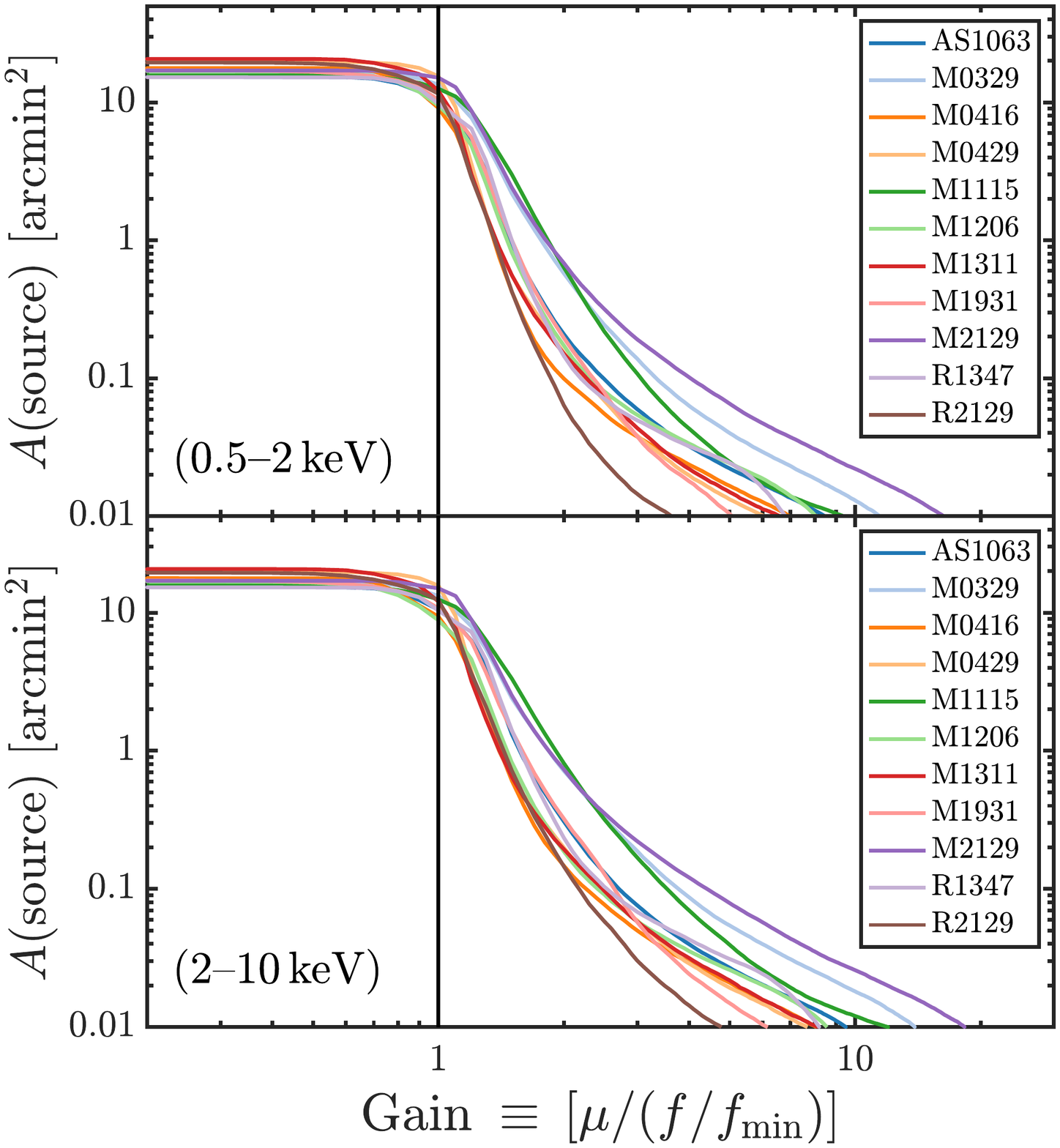}
\caption{Same as in Fig. \ref{gain}, but assuming a constant $R_{\rm ext} = 1\arcsec $ 
everywhere, thus significantly reducing the background effect.}
\label{gain_1arcsec}
\end{center}
\end{figure}

The situation will be similar for future X-ray missions with high angular resolution on a 
large field of view, such as AXIS \citep[see][]{2020Marchesi} or Lynx \citep{2018LynxTeam}.  As an 
exercise, we recomputed Fig. \ref{gain}, but assuming
a flat PSF and a constant $R_{\rm ext}= 1$ arcsec.  This would decrease the
background effect in the most interesting regions, close to the critical lines and therefore embedded 
in the ICM emission.  In other words, $S_{\rm lim}$ would be closer to the
lowest flux limit achievable in the same field $S_{\rm field}$.  The result, shown in Fig. 
\ref{gain_1arcsec}, is that the $Gain$ does not increase significantly. However, this 
simple exercise concerns only the $S/N$ threshold, and does not include the effect of the
improved PSF on the source detection, which has an additional positive effect.  To summarize, 
we expect that the efficiency of source detection in the midst of the ICM would improve 
by a significant factor with the next generation of high-resolution X-ray telescopes, 
but hardly by an order of magnitude.

Clearly, our assessment of the limited effectiveness of strong lensing in the deep exploration of the X-ray sky does not
exclude the possibility that a few highly magnified X-ray source are found, particularly when applied to high-redshift star-forming galaxies or
multiply-imaged bright AGN.  Another possibility is to explore a different mass scale, including
groups (with halo masses $M_{200} < 10^{14} M_\odot$) and galaxies.  The intensity of the
ICM rapidly decreases with mass because the bremsstrahlung emission scales
with the square of the electron density in the (almost) fully ionized ICM.  
While we focused our attention first on massive clusters because their
magnification is far higher and the extent of the critical lines is larger, 
the lower mass range offers several advantages: The magnification
effect becomes more favorable as a result of the weaker foreground and the number density of groups and low-mass clusters is much higher than in massive clusters.
Roughly speaking, the foreground scales with the square of the
mass, but the magnification scales only linearly with mass, so that there will be a sweet spot
where the search for X-ray lensed sources in {\sl Chandra} archival data is more efficient.  
Unfortunately,  {\sl Chandra} archival observations of groups are somewhat limited because most of the
telescope time has been used, understandably, to study massive clusters.  
This approach will be explored in a future paper.


\section{Conclusions}

Strong gravitational lensing by massive clusters has been shown to be very successful
in revealing the properties of very distant, highly magnified galaxies 
in the optical and IR bands, but it was never explored in the X-ray band because of the 
intense diffuse emission of the ICM and the low density of X-ray sources. 
In this work, we performed
a systematic search of X-ray lensed sources in the 
{\sl Chandra} observations of massive clusters for the first time, employing the exquisite optical (imaging 
and spectroscopy) dataset of 11 CLASH targets, as well as our high-precision lens models. Our aim was to identify X-ray emission from 
any unresolved X-ray source magnified by
the foreground massive cluster with a readily identified optical counterpart or
blindly identified in the X-ray images. Our final goal was to investigate 
the efficiency of massive galaxy clusters in acting as cosmic telescopes in the X-ray band.
Our results are summarized below.

We identified nine X-ray emitting sources corresponding to optical counterparts 
of lensed sources identified in HST and MUSE data. Three of these 
(one of which double) are consistent
with being powered by star formation, and the other six have X-ray luminosities in the 
Seyfert range.

We identified 66 X-ray lensed source candidates in the X-ray soft- and 
hard-band images with untargeted detection using {\tt wavdetect} and with the requirement
of having a $S/N>2$ in any of the two bands.

The delensed cumulative number counts are consistent within the uncertainties 
with the log$N$-log$S$  predicted by \citet{2007Gilli} after correcting for 
the sources at $z<z_{\rm cl}+0.1$.  We still find a marginal excess of 20\% at most
(with an average of 10\% on the full flux range probed) in the delensed 
log$N$-log$S$, where the discrepancy is 
highest at about $3\times 10^{-16}$ and $10^{-15}$ erg/s/cm$^2$ in the soft and hard bands, 
respectively.  Nevertheless, this discrepancy is smaller than 
the 1 $\sigma$ combined uncertainty associated with the flux scale and the count
statistics.

The combination of magnification and sensitivity loss in the X-ray band
due to the ICM diffuse emission results in a limited efficiency of massive 
clusters as cosmic telescope. The exploration of the deep X-ray sky at the depth level of the CDFS is not feasible
given the current size and quality of the {\sl Chandra} archive.

Overall, we expect that the detection and characterization of strongly lensed X-ray sources
in the {\sl Chandra} fields of massive clusters is not expected to provide significant
improvements with respect to currently available deep X-ray surveys.
Nevertheless, we plan to extend this work to less massive halos, where the ICM diffuse emission 
is much weaker, employing the high-resolution observations of groups and galaxies in the {\sl Chandra} 
archive, and to simultaneously perform a blind search of AGN-galaxy lensed systems in extragalactic fields.
The rare occurrence of strongly magnified sources may be interesting for cosmological tests
based on variability of lensed sources with multiple counterparts, and it may constitute an important
science case for future high-resolution X-ray telescopes

\begin{acknowledgements}
We thank the anonymous referee for his/her constructive comments that helped improve the paper. We thank Bin Luo for providing the data of the sky coverage in the CDFS.  
A.L., R.G. and P.T. acknowledge financial contribution from the agreement ASI-INAF n.2017-14-H.0.
PB acknowledges financial support from ASI through the agreement ASI-INAF n. 2018-29-HH.0.
GBC thanks the Max Planck Society for support through the Max Planck Research Group for S. H. Suyu and the academic support from the German Centre for Cosmological Lensing.
We acknowledge support from the PRIN MIUR 2017 WSCC32.
\end{acknowledgements}

\bibliography{references_Clusters_PT}

\appendix

\section{X-ray lensed source candidates with multiple counterparts}

We have identified
five multiply lensed sources (labeled ``(m)'' in Table \ref{phot_src}) as X-ray emitter candidates in the optically lensed sources.  
Because in each case only one counterpart has an X-ray signal, we can therefore
check whether the detection is consistent with the aperture photometry at the position of 
the other optical counterparts
(assuming that variability does not play a role for simplicity).  
Source AS1063-46 is detected only in the soft band with $17.5\pm 8.6$ net counts and 
a magnification of $\sim 30$, and it has two other counterparts with a magnification of 3.4 and 17.8.  
The first counterpart would be clearly too faint to be visible, but the second is expected to have 
$\sim 10$ net counts, which does not significantly contradict with the aperture photometry at its
position, which is $2.1 \pm 6.0$ net counts. 
AS1063-95, detected in the soft band with photometry $9.7\pm 5.1$, has two counterparts with 
similar magnification, whose photometry provides
$4.3 \pm 6.9$ and $22.5\pm 16.4$ net counts in the soft band, consistent with the photometry of the 
candidate X-ray emitters. 
Source M0416-221 is detected with $7.3\pm 4.1$ net counts in the soft band and
has two counterparts with photometry $-1.0\pm 9.8$ and $3.0\pm 4.4,$ where 7 and 10 net 
counts were expected, respectively. The nondetection of the two other counterparts
does not create a significant discrepancy with our X-ray detection in this case either.  For 
sources AS1063-46, AS1063-95, and M0416-221, our X-ray detection is therefore consistent with no detection 
at the position of the other counterparts.  

Source M0416-208 is detected in the soft band with $15.6\pm 5.5$ net counts. It has two 
counterparts that are expected to measure 30 and 14 counts in the same band. However, 
the photometry on the two counterparts gives $-2.2\pm 7.4$ and $-0.6\pm 8.4$.
The situation is the same for M0416-241, for which we have a detection with $15.3\pm 7.8$ 
net counts in the hard band.  There is another counterpart of this source, with higher magnification, 
for which about 30 net counts are expected, while its photometry is $2.0\pm 4.7$, which is 
therefore inconsistent at more than $3\sigma$.  We conclude that M0416-208  and M0416-241
should be removed from our candidate list.

\end{document}